\documentclass[prb,twocolumn,lengthcheck,superscriptaddress]{revtex4-1}

\usepackage{graphicx,color}
\usepackage{amsmath,amssymb}
\usepackage{textcomp}
\usepackage{braket,slashed}
\usepackage{hyperref}
\usepackage[english]{babel}

\usepackage{bm}

\usepackage[mathscr]{euscript}
\usepackage{dsfont}

\newcommand{\operator}[1]{\ensuremath{\hat{#1}}}
\newcommand{\voperator}[1]{\ensuremath{\mathds{#1}}}
\newcommand{\manifold}[1]{\ensuremath{\mathcal{#1}}}

\newcommand{\vectorspace}[1]{\ensuremath{\mathbb{#1}}}

\DeclareMathOperator{\order}{\mathscr{O}}
\DeclareMathOperator{\tr}{tr}

\DeclareMathOperator{\End}{\mathbb{L}}

\newcommand{\one}{\mathds{1}}

\renewcommand{\d}{\ensuremath{\mathrm{d}}}
\newcommand{\ic}{\ensuremath{\mathrm{i}}}
\newcommand{\ec}{\ensuremath{\mathrm{e}}}
\newcommand{\defis}{\ensuremath{\triangleq}}

\newcommand{\ham}{\ensuremath{\operator{H}}}

\newcommand{\rket}[1]{\ensuremath{|#1)}}
\newcommand{\rbra}[1]{\ensuremath{(#1|}}
\newcommand{\rbraket}[1]{\ensuremath{(#1)}}

\newcommand{\hilbert}{\ensuremath{\vectorspace{H}}}
\newcommand{\Tplane}{\ensuremath{\vectorspace{T}}}
\newcommand{\varM}{\ensuremath{\manifold{M}}}

\begin{document}
\title{Post-Matrix Product State Methods: To tangent space and beyond}

\author{Jutho Haegeman}
\affiliation{Department of Physics and Astronomy, University of Ghent, Krijgslaan 281 S9, B-9000 Ghent, Belgium}
\author{Tobias J.~Osborne}
\affiliation{Leibniz Universit\"{a}t Hannover, Institute of Theoretical Physics, Appelstrasse 2, D-30167 Hannover, Germany}
\author{Frank Verstraete}
\affiliation{Department of Physics and Astronomy, University of Ghent, Krijgslaan 281 S9, B-9000 Ghent, Belgium}
\affiliation{Faculty of Physics, University of Vienna, Boltzmanngasse 5, A-1090 Wien, Austria}

\begin{abstract}
We develop in full detail the formalism of tangent states to the manifold of matrix product states, and show how they naturally appear in studying time-evolution, excitations and spectral functions. We focus on the case of systems with translation invariance in the thermodynamic limit, where momentum is a well defined quantum number. We present some new illustrative results and discuss analogous constructions for other variational classes. We also discuss generalizations and extensions beyond the tangent space, and give a general outlook towards post matrix product methods.\end{abstract}

\maketitle

\tableofcontents

\section{Introduction}
The last two decades have witnessed a tremendous cross-fertilization between different branches of theoretical physics, including condensed matter physics, quantum information theory and renormalization group theory, with even some ingredients of quantum gravity added to the picture. One of the major breakthroughs along this line of research was the development of the density matrix renormalization group (DMRG) by Steve White\cite{1992PhRvL..69.2863W,1993PhRvB..4810345W}, which has quickly evolved to become the standard numerical tool for finding ground states of one-dimensional quantum spin systems. It was later realized that DMRG corresponds to a variational method that optimizes over a class of states which is known as matrix product states (MPS)\cite{1995PhRvL..75.3537O,1997PhRvB..55.2164R}. Insight from quantum information theory resulted in the development of many extensions of the original DMRG method. For example, more general ansatzes known as tensor network states have been formulated for systems in higher dimensions\cite{2004cond.mat..7066V,2008AdPhy..57..143V,2009JPhA...42X4004C}. Another important development was the formulation of the so-called time-evolving block decimation (TEBD) by Guifre Vidal\cite{2004PhRvL..93d0502V}, for studying real-time evolution of one-dimensional quantum lattice Hamiltonians using the formalism of MPS. 

In condensed matter physics and quantum field theory, most interesting quantum systems, aside from quantum dots, are of macroscopic size in at least one spatial dimension, and the main interest is often in the bulk properties of these systems, without any boundary or finite size influences\footnote{Although the topic of edge excitations is crucial for systems with topological order}. Here too, the formalism of MPS has a nice advantage over alternative numerical approximation methods, as they can be formulated directly in the thermodynamic limit and still depend on only a finite number of parameters when the system under study exhibits translation invariance. These states are now known as uniform MPS (uMPS), but had in fact already been formulated before the development of DMRG under the name finitely correlated states\cite{1992CMaPh.144..443F}, which served as a generalization of the valence bond construction for the spin $1$ model by Affleck, Kennedy, Lieb and Tasaki\cite{1987PhRvL..59..799A,1988CMaPh.115..477A}. The TEBD with imaginary time evolution was among the first methods that was actually able to efficiently optimize over these variational parameters\cite{2007PhRvL..98g0201V}, although other algorithms have now been formulated\cite{2008arXiv0804.2509M,2011PhRvL.107g0601H}. 

The low-energy physics of the bulk is not determined by the ground state alone, but requires knowledge of the spectrum of bulk excitations. These excitations are almost always present in the lab, either due to finite temperature, because the system is being perturbed by probes in spectroscopy experiments or because parameters of the system are being quenched. Recently, new algorithms were formulated for studying time evolution\cite{2011PhRvL.107g0601H} and for describing the elementary excitations\cite{2012PhRvB..85c5130P,2012PhRvB..85j0408H} that depend strongly on the concept of tangent vectors to the original variational manifold of uMPS. Analogous results have also been formulated for generic MPS on finite lattices\cite{2011arXiv1103.2155K,2012PhRvL.109z7203K}. Indeed, it was recently understood that the set of MPS (generic and uniform) with a given bond dimension constitutes a smooth manifold $\varM$ embedded in the full Hilbert space $\hilbert$, where the MPS parameterization can be recognized a principal fiber bundle\cite{2012arXiv1210.7710H}. More precisely, the set of physical states constitutes a K\"{a}hler manifold, which expresses the fact that the complex structure is compatible with the Riemannian geometry that is obtained when inducing the standard Euclidean metric of $\hilbert$ onto $\varM$.\footnote{Since the metric also defines a closed 2-form, the manifold additionally has a symplectic structure.} This allows us to identify the tangent space of $\varM$ with a complex subspace of $\hilbert$, which naturally inherits many properties of $\varM$. Similar constructions were simultaneously developed in the field of numerical analysis, where real-valued tensor networks are also becoming increasingly popular\cite{Holtz:2011cu,Uschmajew:2012ue}. Whereas Ref.~\cite{2012arXiv1210.7710H} focussed on a rigorous study of the differential geometry of $\varM_{\mathrm{MPS}}$ and its tangent space, the current manuscript focuses on the physical relevance of these tangent states. 

In Section~\ref{s:umpsdef}, we summarize the necessary ingredients from Ref.~\onlinecite{2012arXiv1210.7710H} that will be used throughout the remainder of this manuscript. Section~\ref{s:TDVP} explores in full detail the \emph{time-dependent variational principle} (TDVP) which was introduced in Ref.~\onlinecite{2011PhRvL.107g0601H}. The TDVP is a general prescription for the optimal way to approximate a time-evolving quantum state within a given variational manifold. As such, it does not yet take us outside of the manifold of uMPS. Section~\ref{s:excitations} then explains why the tangent space itself is also useful as a variational subspace for the study of \emph{elementary bulk excitations} of a system, and relates this to linear response theory. We also show how the tangent space can be used to get a quick estimate of spectral functions. The algorithms presented in this section should thus be considered as \emph{post-MPS} methods that allow us extract dynamical information from new variational subspaces that are beyond ---but based on--- the original MPS manifold. We then illustrate the power of these methods by considering some explicit examples in Section~\ref{s:examples}. We briefly compare our methods to analogous constructions that were developed the context of other variational classes, in particular mean field theory, in Section~\ref{s:analogies}. Finally, Section~\ref{s:extensions} introduces extensions and generalizations that even takes us beyond the MPS tangent space and gives an outlook on how this will eventually result in the development of a new \emph{Fock space} on top of the MPS vacuum. 

\section{Manifold of uniform matrix product states in the thermodynamic limit}
\label{s:umpsdef}
This section summarizes the definition and key properties of the variational manifold of uMPS, which were derived in full detail in Ref.~\onlinecite{2012arXiv1210.7710H}. Readers familiar with that paper can safely skip this section. A proper treatment of quantum states in the thermodynamic limit requires a description in terms of $C^{\ast}$-algebras, which is how finitely correlated states were originally constructed \cite{1992CMaPh.144..443F,1991JPhA...24L.185F}. We employ the results from these papers, but adopt a physics-style notation which is to be understood as a limiting procedure of finite lattices on which our definitions are meaningful. Even then this thermodynamic limit requires careful attention due to intricate effects such as infrared divergences and the orthogonality catastrophe\cite{Anderson:1967aa}. We will emphasize the necessary steps required to avoid these intricacies and obtain consistent results.

\subsection{Definition and properties of the manifold}
\label{ss:manifold}
Consider a one-dimensional lattice $\mathcal{L}\subset \mathbb{Z}$ with $\lvert\mathcal{L}\rvert=2N+1$ sites labeled by the integer $n\in\mathcal{L}=\{-N,\ldots,N\}$. Every site $n$ contains a $d$-dimensional quantum variable, so that local Hilbert space $\hilbert_{n}\cong \mathbb{C}^{d}$ is spanned by a basis $\{\ket{s_{n}}\mid s_{n}=1,\ldots,d\}$. The total Hilbert space is given by $\hilbert=\bigotimes_{n=-N}^{N} \hilbert_{n}$ and is spanned by the product basis
\begin{equation}
\ket{\bm{s}}=\ket{s_{-N}}_{-N}\otimes\cdots \otimes\ket{s_{0}}_{0}\otimes \cdots \otimes \ket{s_{N}}_{N}.
\end{equation}
In the thermodynamic limit $N\to\infty$, a uniform matrix product state $\ket{\Psi(A)}\in \hilbert$ is defined as
\begin{equation}
\ket{\Psi(A)}\defis\sum_{\{s_n\}=1}^{d} \tr[V\cdots A^{s_{-1}}A^{s_{0}}A^{s_{+1}}\cdots ] \ket{\bm{s}},\label{eq:defumps}
\end{equation}
and is parameterized by a tensor $A\in\mathbb{A}\cong\mathbb{C}^{D\times d\times D}$, or, in an alternative interpretation, by a set of complex $D\times D$ matrices $A^s$, for $s=1,\ldots,d$. The matrix dimension $D$ is known as the bond dimension. The $D\times D$ matrix $V$ encodes the boundary conditions, namely $V=\openone$ corresponds to periodic boundary conditions whereas $V$ should be a rank $1$ object (\textit{i.e.}~$V=\bm{v}_{\mathrm{R}}\bm{v}_{\mathrm{L}}^{\dagger}$ with  $\bm{v}_{\mathrm{L},\mathrm{R}}\in\mathbb{C}^{D}$) for a system with open boundary conditions. Based on the results from Ref.~\onlinecite{1992CMaPh.144..443F,1991JPhA...24L.185F}, it was shown in Ref.~\onlinecite{2012arXiv1210.7710H} that the expectation value of local observables is well defined and independent of the boundary matrix $V$ when $A$ is an element from the open subset $\mathcal{A}\subset\mathbb{A}$ which is known as the set of \emph{injective} MPS or of pure finitely correlated states. Equivalently, the set $\mathcal{A}$ corresponds to all tensors $A$ for which the transfer matrix $\voperator{E}=\sum_{s=1}^{d} A^{s} \otimes \overline{A}^s$ has a single eigenvalue $\omega^{(1)}=\rho(\voperator{E})$ with $\rho(\voperator{E})$ the spectral radius of $\voperator{E}$, whereas all other eigenvalues $\omega^{(i)}$, $i=2,\ldots,D^2$ satisfy $\lvert\omega^{(i)}\rvert< \rho(\voperator{E})$. Furthermore, the left and right eigenvectors $\rbra{l}$ and $\rket{r}$ corresponding to $\omega^{(1)}$, when written as positive semidefinite Hermitian matrices, should have full rank, \textit{i.e.}~they should be strictly positive definite. To obtain a normalizable state, the tensor $A\in\mathcal{A}$ should then be 'renormalized' as $A/\sqrt{\omega^{(1)}}$ so that the spectral radius of the transfer matrix becomes $1$.

Let now $\varM\subset\hilbert$ denote the set of states $\{\ket{\Psi(A)}|A\in\manifold{A}\}$. The state $\ket{\Psi(A)}$ is invariant under a reparameterization $A\leftarrow A_G$ where $A_G^s=G^{-1} A^s G$ for any invertible matrix $G\in\mathsf{GL}(D,\mathbb{C})$. This invariance is known as gauge invariance, and the map $(A,G)\to A_G$ represents the (right) group action of the gauge group $\mathsf{G}\cong\mathsf{PGL}(D,\mathbb{C})$, where we had to define the gauge group $\mathsf{G}$ as the projective linear group $\mathsf{PGL}(D,\mathbb{C})$ obtained by taking the quotient of $\mathsf{GL}(D,\mathbb{C})$ with its center subgroup $\mathsf{GL}(1,\mathbb{C})$ of matrices $G$ which are proportional to the unit matrix, since these choices have the trivial effect $A_G=A$. By restricting to the open subset $\manifold{A}\subset\mathbb{A}$, one can then show that the group action is free and proper. These properties express that the group action is sufficiently nice to obtain a smooth quotient space $\mathcal{A}/\mathsf{G}$. The injectivity property of the MPS ensure that this quotient space is diffeomorphic (and in fact biholomorphic) to the set of states $\varM$, thus also turning the latter one into a smooth (complex) manifold. The MPS representation that maps the tensor $A\in\manifold{A}$ to the physical state $\ket{\Psi(A)}\in\varM$ and exhibits invariance under the action of $\mathsf{G}$ can thus be given the structure of a principal fiber bundle, which is a useful identification when also studying the tangent space of $\manifold{A}$ and $\varM$ in the next subsection.

We now introduce some notations that are used throughout the remainder of this manuscript. Since we always assume that $A\in\manifold{A}$, the transfer matrix $\voperator{E}=\voperator{E}^{A}_{A}$ has a unique eigenvalue $\omega^{(1)}$ that can scaled to be exactly one with corresponding left and right eigenvectors $\rbra{l}$ and $\rket{r}$ corresponding to strictly positive definite matrices $l$ and $r$ that we assume to be normalized as $\rbraket{l|r}=\tr[lr]=1$. All other eigenvalues $\omega^{(k)}$, $k>1$ lie strictly within the unit circle. We also define $\voperator{S}=\rket{r}\rbra{l}$ as a projector onto the eigenspace of eigenvalue $1$, and its complement $\voperator{Q}=\voperator{\one}-\voperator{S}$. Note that the right action of the transfer matrix on a vector $\rket{x}$ can also be encoded as a completely positive map
\begin{equation}
\mathscr{E}:\mathbb{C}^{D}\to\mathbb{C}^{D}:x\mapsto \mathscr{E}(x)=\sum_{s=1}^{d} A^{s} x A^{s\dagger}.\label{eq:defEpsilon}
\end{equation}
The dual map
\begin{equation}
\widetilde{\mathscr{E}}:\mathbb{C}^{D}\to\mathbb{C}^{D}:y\mapsto \widetilde{\mathscr{E}}(y)=\sum_{s=1}^{d}  A^{s\dagger}yA^{s}\label{eq:defEpsilontilde}
\end{equation}
encodes the left action on a vector $\rbra{y}$. The actions of these maps can be computed with computing time that scales $\order(D^{3})$, so that $l$ and $r$ (eigenvectors corresponding to the largest eigenvalue) can efficiently be computed using an iterative eigensolver. While we do not assume that the tensor $A$ satisfies any gauge fixing condition in any formula in this manuscript, it is often convenient to use the left ($l=\one$, $\sum_{s} A^{s\dagger} A^s=\one$) or right ($r=\one$, $\sum_{s} A^sA^{s\dagger} =\one$) gauge fixing condition in numerical implementations.

Given a set of local operators $\operator{O}^\alpha$, we can use these definitions to compute the 2-point connected correlation function as 
\begin{equation}
\begin{split}
\Gamma^{(\alpha,\beta)}(n)&=\rbraket{l|\voperator{E}_{O^{\alpha}}\voperator{E}^{n-1}\voperator{E}_{O^{\beta}}|r}-\rbraket{l|\voperator{E}_{O^{\alpha}}|r}\rbraket{l|\voperator{E}_{O^{\beta}}|r}\\
&=\rbraket{l|\voperator{E}_{O^{\alpha}}\voperator{Q}\big(\voperator{Q}\voperator{E}\voperator{Q}\big)^{n-1}\voperator{Q}\voperator{E}_{O^{\beta}}|r}.
\end{split}
\end{equation}
where we have used
\begin{displaymath}
\voperator{E}^{n}-\voperator{S}=\voperator{Q}\voperator{E}^n\voperator{Q}=\voperator{Q}\big(\voperator{Q}\voperator{E}\voperator{Q}\big)^n\voperator{Q}.
\end{displaymath}
The correlation length $\xi$ is then determined by the largest eigenvalue of $\voperator{Q}\voperator{E}\voperator{Q}$ as
\begin{equation}
\xi=-\frac{1}{\log \left[\rho(\voperator{Q}\voperator{E}\voperator{Q})\right]}.\label{eq:defcorr}
\end{equation}
Under the given assumption, $\rho(\voperator{Q}\voperator{E}\voperator{Q})<1$ and the correlation length $\xi$ is finite. Hence, all injective uMPS are exponentially clustering\cite{1992CMaPh.144..443F}. The correlation length is determined by $\rho(\voperator{Q}\voperator{E}\voperator{Q})$, which is equal to the eigenvalue of the transfer matrix $\voperator{E}$ that is second largest in absolute value.

\begin{widetext}
\subsection{Tangent space and momentum eigenstates}
\label{ss:tangentspace}
We now discuss the structure of the tangent space of the manifold of uMPS $\varM$ and the parametrization thereof. We first introduce the notation
\begin{equation}
\ket{\Phi^{(A)}(B)}=\ket{\Phi(B;A)}=B^{i}\partial_{i} \ket{\Psi(A)}=\sum_{n\in\mathbb{Z}}\sum_{\{s_n\}=1}^{d} \bm{v}_{\mathrm{L}}^{\dagger}\left[\left(\prod_{m<n} A^{s_{m}}\right) B^{s_{n}} \left(\prod_{m'>n} A^{s_{m'}}\right)\right]\bm{v}_{\mathrm{R}} \ket{\bm{s}},
\label{eq:defumpstangent}
\end{equation}
where $\partial_{i}=\partial\ /\partial A^{i}$ and the index $i$ is shorthand for a collective index $(\alpha,\beta,s)$ combining the two virtual indices and the physical index of tensor $A$, \textit{i.e.}~$A^{(\alpha,\beta,s)}=A^{s}_{\alpha,\beta}$. The tensor $B$ can take values in the tangent space of parameter space $T_A \manifold{A}$ at the point $A$. Since the subset of injective MPS $\manifold{A}$ is an open subset of the full parameter space $\mathbb{A}=\mathbb{C}^{D\times d\times D}$, we obtain $T_A \manifold{A}\equiv \mathbb{A}$. The physical states $\ket{\Phi(B;A)}$ for $B\in\mathbb{A}=\mathbb{C}^{D\times d\times D}$ span the tangent space $T_{A}\varM$. More completely, we can interpret this whole construction as mapping elements $(B;A)$ of the tangent bundle $T \manifold{A}$ to elements $(\ket{\Phi(B;A)},\ket{\Psi(A)})$ of the tangent bundle $T\varM\subset \hilbert\times\hilbert$ , where $\Phi$ should be recognized as the tangent map of $\Psi$, which is necessarily linear in its first argument.

For the application of the variational principle to the study of translation invariant phenomena, this tangent space ---consisting completely out of translation invariant states--- is sufficient. However, we can also interpret $\ket{\Psi(A)}$ as a special point in the larger class of general MPS with site dependent matrices, and define generalized tangent vectors
\begin{equation}
\ket{\Phi^{(A)}[\bm{B]}}=\sum_{n\in\mathbb{Z}}\sum_{\{s_n\}=1}^{d} \bm{v}_{\mathrm{L}}^{\dagger}\left[\left(\prod_{m<n} A^{s_{m}}\right) B^{s_{n}}(n) \left(\prod_{m'>n} A^{s_{m'}}\right)\right]\bm{v}_{\mathrm{R}} \ket{\bm{s}}
\end{equation}
where we use square brackets to denote a `functional dependence' on a set of site-dependent tensors $\bm{B}=\{B(n)\}_{n\in\mathbb{Z}}$. This larger tangent space is denoted as $\mathbb{T}^{\ket{\Psi(A)}}$ and turns out to be important when studying excited states, for which translation invariance is no longer a good assumption. However, when both the Hamiltonian and its ground state are translation invariant, we know that we can label the excited states by a momentum quantum number $p\in [-\pi,\pi)$. The momentum $p$ sector $\mathbb{T}^{\ket{\Psi(A)}}_{p}$ of the larger tangent space $\mathbb{T}^{\ket{\Psi(A)}}$ is obtained by choosing $B^{s}(n)=B^{s}\ec^{\ic p n}$, and we define
\begin{equation}
\ket{\Phi_{p}(B;A)}=\ket{\Phi^{(A)}_{p}(B)}=\sum_{n\in\mathbb{Z}}\ec^{\ic p n}\sum_{\{s_n\}=1}^{d} \bm{v}_{\mathrm{L}}^{\dagger}\left[\left(\prod_{m<n} A^{s_{m}}\right) B^{s_{n}} \left(\prod_{m'>n} A^{s_{m'}}\right)\right]\bm{v}_{\mathrm{R}} \ket{\bm{s}},\label{eq:defumpstangentp}
\end{equation}
with thus $\ket{\Phi_{0}(B;A)}=\ket{\Phi(B;A)}$. Hence, $\Phi_{p}^{(A)}$ represents a linear map from $\mathbb{A}$ to the momentum $p$ sector of the tangent space $\Tplane^{\ket{\Psi(A)}}\subset\hilbert$ at the translation invariant point $\ket{\Psi(A)}$. The full tangent space is obtained as $\Tplane^{\ket{\Psi(A)}}=\int_{p\in[-\pi,\pi)}^{\oplus}\Tplane^{\ket{\Psi(A)}}_p$.

We now repeat in detail some calculations using the tangent states $\ket{\Phi_{p}(B;A)}$ from Ref.~\onlinecite{2012arXiv1210.7710H}, to point out possible divergences that can occur. We introduce the generalized notation $\voperator{E}^{A}_{B}=\sum_{s=1}^{d} A^s\otimes \overline{B}^s$, which will be used extensively when evaluating expectation values with tangent vectors. We first compute the overlap between a tangent vector and the original uMPS $\ket{\Psi(A)}$ and obtain
\begin{equation}
\braket{\Psi(\overline{A})|\Phi_{p}(B;A)}=\sum_{n\in\mathbb{Z}} \ec^{\ic p n}\rbraket{l|\voperator{E}^{B}_{A}|r}= 2\pi \delta(p) \rbraket{l|\voperator{E}^{B}_{A}|r}\label{eq:ioverlapwithmps}
\end{equation}
so that all states $\ket{\Phi_{p}(B)}$ with $p\neq 0$ are automatically orthogonal to $\ket{\Psi(A)}$ due to the orthogonality of the different momentum sectors. For $p=0$, $\braket{\Psi(\overline{A})|\Phi_{0}(B;A)}$ is proportional to $\rbraket{l|\voperator{E}^{B}_{A}|r}$, with a diverging proportionality factor $2\pi \delta(0)=\lvert \mathbb{Z}\rvert$ where the cardinality $\lvert\mathbb{Z}\rvert$ represents the diverging number of lattice sites ($\mathcal{L}=\mathbb{Z}$). It is useful to define an orthogonal complement $\Tplane_{p}^{(A)\perp}$, which is equal to $\Tplane_{p}^{\ket{\Psi(A)}}$ if $p\neq 0$ and only contains the tangent vectors for which the tensor $B$ satisfies the linear relation $\rbraket{l|\voperator{E}^{B}_{A}|r}=0$ if $p=0$. We will discuss the physical relevance of restricting to this orthogonal subspace in the following sections.

Next, we compute the overlap between two tangent vectors $\braket{\Psi_{p'}(\overline{B'};\overline{A})|\Psi_{p}(B;A)}$. We have to be very careful with the infinite sums over the positions $n\in \mathbb{Z}$ and $n'\in\mathbb{Z}$ of $B$ and $B'$. When a diverging result is obtained, it is easily possible to make errors by miscounting. Only when the result is guaranteed to be finite can we freely use index substitutions. We therefore replace every occurrence of $\voperator{E}^{n}$ by a `regularized' operator $\voperator{Q}\voperator{E}^{n}\voperator{Q}=\voperator{E}^{n}\voperator{Q}=\voperator{Q}\voperator{E}^{n}=\voperator{E}^{n}-\voperator{S}=\voperator{Q}(\voperator{Q}\voperator{E}\voperator{Q})^{n}\voperator{Q}$ with $\rho(\voperator{Q}\voperator{E}\voperator{Q})<1$ and a `singular' part $\voperator{S}=\rket{r}\rbra{l}$. The reason of this notation becomes clear if we now evaluate $\braket{\Psi_{p}(\overline{B};\overline{A})|\Psi_{p'}(B';A)}$ as
\begin{equation*}
\begin{split}
\langle\Phi_{p}(\overline{B};\overline{A})&\mid\Phi_{p'}(B';A)\rangle= \overline{B}^{\overline{\imath}} N_{\overline{\imath},j}(p,p'; \overline{A},A) {B'}^j\\
\qquad=& \sum_{n=-\infty}^{+\infty}\sum_{n'=-\infty}^{+\infty}\ec^{+\ic p' n' - \ic p n}\left[\theta(n=n')\rbraket{l|\voperator{E}^{B'}_{B}|r}\right.\\
&\qquad\qquad\qquad\qquad\qquad\left.+ \theta(n'>n) \rbraket{l|\voperator{E}^{A}_{B} (\voperator{E})^{n'-n-1}\voperator{E}^{B'}_{A}|r}+ \theta(n'<n) \rbraket{l|\voperator{E}^{B'}_{A} (\voperator{E})^{n-n'-1}E^{A}_{B}|r}\right]\\
=&\sum_{n_{0}=-\infty}^{+\infty}\ec^{\ic (p'-p)n_{0}}\sum_{\Delta n=-\infty}^{+\infty}\ec^{\ic p \Delta n}\left[\theta(\Delta n=0)\rbraket{l|\voperator{E}^{B'}_{B}|r}\right.\\
&\qquad\qquad\qquad\qquad\qquad\left.+\theta(\Delta n > 0) \rbraket{l|\voperator{E}^{A}_{B} \voperator{Q}\voperator{E}^{\Delta n-1}\voperator{Q}\voperator{E}^{B'}_{A}|r}+\theta(\Delta n<0) \rbraket{l|\voperator{E}^{B'}_{A} \voperator{Q}\voperator{E}^{-\Delta n-1}\voperator{Q}\voperator{E}^{A}_{B}|r}\right]\\
&+\rbraket{l|\voperator{E}^{A}_{B}|r}\rbraket{l|\voperator{E}^{B'}_{A}|r} \sum_{n=-\infty}^{+\infty}\sum_{n'=-\infty}^{n-1}\ec^{\ic p' n' - \ic p n}+\rbraket{l|\voperator{E}^{B'}_{A}|r}\rbraket{l|\voperator{E}^{A}_{B}|r} \sum_{n=-\infty}^{+\infty}\sum_{n'=n+1}^{+\infty}\ec^{\ic p' n' - \ic p n}.
\end{split}
\end{equation*}
In these expression, we have introduced a ``discrete'' Heaviside function $\theta$ taking a logical expression as argument and resulting $1$ if the argument is true.  We have denoted the matrix elements of this overlap with respect to $B$ and $B'$ as $N_{\overline{\imath},j}(p,p'; \overline{A},A)$, where $[N_{\overline{\imath},j}]$ is referred to as the \emph{effective norm matrix}. The explicit dependence on both $\overline{A}$ and $A$ is used to indicate that it is not a holomorphic function of $A$ alone. There is no point in trying to compute the overlap between two tangent vectors $\ket{\Phi_{p}(B;A)}$ and $\ket{\Phi_{p'}(B';A')}$ at different gauge-inequivalent points $A$ and $A'$, as this overlap is automatically zero, which can be considered as a generalization of Anderson's orthogonality catastrophe\cite{Anderson:1967aa}. By using the well known result for the geometric series of an operator with spectral radius smaller than one, we obtain
\begin{equation}
\sum_{n=0}^{+\infty}\voperator{Q}\voperator{E}^{n}\voperator{Q}=\sum_{n=0}^{+\infty}\voperator{Q}(\voperator{Q}\voperator{E}\voperator{Q})^{n}\voperator{Q}=\voperator{Q}(\voperator{\one}-\voperator{Q}\voperator{E}\voperator{Q})^{-1}\voperator{Q}
\end{equation}
and thus
\begin{equation}
\begin{split}
\braket{\Phi_{p}(\overline{B};\overline{A})|\Phi_{p'}(B';A)}=& \overline{B}^{\overline{\imath}} N_{\overline{\imath},j}(p,p'; \overline{A},A) {B'}^j=2\pi \delta(p-p')\overline{B}^{\overline{\imath}} N_{\overline{\imath},j}(p; \overline{A},A) {B'}^j\\
=&2\pi\delta(p'-p)\left[\rbraket{l|\voperator{E}^{B'}_{B}|r}+\rbraket{l|\voperator{E}^{A}_{B} \voperator{Q} (\voperator{\one}-\ec^{\ic p}\voperator{Q}\voperator{E}\voperator{Q})^{-1}\voperator{Q}\voperator{E}^{B'}_{A}|r}\right.\\
&\qquad\left.+\rbraket{l|\voperator{E}^{B'}_{A} \voperator{Q}(\voperator{\one}-\ec^{-\ic p}\voperator{Q}\voperator{E}\voperator{Q})^{-1}\voperator{Q}\voperator{E}^{A}_{B}|r}
+(2\pi\delta(p)-1)\rbraket{l|\voperator{E}^{B'}_{A}|r}\rbraket{l|\voperator{E}^{A}_{B}|r}\right]
\end{split}
\label{eq:psipoverlap}
\end{equation}
As expected, momentum eigenstates cannot be normalized to unity in an infinitely large system, but rather satisfy a $\delta$ normalization. For equal momenta, $N_{\overline{\imath},j}(p,p; \overline{A},A)$ contains the diverging prefactor $2\pi\delta(0)=\lvert\mathbb{Z}\rvert$. The remaining part has been denoted as $N_{\overline{\imath},j}(p; \overline{A},A)$ and can be extracted from the terms inside the square brackets. Inside these brackets, the regular part $\voperator{Q}\voperator{E}\voperator{Q}$ produces a finite contribution where $B$ and $B'$ are strongly connected. We therefore also refer to these terms as the \emph{connected contribution}. For $p=0$, the product $\voperator{Q}(\voperator{\one}-\ec^{\pm \ic p}\voperator{Q}\voperator{E}\voperator{Q})^{-1}\voperator{Q}$ can be interpreted as the pseudo-inverse of the singular superoperator $\voperator{\one}-\voperator{E}$, which has an eigenvalue zero associated to the left and right eigenvectors $\rbra{l}$ and $\rket{r}$. We henceforth define $(\voperator{\one}-\ec^{\pm \ic p}\voperator{E})^{\mathsf{P}}=\voperator{Q}(\voperator{\one}-\ec^{\pm \ic p}\voperator{Q}\voperator{E}\voperator{Q})^{-1}\voperator{Q}$, so that $(\voperator{\one}-\voperator{E})^{\mathsf{P}} (\voperator{\one}-\voperator{E})=(\voperator{\one}-\ec^{\pm \ic p}\voperator{E})(\voperator{\one}-\ec^{\pm \ic p}\voperator{E})^{\mathsf{P}}=\voperator{Q}=\voperator{\one}-\rket{r}\rbra{l}$. Only for zero momentum does $(\voperator{\one}-\ec^{\pm \ic p}\voperator{E})^{\mathsf{P}}$ denote a true pseudo-inverse. For momentum zero, there is an additional divergence inside the square brackets coming from the singular part $\voperator{S}$. Here $B$ and $B'$ appear in two separate factors, and this term is henceforth referred to as the \emph{disconnected contribution}. Using Eq.~\eqref{eq:ioverlapwithmps}, this term can be traced back to the non-zero overlap with the original uMPS $\ket{\Psi(A)}$. It disappears for tangent vectors in $\Tplane_{p}^{(A)\perp}$. Note also that for momentum zero, $N_{\overline{\imath},j}(0,0,\overline{A},A)$ can be identified with the metric of the uMPS manifold.
\end{widetext}

The parameterization of tangent vectors inherits the gauge invariance of the MPS $\ket{\Psi(A)}$, whereby $\ket{\Phi_p(B;A)}=\ket{\Phi_p(B_G;A_G)}$. By fixing the representation $A$ of the base point $\ket{\Psi(A)}$, this `multiplicative' gauge invariance is also fixed. Nevertheless, the linear map $\Phi_p^{(A)}:\mathbb{A}\mapsto \Tplane_{p}^{\ket{\Psi(A)}}$ still has a non-trivial null space which can be associated with infinitesimal gauge transformations of the MPS $\ket{\Psi(A)}$. Since the full tangent space $\Tplane^{\ket{\Psi(A)}}$ was obtained by interpreting the uMPS $\ket{\Psi(A)}$ as a special point in the space of generic MPS, we have to consider site-dependent gauge transformations that take $A^s(n)=A^s$ to $A^s_G(n)=G(n-1,\eta)^{-1} A^s G(n,\eta)$ for a one-parameter family of gauge transformations $G(n,\eta)=\exp(\eta x(n))$. Expressing the invariance of the MPS at first order in $\eta$ for a choice $x(n)=x\exp(\ic p n)$ allows to conclude that $\ket{\Phi^{(A)}_{p}(B)}=0$ for $B=\mathscr{N}^{(A)}_{p}(x)$, where the action of $\mathscr{N}^{(A)}_{p}$ is given by
\begin{equation}
\mathscr{N}^{(A)}_{p}: x\mapsto \mathscr{N}^{(A)s}_{p}(x)= A^{s} x-\ec^{-\ic p} x A^{s}, \forall s=1,\ldots,d.\label{eq:defnphip}
\end{equation}
The map $\mathscr{N}^{(A)}_{p}$ establishes an isomorphism between the null space $\mathbb{N}^{(A)}_{p}$ of the map $\Phi^{(A)}_{p}$ and the Lie algebra of the gauge group, which is equal to $\mathfrak{gl}(D,\mathbb{C})=\mathbb{C}^{D\times D}$ for $p\neq 0$ and to $\mathfrak{pgl}(D,\mathbb{C})=\{x\in\mathbb{C}^{D\times D}| \tr[x]=0\}$ for $p=0$. It can easily be checked that this is an isomorphism by trying to determine the null space of the map $\mathscr{N}_{p}$. By multiplying $\mathscr{N}^{s}_{\Phi_{p}}(x)=0$ ($\forall s=1,\ldots,d)$ to the left with $(A^s)^\dagger$ and summing over $s$, we obtain the requirement
\begin{equation}
\voperator{E}\rket{xr}=\ec^{-\ic p} \rket{xr}.
\end{equation}
For $p\neq 0$, this equation has no solutions, whereas for $p=0$, the only solution are matrices $x$ proportional to the unit matrix $\one_{D}$, which is not within the algebra $\mathfrak{pgl}(D,\mathbb{C})$ of traceless matrices. For nonzero $p$, we thus obtain $\dim \Tplane_{p}=\dim \mathbb{A}-\dim \mathbb{N}^{(A)}_{p}=(d-1)D^2$, while we obtain $\dim \Tplane_{0}=(d-1)D^2 +1$ for zero momentum. But of course, $\ket{\Psi(A)}\in \Tplane_{0}^{\ket{\Psi(A)}}$, and restricting to the part $\Tplane_{0}^{\perp}$ that is orthogonal to $\ket{\Psi(A)}$ also reduces the dimension to $\dim \Tplane_{0}^{\perp}=(d-1)D^{2}$.  Within the language of principal fiber bundles, the null space $\mathbb{N}^{(A)}_{p}$ is known as the vertical subspace, and for any $B\in\mathbb{A}$, $B'\in\mathbb{N}_{p}^{(A)}$ we obtain $\ket{\Phi_{p}^{(A)}(B+B')}=\ket{\Phi_{p}^{(A)}(B)}$. Therefore, we sometimes refer to this as an \emph{additive gauge freedom} in the parameterization of MPS tangent vectors. Since the vectors $B=\mathscr{N}^{(A)}_{p}(x)$ in the null space $\mathbb{N}^{(A)}_{p}$ are also eigenvectors with zero eigenvalue of the effective normalization matrix $N_{\overline{\imath},j}(p)$, as well as of any other effective operator $O_{\overline{\imath},j}(p)$ that we obtain by restricting a physical operator $\hat{O}$ to the tangent space $\mathbb{T}_{p}^{\ket{\Psi(A)}}$, we also refer to them as \emph{null modes}. In order to associate a unique parametrization to every tangent vector $\ket{\Phi^{(A)}_{p}(B)}\in\Tplane_{p}$, we should define a complementary space $\mathbb{B}_{p}^{(A)}$ such that $\mathbb{A}=\mathbb{B}_{p}^{(A)}\oplus \mathbb{N}_{p}^{(A)}$ and restrict to parametrizations $B\in\mathbb{B}^{(A)}_{p}\oplus _{p}$. While there is no unique definition for $\mathbb{B}^{(A)}_{p}$ ---referred to as the horizontal subspace---, a gauge covariant description would be such that $B_G\in\mathbb{B}^{(A_G)}_{p}$ for any $B\in\mathbb{B}^{(A)}_{p}$. This is easily accomplished using the machinery of principal bundle connections\cite{2012arXiv1210.7710H}. We only mention the result. Two different choices for $\mathbb{B}^{(A)}_{p}$ are obtained as the subspace of solutions of one of the two following linear homogeneous set of equations, which we can then call gauge fixing conditions for the additive gauge freedom:
\begin{itemize}
\item \emph{Left-gauge fixing condition}:
\begin{equation}
\sum_{s=1}^{d} {A^{s}}^{\dagger} l B^{s}=0\qquad\Leftrightarrow\qquad\rbra{l}\voperator{E}^{B}_{A}=0.\label{eq:leftgaugeuB}
\end{equation}
\item \emph{Right-gauge fixing condition}:
\begin{equation}
\sum_{s=1}^{d} B^{s} r {A^{s}}^{\dagger}=0\qquad\Leftrightarrow\qquad\voperator{E}^{B}_{A}\rket{r}=0.\label{eq:rightgaugeuB}
\end{equation}
\end{itemize}
Since these conditions are $D^{2}$ dimensional, they fix all $D^{2}$ linearly independent gauge transformations in $\mathbb{N}_{p}^{(A)}$ for $p\neq 0$. For $p=0$, there are only $D^{2}-1$ linearly independent gauge transformations, and these $D^{2}$ gauge fixing conditions also include norm preservation, \textit{i.e.}~they imply $\braket{\Psi(A)|\Phi_{0}(B)}=0$ or thus $\ket{\Phi_{0}(B)}\in\Tplane_{{0}}^{\ket{\Psi(A)}\perp}$. Note that the horizontal subspaces $\mathbb{B}^{(A)}_{p}$ defined by these equations are momentum-independent, even though the vertical subspace $\mathbb{N}^{(A)}_{p}$ did explicitly depend on the momentum $p$. Since we will explicitly be using either of the two conditions from Eq.~\eqref{eq:leftgaugeuB} and Eq.~\eqref{eq:rightgaugeuB}, the horizontal subspace is henceforth denoted as $\mathbb{B}^{(A)}$ without any reference to the momentum $p$.

Either choice for $\mathbb{B}^{(A)}$ results in a considerable simplification of the effective norm matrix, since all non-local terms in Eq.~\eqref{eq:psipoverlap} cancel:
\begin{equation}
\braket{\Phi_{p}(\overline{B};\overline{A})|\Phi_{p}(B;A)}=2\pi\delta(p'-p) \rbraket{l|\voperator{E}^{B'}_{B}|r}\label{eq:psikoverlapsimp}.
\end{equation}
However, this simplification is only really useful if we can directly parametrize tensors $B\in\mathbb{B}^{(A)}$. A linear parameterization $B=\mathscr{B}^{(A)}(X)$ depending on a $D\times D(d-1)$ matrix $X$ that satisfies Eq.~\eqref{eq:leftgaugeuB} can be constructed. We first define the $D\times Dd$ matrix $L$ as
\begin{equation}
[L]_{\alpha;(\beta,s)}= [ {A^{s}}^{\dagger}l^{1/2}]_{\alpha,\beta}\label{eq:utdvpdefR}
\end{equation}
and then construct a $dD\times (d-1)D$ matrix $V_{L}$ which contains an orthonormal basis for the null space of $L$, \textit{i.e.}~$L V_{L}=0$ and $V_{L}^{\dagger} V_{L}=\one_{(d-1)D}$. Setting $[V^{s}_{L}]_{\alpha,\beta}=[V_{L}]_{(\alpha,s);\beta}$, we then define the representation $\mathscr{B}^{(A)s}(x)$ as
\begin{equation}
\mathscr{B}^{(A)s}(X)= l^{-1/2} V_L^s X  r^{-1/2}\label{eq:defBrepresentation}
\end{equation}
in order to obtain
\begin{equation}
\braket{\Phi_{p'}(\overline{\mathscr{B}}^{(A)}(\overline{X}))|\Phi_{p}(\mathscr{B}^{(A)}(Y))}=2\pi\delta(p-p')\tr\left[ X^{\dagger} Y\right],\label{eq:effnormrepresentation}
\end{equation}
in combination with the left gauge fixing condition $\sum_{s=1}^{d} {A^{s}}^{\dagger} l \mathscr{B}^{(A)s}(X) =0$. An alternative representation for tensors $B$ satisfying the left gauge fixing conditions follows similarly. 

\section{Time-dependent variational principle}
\label{s:TDVP}
After White's formulation of the DMRG, another major breakthrough was the formulation of DMRG-inspired algorithms to study dynamic properties of quantum spin chains. These algorithms can be divided into two classes: some algorithms directly probe spectral functions, whereas other algorithms aim to approximate the full time-evolving wave function. These methods are reviewed by \citet{Schollwoeck:2006fk} and \citet{2008arXiv0808.2620J}. Today's best known and most powerful method for approximating the time-evolving wave function within the MPS manifold is the time-evolving block decimation (TEBD), which was developed by \citet{2004PhRvL..93d0502V}. It was later reformulated in order to be compatible with traditional DMRG implementations \cite{2004PhRvL..93g6401W,2004JSMTE..04..005D}. The TEBD is based on an iterative application of a Lie-Trotter-Suzuki decomposition \cite{Trotter:1959aa,Suzuki:1976aa} of the exact evolution operator for a small time step $\d t$ as
\begin{equation}
\exp(\ic \ham \d t)= \exp(\ic \ham^{(A)} \d t) \exp(\ic \ham^{(B)} \d t) + \order(\d t^{2}).
\end{equation}
Higher order decompositions with an error of $\order(\d t^{p})$ are also possible \cite{Suzuki:1990aa}. $\ham^{(A)}$ and $\ham^{(B)}$ provide a decomposition of the (possibly time-dependent) Hamiltonian $\ham=\ham^{(A)}+\ham^{(B)}$, such that $\ham^{(A)}$ and $\ham^{(B)}$ separately contain local terms that all commute. If necessary, a decomposition into more than two parts is also possible. For a nearest-neighbor Hamiltonian $\ham=\sum_{n\in\mathbb{Z}} \operator{h}_{n,n+1}$, a possible decomposition scheme is into even and odd terms: $\ham^{(A)}=\sum_{n\in\mathbb{Z}} \operator{h}_{2n,2n+1}$ and $\ham^{(B)}=\sum_{n\in\mathbb{Z}}\operator{h}_{2n+1,2n+2}$. The individual operators $\exp(\ic \operator{H}^{(A)})$ and $\exp(\ic \operator{H}^{(B)})$ then split into a product of local unitaries, that can be dealt with in a parallelized and efficient way. When applied to a generic MPS, the individual evolution operators take the state outside the original manifold, since they have the effect of increasing the virtual bond dimension. Once a given maximum bond dimension has been reached, one then approximates the newly obtained state by an MPS with this maximal bond dimension. The best strategy for truncating a single bond dimension is obtained by discarding the smallest Schmidt values. However, since the Hamiltonian evolution acts on all bonds,  several bond dimensions have to be simultaneously truncated. The strategy based on discarding the smallest Schmidt values still serves as a good initial guess but is not optimal. For lattices $\mathcal{L}$ of finite size, an optimal MPS representation with given maximal bond dimension can be obtained by minimizing the norm difference using algorithms inspired by the sweeping process of the finite-size algorithm of the DMRG\cite{2004PhRvL..93t7204V}. 

The TEBD can also be applied to translation invariant systems in the thermodynamic limit. In combination with imaginary time evolution, this allowed for the first time to find the best ground state approximation within the class of uMPS\cite{2007PhRvL..98g0201V,2008PhRvB..78o5117O}. However, no optimal strategy for truncation of the bond dimension is known in the case of infinite lattices. As a variational strategy, the infinite size time evolving block decimation then requires a scaling of $\d t\to 0$ as the optimal approximation is approached, in order to correct for the truncation error. Since the exact imaginary time evolution automatically slows down in the neighborhood of the best ground state (approximation), the need for a decreasing time step induces an additional unfavorable slowing down.

In addition, both for finite and infinite systems, some symmetries of the Hamiltonian $\ham$ might not be inherited by the individual Trotter evolution operators $\exp(\ic \ham^{(A)} \d t)$ and $\exp(\ic \ham^{(B)}\d t)$. In itself, the Lie-Trotter-Suzuki decomposition is symplectic and under an exact iterative application of the Trotter operators, errors resulting from these broken symmetries would be strongly bound. However, the additional truncation after every evolution step ruins the symplecticity and drifting errors are possible. In particular, for a time-independent Hamiltonian $\ham$, the expectation value of the Hamiltonian is a constant of motion, but will drift away in a simulation based on the TEBD. One last downside of TEBD is that it cannot cope with Hamiltonians containing long-range interaction terms. 

Recently, a new algorithm for approximating time-evolving quantum states with MPS was proposed\cite{2011PhRvL.107g0601H}, based on the TDVP of Dirac\cite{Dirac:1930aa,Frenkel:1934aa,Kramer:1981aa}. The TDVP is a general method that can be formulated for any variational manifold and any Hamiltonian, both with short- and long-range interactions. In combination with the manifold of MPS, the TDVP allows one to overcome the aforementioned short-comings of the TEBD. Unlike TEBD, which is necessarily formulated using discrete time steps $\d t$, the TDVP is naturally formulated in continuous time. The TDVP transforms the linear Schr\"{o}dinger equation in the full Hilbert space $\hilbert$ into a non-linear set of symplectic differential equations in the parameter space of the variational manifold. If we could solve these non-linear differential equations exactly in continuous time, the only source of errors would be the restriction to the manifold itself. The TDVP describes the best direction in which the quantum state can evolve without leaving the variational manifold in order to approximate the time-dependent Schr\"{o}dinger equation. Hence, no truncation of any kind is necessary. In addition, there is no need for a Trotter decomposition and thus no corresponding Trotter error. This approach is also perfectly applicable in case of imaginary time evolution (there is no symplecticity of course). It can also be applied to Hamiltonians with long-range interactions and to generic MPS on finite lattices\cite{2012PhRvL.109z7203K,2013arXiv1304.7725H}, or to a finite subsystem of an infinite lattice\cite{2012arXiv1207.0691M}. Most importantly, the TDVP can be implemented for the case of MPS with an efficiency that is comparable to other methods such as DMRG and TEBD (namely $\order(D^3)$). Of key importance for such an efficient implementation is the use of the gauge fixing conditions for tangent vectors, which was introduced in the previous section.

A numerical integration of the TDVP equation does of course require a discretization of the time variable, resulting in additional errors due to the final time step. However, since many standard numerical integrators can be used, these errors are well controlled and well understood from the general theory of (symplectic) differential equations (on manifolds) \cite{Hairer:2004aa}. We now derive the TDVP equation based on an action formalism for the case of uMPS in the thermodynamic limit. We then compare the resulting equations to the geometric argument that was used in the original publication\cite{2011PhRvL.107g0601H}. While the geometric construction might provide a better visual insight into the approximation made by the TDVP, the action formalism is better suited to derive the properties of the resulting non-linear differential equation. We discuss the symplectic properties of real time evolution and discuss convergence and error measures that can be used to assess the approximation error made by confining the evolution to $\varM$. Finally, we outline the details of a simple first-order Euler based algorithm. The numerical implementation of more advanced integration schemes are described elsewhere\footnote{J.~Haegeman \textit{et. al.}, in preparation}.

\begin{widetext}
\subsection{Principle of least action}
\label{ss:tdvp:action}
The dynamics of isolated quantum systems are governed by the time-dependent Schr\"{o}\-dinger equation (TDSE)
\begin{equation}
\ic \frac{\d\ }{\d t}\ket{\Psi(t)}=\ham(t)\ket{\Psi(t)},\label{eq:schrodinger}
\end{equation}
which is a linear first-order differential equation in $\hilbert$. Note that we allow for the Hamiltonian $\ham$ to be time-dependent. The TDSE can be derived by applying the variational principle of least action to the action functional
\begin{equation}
S_\hilbert [\overline{\Psi},\Psi]=\int_{-\infty}^{+\infty}\bigg(\frac{\ic}{2}\braket{\Psi(t)|\dot{\Psi}(t)}-\frac{\ic}{2}\braket{\dot{\Psi}(t)|\Psi(t)} -\braket{\Psi(t)|\ham(t)|\Psi(t)}\bigg)\,\d t.\label{eq:actionfull}
\end{equation}
We now focus on one-dimensional quantum lattice systems with translation invariance. For notational simplicity we assume that the Hamiltonian contains nearest-neighbor terms only,\textit{i.e.}~$\ham=\sum_{n\in\mathbb{Z}} \operator{T}^{n}\operator{h}\operator{T}^{-n}$ where $\operator{T}$ is the translation operator that shifts the system by one site and $\operator{h}$ has non-trivial support only on sites zero and one. The generalization to interaction terms on a larger number of neighboring sites or even long-range interaction terms is straightforward, since no Trotter-like decomposition of the Hamiltonian is required. An initial state $\ket{\Psi_0}$ that can be encoded as a uMPS $\ket{\Psi(A_0)}$ will in general leave the manifold of uMPS $\varM$ under exact time evolution. In order to confine the dynamics to $\varM$ we can analogously apply the principle of least action to
\begin{equation}
S_{\varM}[\overline{A},A]=\int_{-\infty}^{+\infty} \bigg(\frac{\ic}{2}\big[\dot{A}^j(t)\partial_j-\dot{\overline{A}}^{\overline{\jmath}}(t)\partial_{\overline{\jmath}} \big]\braket{\Psi(\overline{A}(t))|\Psi(A(t))}-\braket{\Psi(\overline{A}(t))|\ham(t)|\Psi(A(t))}\bigg)\,\d t,\label{eq:mpsaction}
\end{equation}
where the square brackets indicate the functional dependence on the full evolution $A(t)$ and its complex conjugate. Whereas the Schr\"{o}dinger equation describes a unitary process (norm-preserving) that makes the integrand of $S_\hilbert$ exactly zero, the extremization of $S_{\varM}$ might result in a flow equation for $\ket{\Psi(A(t))}$ that is not norm-preserving and minimizes $S_{\varM}$ by converging to a state with zero norm. It is therefore better to define a modified, normalized action which does result in norm-independent dynamics as
\begin{equation}
\begin{split}
\widetilde{S}_{\varM}[\overline{A},A]=&\int_{-\infty}^{+\infty} \frac{\frac{\ic}{2}\big[\dot{A}^j(t)\partial_j-\dot{\overline{A}}^{\overline{\jmath}}(t)\partial_{\overline{\jmath}} \big]\braket{\Psi(\overline{A}(t))|\Psi(A(t))} -\braket{\Psi(\overline{A}(t))|\ham|\Psi(A(t))}}{\braket{\Psi(\overline{A}(t))|\Psi(A(t))}}\,\d t\\
=&\int_{-\infty}^{+\infty}\bigg( \frac{\ic}{2}\big[\dot{A}^j(t)\partial_j-\dot{\overline{A}}^{\overline{\jmath}}(t)\partial_{\overline{\jmath}} \big]\ln N(\overline{A}(t),A(t)) -H(\overline{A}(t),A(t))\bigg)\,\d t,
\end{split}\label{eq:actionmod}
\end{equation}
where
\begin{align}
N(\overline{A},A)&=\braket{\Psi(\overline{A})|\Psi(A)},&H(\overline{A},A)&=\frac{\braket{\Psi(\overline{A})|\ham|\Psi(A)}}{\braket{\Psi(\overline{A})|\Psi(A)}}.
\end{align}
Note that we henceforth omit the explicit time-dependence of the Hamiltonian $\ham$ for the sake of simplicity. For a time-dependent Hamiltonian $\ham(t)$, the energy function $H$ would have an explicit time-dependence [\textit{i.e.}~$H(\overline{A},A,t)$], but this has no further effect on the resulting expressions. When working with a uMPS $\ket{\Psi(A)}$, we always assume that is has been properly normalized such that $\rho(\voperator{E}^{A}_{A})=1$ and we can set $N(\overline{A},A)=1$. However, this is only true after a normalization of the matrices $A^s$ and in general $N=\braket{\Psi(\overline{A})|\Psi(A)}$ depends on $A$. For the translation-invariant nearest-neighbor Hamiltonian $\ham=\sum_{n\in\mathbb{Z}} \operator{T}^{n}\operator{h}\operator{T}^{-n}$, we obtain
\begin{equation}
H(\overline{A},A)=\lvert \mathbb{Z}\rvert h(\overline{A},A)
\end{equation}
where, for a properly normalized uMPS $\ket{\Psi(A)}$,
\begin{equation}
h(\overline{A},A)=\frac{\braket{\Psi(\overline{A})|\operator{h}|\Psi(A)}}{\braket{\Psi(\overline{A})|\Psi(A)}}=\rbraket{l|\voperator{H}^{AA}_{AA}|r}
\end{equation}
and we have defined a new superoperator
\begin{equation}
\voperator{H}^{A_{1}A_{2}}_{A_{3}A_{4}}=\sum_{s,t,u,v=1}^{d} \braket{u,v|\operator{h}|s,t} A_{1}^{s}A_{2}^{t}\otimes \overline{A_{3}^{u}A_{4}^{v}}.\label{eq:defsupopH}
\end{equation}

If we define the norm-independent action with normalized integrand for the full Hilbert space $\hilbert$ as $\widetilde{S}_{\hilbert}$, then the stability of $\widetilde{S}_{\hilbert}$ with respect to variations $\bra{\Psi(t)}\mapsto \bra{\Psi(t)}+\bra{\delta\Psi(t)}$ requires 
\begin{equation}
\bigg(\operator{\one}-\frac{\ket{\Psi(t)}\bra{\Psi(t)}}{\braket{\Psi(t)|\Psi(t)}}\bigg)\bigg(\ic\frac{\d\ }{\d t}\ket{\Psi(t)}-\ham \ket{\Psi(t)}\bigg)=0.\label{eq:schrodingermod}
\end{equation}
Applying the variational principle to the modified action $\widetilde{S}_{\hilbert}$ thus imposes the Schr\"{o}\-ding\-er equation in the plane orthogonal to the vector $\ket{\Psi(t)}$, whereas it leaves the evolution in the direction of the current vector $\ket{\Psi(t)}$ unspecified. Since a nonzero parallel component of the evolution vector ($\braket{\Psi(t)|\dot{\Psi}(t)} \neq 0$) results in norm or phase changes, the use of the modified action unties the restriction to a specific choice of phase and normalization of the state. Norm-independent dynamics within the variational manifold of uMPS are described by the Euler-Lagrange equations of $\widetilde{S}_{\text{uMPS}}$, which are given by
\begin{equation}
+\ic \widetilde{N}_{\overline{\imath},j}(\overline{A}(t),A(t)) \dot{A}^{j}(t)=H_{\overline{\imath}}(\overline{A}(t),A(t))\label{eq:tdvpeq}
\end{equation}
and its complex conjugate. Here, we have introduced the gradient
\begin{equation}
H_{\overline{\imath}}(\overline{A},A)=\partial_{\overline{\imath}}H(\overline{A},A)=\frac{\braket{\partial_{\overline{\imath}}\Psi(\overline{A})|\operator{H}|\Psi(A)}}{N(\overline{A},A)}-\frac{\braket{\partial_{\overline{\imath}}\Psi(\overline{A})|\Psi(A)}\braket{\Psi(\overline{A})|\operator{H}|\Psi(A)}}{N(\overline{A},A)^2}
\label{eq:gradient}
\end{equation}
and the modified metric 
\begin{equation}
\widetilde{N}_{\overline{\imath},j}(\overline{A},A)=\partial_{\overline{\imath}}\partial_{j}\ln N(\overline{A},A)=\frac{N_{\overline{\imath},j}(0,0;\overline{A},A)}{N(\overline{A},A)}-\frac{\braket{\partial_{\overline{\imath}}\Psi(\overline{A})|\Psi(A)}\braket{\Psi(\overline{A})|\partial_{j}\Psi(A)}}{N(\overline{A},A)^2}.\label{eq:metricmod}
\end{equation}
Whenever we need to compute these quantities, we can properly normalize the uMPS $\ket{\Psi(A)}$ and assume that $N(\overline{A},A)=1$. The metric $N_{\overline{\imath},j}(0,0;\overline{A},A)=\braket{\partial_{\overline{\imath}} \Psi(\overline{A})|\partial_j \Psi(A)}$ was implicitly defined in Eq.~\eqref{eq:psipoverlap}. It can easily be seen that the second term in Eq.~\eqref{eq:metricmod} contains a double divergence ---it is proportional to $\lvert\mathbb{Z}\rvert^2$--- and cancels exactly with the divergent term within the square brackets in Eq.~\eqref{eq:psipoverlap}. We thus obtain
\begin{displaymath}
\overline{B}^{\overline{\imath}} \widetilde{N}_{\overline{\imath},j}(\overline{A},A) (B')^j =  \lvert\mathbb{Z}\rvert \left[\rbraket{l|\voperator{E}^{B'}_{B}|r}+\rbraket{l|\voperator{E}^{A}_{B} (\voperator{\one}-\voperator{E})^{\mathsf{P}}\voperator{E}^{B'}_{A}|r}+\rbraket{l|\voperator{E}^{B'}_{A}(\voperator{\one}-\voperator{E})^{-1}\voperator{E}^{A}_{B}|r}
-\rbraket{l|\voperator{E}^{B'}_{A}|r}\rbraket{l|\voperator{E}^{A}_{B}|r}\right].
\end{displaymath}

 We can also evaluate the first term in Eq.~\eqref{eq:gradient} using the same techniques as in the previous section. We prefer to evaluate the more general expression $\braket{\Phi_{p}(\overline{B})|\ham|\Psi(A)}$, which defines $\braket{\partial_{\overline{\imath}}\Psi(\overline{A})|\operator{H}|\Psi(A)}$ through $\overline{B}^{\overline{\imath}}\bra{\partial_{\overline{\imath}}\Psi(\overline{A})}=\bra{\Phi_0(B)}$. Using the translation-invariant Hamiltonian with nearest-neighbor interactions $\ham=\sum_{n\in\mathbb{Z}} \operator{T}^{n}\operator{h}\operator{T}^{-n}$, we obtain
\begin{multline*}
\braket{\Phi_{p}(\overline{B})|\ham|\Psi(A)}= \sum_{n=-\infty}^{+\infty}\sum_{n'=-\infty}^{+\infty}\ec^{- \ic p n}\left[\theta(n=n')\rbraket{l|\voperator{H}^{A A}_{B A}|r}+\theta(n=n'+1)\rbraket{l|\voperator{H}^{A A}_{A B}|r}\right.\\
\left.+ \theta(n>n'+1) \rbraket{l|\voperator{H}^{AA}_{AA} (\voperator{E})^{n-n'-2}\voperator{E}^{B}_{A}|r}+ \theta(n<n') \rbraket{l|\voperator{E}^{B}_{A} (\voperator{E})^{n'-n-1}\voperator{H}^{AA}_{AA}|r}\right].
\end{multline*}
Repeating the same tricks as for the evaluation of $\braket{\Phi_{p'}(B')|\Phi_{p}(B)}$ leads to
\begin{multline}
\braket{\Phi_{p}(\overline{B};\overline{A})|\ham|\Psi(A)}=2\pi\delta(p)\left[\rbraket{l|\voperator{H}^{A A}_{B A}|r}+\rbraket{l|\voperator{H}^{A A}_{A B}|r}+\rbraket{l|\voperator{H}^{AA}_{AA} (\voperator{\one}-\voperator{E})^{\mathsf{P}}\voperator{E}^{A}_{B}|r}\right.\\\left.+\rbraket{l|\voperator{E}^{A}_{B} (\voperator{\one}-\voperator{E})^{\mathsf{P}}\voperator{H}^{AA}_{AA}|r}+(\lvert\mathbb{Z}\rvert-2) \rbraket{l|\voperator{H}^{AA}_{AA}|r}\rbraket{l|\voperator{E}^{A}_{B}|r}\right].\label{eq:grad}
\end{multline}
As expected, the translation invariant state $\ham \ket{\Psi(A)}$ has zero overlap with momentum eigenstates with $p\neq 0$. Indeed, this is a necessary condition in order to be able to approximate time-evolution within the translation-invariant manifold of uMPS. Hence, only the tangent vectors with zero momentum feature in the TDVP. For $p=0$, the overlap is proportional to $2\pi\delta(0)=\lvert\mathbb{Z}\rvert$, which matches with the same factor in $\widetilde{N}_{\overline{\imath},j}(\overline{A},A)$. As for the metric, there is an additional divergence inside the brackets, which is related to the non-zero overlap of the tangent vector $\ket{\Phi_0{B}}$ with the original  uMPS $\ket{\Psi(A)}$ (\textit{i.e.}~$\rbraket{l|\voperator{E}^{A}_{B}|r}\neq 0$). Here too, this additional divergence cancels with the second term in Eq.~\eqref{eq:gradient}, resulting in
\begin{displaymath}
\overline{B}^{\overline{\imath}} H_{\overline{\imath}}(\overline{A},A)=\lvert\mathbb{Z}\rvert\left[\rbraket{l|\voperator{H}^{A A}_{B A}|r}+\rbraket{l|\voperator{H}^{A A}_{A B}|r}+\rbraket{l|\voperator{H}^{AA}_{AA} (\voperator{\one}-\voperator{E})^{\mathsf{P}}\voperator{E}^{A}_{B}|r}+\rbraket{l|\voperator{E}^{A}_{B} (\voperator{\one}-\voperator{E})^{\mathsf{P}}\voperator{H}^{AA}_{AA}|r}-2\rbraket{l|\voperator{H}^{AA}_{AA}|r}\rbraket{l|\voperator{E}^{A}_{B}|r}\right].
\end{displaymath}
\end{widetext}

A translation-invariant time-evolving quantum state for a one-dimensional lattice system can thus be approximated with a time-evolving uMPS $\ket{\Psi(A(t))}$ where the evolution of the parameterization $A(t)$ is specified by Eq.~\eqref{eq:tdvpeq}. However, this equation does not fully specify the time-evolution of all degrees of freedom, since the modified metric $\widetilde{N}_{\overline{\imath},j}$ has a number of null modes (eigenvectors with zero eigenvalues). Firstly, since $B^i \ket{\partial_i \Psi(A)}=\ket{\Phi_0(B)}=0$ for any $B=\mathscr{N}_{0}^{(A)}(x)$ with $x\in \mathfrak{pgl}(D,\mathbb{C})$, the modified metric $\widetilde{N}_{\overline{\imath},j;}(\overline{A},A)$ inherits the $D^2-1$ linearly independent null modes of $N_{\overline{\imath},j}(0,0;\overline{A},A)$. These null modes do not render the linear system for $\dot{A}$ in Eq.~\eqref{eq:tdvpeq} unsolvable, since any null mode $B=\mathscr{N}_{0}^{(A)}(x)$ also satisfies $\overline{B}^{\overline{\imath}} H_{\overline{\imath}}(\overline{A},A)=0$. These modes result in gauge transformations of the parameterization without influencing the physical state. Hence, they are not determined by the dynamics and will have to be fixed by a gauge fixing prescription, which is mathematically equivalent to choosing one of the infinitely many solutions of the linear system for $\dot{A}$ and thus to define a pseudo-inverse $\widetilde{N}^{i,\overline{\jmath}}(\overline{A},A)$ of the modified metric.

However, before being able to do so, we have to take into account all null modes of the modified metric $\widetilde{N}_{\overline{\imath},j}(\overline{A},A)$. Since $A^i \ket{\partial_i \Psi(A)}=\ket{\Phi_0(A)}\sim \ket{\Psi(A)}$, we have $B=A$ as an additional, linearly independent null mode of $\widetilde{N}_{\overline{\imath},j}(\overline{A},A)$. Note that also $\overline{A}^{\overline{\imath}} H_{\overline{\imath}}(\overline{A},A)=0$. The mode $B=A$ was responsible for the divergences of order $\lvert\mathbb{Z}\rvert^2$ in $\braket{\partial_{\overline{\imath}}\Psi(\overline{A})|\ham|\Psi(A)}$ and $\braket{\partial_{\overline{\imath}} \Psi(\overline{A})|\partial_j\Psi(A)}$, which were canceled by the second term in Eq.~\eqref{eq:gradient} and Eq.~\eqref{eq:metricmod} respectively. This mode results in physical changes in norm or phase, which are not fixed by the Euler-Lagrange equation of the norm-independent action.This can also be seen from Eq.~\eqref{eq:schrodingermod}. Hence, we need an additional constraint to be able to invert $\widetilde{N}_{\overline{\imath},j}$ and to fully fix the time-evolution of $A(t)$. This boils down to fixing the value of $\braket{\Psi(A(t))|\frac{\d\ }{\d t}|\Psi(A(t))}=\braket{\Psi(A(t))|\Phi(\dot{A}(t))}$, which was left unspecified by the norm-independent dynamics. Since we would like to keep the norm of the time-evolving uMPS $\ket{\Psi(A(t))}$ fixed to one, we need to impose at least
\begin{equation}
\frac{\d\ }{\d t} \braket{\Psi(A(t))|\Psi(A(t))} = 2\Re\left[\braket{\Psi(A(t))|\Phi(\dot{A}(t))}\right]=0.
\end{equation}
If we also fix the freedom in phase by 
\begin{equation}
\Im\left[\braket{\Psi(A(t))|\Phi_0(\dot{A}(t))}\right]=0,
\end{equation}
then we effectively restrict to tangent vectors $\ket{\Phi_0(\dot{A}(t))}\in \Tplane_{0}^{\ket{\Psi(A)}\perp}$. Hence, we are allowed to restrict to a parameterization $\dot{A}(t)\in\mathbb{B}^{(A)}$ by imposing either the left or right gauge fixing prescriptions defined in Eq.~\eqref{eq:leftgaugeuB} and Eq.~\eqref{eq:rightgaugeuB}, which allows us to cancel many terms in the expressions for $\widetilde{N}_{\overline{\imath},j}(\overline{A},A)$ and $H_{\overline{\imath}}(\overline{A},A)$. In addition, we can then define a pseudo-inverse metric $\tilde{N}^{i,\overline{\jmath}}(\overline{A},A)$ such that
\begin{equation}
\widetilde{N}^{i,\overline{\jmath}}(\overline{A},A) \widetilde{N}_{\overline{\jmath},k}(\overline{A},A) = (P_{\mathbb{B}^{(A)}})^{i}_{\ k}.\label{eq:defPB}
\end{equation}
For this gauge fixing prescription, the TDVP equations can then be rewritten as
\begin{equation}
\dot{A}^{i}=\widetilde{N}^{i,\overline{\jmath}}(\overline{A},A)H_{\overline{\jmath}}(\overline{A},A).
\end{equation}

\subsection{Geometric construction}
\label{ss:tdvp:geometric}
As for the evolution produced by $\widetilde{S}_{\hilbert}$ in the full Hilbert space, one can now define the orthogonal projector onto the space orthogonal to the uMPS $\ket{\Psi(A)}$ as
\begin{equation}
\operator{P}_{0}(\overline{A},A)=\operator{\one}-\frac{\ket{\Psi(A)}\bra{\Psi(\overline{A})}}{N(\overline{A},A)}
\end{equation}
and observe that
\begin{align*}
\widetilde{N}_{\overline{\imath},j}(\overline{A},A)&=\frac{\braket{\partial_{\overline{\imath}}\Psi(\overline{A})|\operator{P}_0(\overline{A},A)|\partial_{j}\Psi(A)}}{N(\overline{A},A)},\\
H_{\overline{\imath}}(\overline{A},A)&=\frac{\braket{\partial_{\overline{\imath}}\Psi(\overline{A})|\operator{P}_0(\overline{A},A)\operator{H}|\Psi(A)}}{N(\overline{A},A)}.
\end{align*}
The TDVP equation [Eq.~\eqref{eq:tdvpeq}] can thus be rewritten as
\begin{multline*}
\braket{\partial_{\overline{\imath}}\Psi(\overline{A}(t))|\operator{P}_0(\overline{A}(t),A(t))|\partial_j\Psi(A(t))} \dot{A}^{j}(t) \\
= \braket{\partial_{\overline{\imath}}\Psi(\overline{A}(t))|\operator{P}_0(\overline{A}(t),A(t))\operator{H}|\Psi(A(t))}.
\end{multline*}
which is the same solution that is obtained if one tries to express that $\dot{A}$ minimizes the norm of the difference between both sides of the Schrodinger equation
\begin{displaymath}
\left\lVert \hat{P}_0(\overline{A},A)\left[ \ic \ket{\partial_j\Psi(A)}\dot{A}^{j}-\ham \ket{\Psi(A)}\right]\right\rVert.\label{eq:schrodingernormdifference}
\end{displaymath}
where one only considers the components orthogonal to the current state $\ket{\Psi(A)}$. Note that, as before, this problem does not have a unique minimum and there are many choices $\dot{A}$ that lead to the same physical state $\frac{\d\ }{\d t}\ket{\Psi(A(t))}$. Restricting to $\dot{A}\in\mathbb{B}^{(A)}$ selects a unique solution. Equivalently, one can just minimize $\lVert  \ic \ket{\partial_j\Psi(A)}\dot{A}^{j}-\ham \ket{\Psi(A)}\rVert$ for all choices $\dot{A}\in\mathbb{B}^{(A)}$, in which case the orthogonality with respect $\ket{\Psi(A)}$ is contained in the parametrization and does not have to be included explicitly.

We can also define a projector in Hilbert space that projects onto $\mathbb{T}^{\ket{\Psi(A)}\perp}_{0}$ as
\begin{equation}
\hat{P}_{\mathbb{T}_{0}^{\perp}}=\ket{\partial_i\Psi(A)}\widetilde{N}^{i,\overline{\jmath}}(\overline{A},A)\bra{\partial_{\overline{\jmath}} \Psi(\overline{A})}.
\end{equation}
Note that this projector is in fact independent of the gauge fixing prescription that was used to define $\widetilde{N}^{i,\overline{\jmath}}$. One can check that a sufficient condition such that $\hat{P}_{\mathbb{T}_{0}^{\perp}}$ acts as a projector is that $\widetilde{N}^{i,\overline{\jmath}} \widetilde{N}_{\overline{\jmath},k} \widetilde{N}^{k,\overline{l}}=\widetilde{N}^{i,\overline{l}}$, where we assume that only tangent vectors $\ket{\partial_{k}\Psi(A)}\perp \ket{\Psi(A)}$ are involved and $\braket{\Psi(\overline{A})|\Psi(A)}=1$ such that $\braket{\partial_{\overline{j}}\Psi(\overline{A})|\partial_{k}\Psi(A)}=N_{\overline{j},k}=\widetilde{N}_{\overline{j},k}$.
The TDVP equations can now be written in Hilbert space as
\begin{equation}
\ic\frac{\d\ }{\d t}\ket{\Psi(A(t))}=\hat{P}_{\mathbb{T}_{0}^{\perp}}\hat{H}\ket{\Psi(A(t))}.\label{eq:TDVPinH}
\end{equation}

\subsection{Symplectic properties of real-time evolution}
\label{ss:tdvp:symplectic}
We now associate to any time-independent operator $\operator{F}\in\End(\hilbert)$ the function
\begin{equation}
F: (\overline{A},A) \mapsto \frac{\braket{\Psi(\overline{A})|\operator{F}|\Psi(A)}}{\braket{\Psi(\overline{A})|\Psi(A)}}
\end{equation}
that maps the coordinates $(\overline{A},A)$ of a uMPS $\ket{\Psi(A)}$ in the manifold $\varM$ to its expectation value. In addition, for any two functions $F(\overline{A},A)$ and $G(\overline{A},A)$, we define a Poisson bracket $\{F,G\}$ as
\begin{multline}
\{F,G\}(\overline{A},A)=-\ic\partial_{i}F(\overline{A},A) \widetilde{N}^{i,\overline{\jmath}}(\overline{A},A) \partial_{\overline{\jmath}}G(\overline{A},A)\\
+\ic\partial_{i}G(\overline{A},A) \widetilde{N}^{i,\overline{\jmath}}(\overline{A},A)\partial_{\overline{\jmath}}F(\overline{A},A).\label{eq:poisson}
\end{multline}
We can then write the evolution of the expectation value $O(\overline{A}(t),A(t))$ for a solution $A(t)$ of the TDVP equations as
\begin{equation}
\frac{\d\ }{\d t}O(\overline{A}(t),A(t))=\{O,H\}(\overline{A}(t),A(t)).
\end{equation}
The generalization for time-dependent operators $\operator{O}(t)$ is straightforwardly given by
\begin{multline}
\frac{\d\ }{\d t}O(\overline{A}(t),A(t),t)=\{O,H\}(\overline{A}(t),A(t),t)\\
+\frac{\partial O}{\partial t} (\overline{A}(t),A(t),t).
\end{multline}
However, we can only really compare this to the symplectic structure of classical Hamiltonian dynamics if the definition in Eq.~\eqref{eq:poisson} satisfies all properties of the Poisson bracket, not only the antisymmetry in its arguments but also the Jacobi identity
\begin{equation}
\{F,\{G,H\}\}+\{G,\{H,F\}\}+\{H,\{F,G\}\}=0.
\end{equation}
Writing out this lengthy equation and doing some index substitutions, the non-trivial part of this identity that needs to be checked is whether
\begin{displaymath}
F_i G_{k} H_{\overline{l}} \left(\widetilde{N}^{i,\overline{\jmath}} \partial_{\overline{\jmath}} \widetilde{N}^{k,\overline{l}}-\widetilde{N}^{k,\overline{\jmath}} \partial_{\overline{\jmath}} \widetilde{N}^{i,\overline{l}}\right)=0
\end{displaymath}
together with its cyclic permutations. Let us first explain the usual derivation, which is not applicable here as we explain afterwards. If $\widetilde{N}^{k,\overline{l}}$ would be a proper inverse of $\widetilde{N}_{\overline{l},m}$, we could write
\begin{displaymath}
\partial_{\overline{\jmath}}\widetilde{N}^{k,\overline{l}}=-\widetilde{N}^{k,\overline{m}}\partial_{\overline{\jmath}} \widetilde{N}_{\overline{m},n} \widetilde{N}^{n,\overline{l}}
\end{displaymath}
and use $\partial_{\overline{\jmath}} \widetilde{N}_{\overline{m},n}=\partial_{\overline{m}} \widetilde{N}_{\overline{\jmath},n}=\partial_{\overline{\jmath}}\partial_{\overline{m}}\partial_{n}\log(N)$ to show that the expression inside the brackets is identically zero. However, because $\widetilde{N}^{k,\overline{l}}$ is only a pseudo-inverse, we cannot use the previous expression for $\partial_{\overline{\jmath}}\widetilde{N}^{k,\overline{l}}$. In general, the terms in the round brackets are not zero. It is only when they act on covariant vectors $F_{i}G_{k}H_{\overline{l}}$ that originate from functions which are associated with physical operators and which inherit the gauge invariance of the uMPS, that we can show that that this expression vanishes. The pseudo-inverse $\widetilde{N}^{i,\overline{\jmath}}$ is such that $H_{\overline{l}}=\widetilde{N}_{\overline{l},m}\widetilde{N}^{m,\overline{n}}H_{\overline{n}}$ for gauge invariant functions $H$. If we insert this explicitly in the expression above and use $\widetilde{N}^{k,\overline{l}}\widetilde{N}_{\overline{l},m}=(P_{\mathbb{B}})^{k}_{\ m}$ [Eq.~\eqref{eq:defPB}], we obtain
\begin{multline*}
(\partial_{\overline{\jmath}}\widetilde{N}^{k,\overline{l}}) \widetilde{N}_{\overline{l},m} \widetilde{N}^{m,\overline{n}}=-\widetilde{N}^{k,\overline{l}}\partial_{\overline{\jmath}} \widetilde{N}_{\overline{l},m} \widetilde{N}^{m,\overline{n}}\\
+(\partial_{\overline{\jmath}} (P_{\mathbb{B}})^{k}_{\ m} )\widetilde{N}^{m,\overline{n}}
\end{multline*}
and thus
\begin{align*}
\Big(\widetilde{N}^{i,\overline{\jmath}}&\partial_{\overline{\jmath}}\widetilde{N}^{k,\overline{l}}-\widetilde{N}^{k,\overline{\jmath}}\partial_{\overline{\jmath}}\widetilde{N}^{i,\overline{l}}\Big) \widetilde{N}_{\overline{l},m} \widetilde{N}^{m,\overline{n}}\\
=&-\left(\widetilde{N}^{i,\overline{\jmath}}\widetilde{N}^{k,\overline{l}}\partial_{\overline{\jmath}} \widetilde{N}_{\overline{l},m} -\widetilde{N}^{k,\overline{\jmath}}\widetilde{N}^{i,\overline{l}}\partial_{\overline{\jmath}} \widetilde{N}_{\overline{l},m}\right)\widetilde{N}^{m,\overline{n}}\\
&+\left(\widetilde{N}^{i,\overline{\jmath}}\partial_{\overline{\jmath}} (P_{\mathbb{B}})^{k}_{\ m}-\widetilde{N}^{k,\overline{\jmath}}\partial_{\overline{\jmath}} (P_{\mathbb{B}})^{i}_{\ m}\right)\widetilde{N}^{m,\overline{n}}\\
=&\left(\widetilde{N}^{i,\overline{\jmath}}\partial_{\overline{\jmath}} (P_{\mathbb{B}})^{k}_{\ m}-\widetilde{N}^{k,\overline{\jmath}}\partial_{\overline{\jmath}} (P_{\mathbb{B}})^{i}_{\ m}\right)\widetilde{N}^{m,\overline{n}}
\end{align*}
where we have now properly used $\partial_{\overline{\jmath}} \widetilde{N}_{\overline{m},n}=\partial_{\overline{m}} \widetilde{N}_{\overline{\jmath},n}$. To evaluate the remaining expression, we need the derivative of the projector $P_{\mathbb{B}}$ onto the horizontal subspace. Note that for any vector $B$, the action of $P_{\mathbb{B}}$ on $B$ is to replace it by $B+\mathscr{N}^{(A)}_0(X)$ for some $X$ such that $B+\mathscr{N}^{(A)}_0(X)\in\mathbb{B}$. The map $\mathscr{N}^{(A)}_{p}$ depends only holomorphically on $A$, and the only dependence on $\overline{A}$ can be in the specific $X$ that was used to make the output satisfy the gauge fixing condition. This implies that for any $B$, we obtain
\begin{multline*}
[\partial_{\overline{\jmath}} (P_{\mathbb{B}})^{k}_{\ m} ]B^{m} = \partial_{\overline{\jmath}} [(P_{\mathbb{B}})^{k}_{\ m} B^{m}]\\
=\partial_{\overline{\jmath}} [B^{k}+\mathscr{N}^{(A)k}_0(X)]=\mathscr{N}^{(A)k}_0(\partial_{\overline{\jmath}}X)
\end{multline*}
from which we can infer that the range of $\partial_{\overline{\jmath}} (P_{\mathbb{B}})^{k}_{\ m}$ is always in the vertical subspace $\mathbb{N}^{(A)}_{0}$. The remaining two terms cancel because $F_{i} \partial_{\overline{\jmath}} (P_{\mathbb{B}})^{i}_{\ m} =0 $ and $G_{k}\partial_{\overline{\jmath}} (P_{\mathbb{B}})^{k}_{\ m}=0$. In conclusion, the Jacobi-identity is only satisfied when using gauge invariant functions. In that case, we are really working on the physical manifold $\varM\subset\hilbert$, which unfortunately we only know how to parametrize globally using the overcomplete parametrization in terms of the tensor $A\in\manifold{A}$. 

Indeed, it was shown in Ref.~\onlinecite{2012arXiv1210.7710H} that the manifold of uMPS $\varM$ is a K\"{a}hler manifold, which implies that its metric also defines a symplectic structure, \textit{i.e.}~a real two-form $\omega=\ic \tilde{N}_{\overline{\imath},j} \d z^{j}\wedge \d\overline{z}^{\overline{\imath}}$ that is closed. The fact that it is closed ($\d \omega=0$) relies on $\partial_{\overline{k}}\widetilde{N}_{\overline{\imath},j}=\partial_{\overline{\imath}}\widetilde{N}_{\overline{k},j}$ and thus expresses the essential property for having the Jacobi identity. We can complete the relationship between $\omega$ and the Poisson bracket by defining for every (gauge-invariant) function $F$ the \emph{Hamiltonian vector field}
\begin{multline}
X_{F}(\overline{A},A)= -\ic F_{\overline{\imath}}(\overline{A},A) \tilde{N}^{\overline{\imath}j}(\overline{A},A)\partial_{j}\\
+\ic F_{j}(\overline{A},A)\tilde{N}^{\overline{\imath}j}(\overline{A},A)\partial_{\overline{\imath}}
\end{multline}
and check that
\begin{equation}
\{F,G\}=\d F(X_G)=\omega(X_F,X_G).
\end{equation}
These relations are familiar from classical Hamiltonian mechanics, the only less common ingredient being that the most natural description of phase space is in terms of complex coordinates. 

Under exact integration of the TDVP equations for a time-independent Hamiltonian, the antisymmetry of the Poisson bracket results in $\dot{H}=\{H,H\}=0$, which implies that the energy expectation value $H(\overline{A}(t),A(t))$ is an exact conserved quantity of the TDVP equations. The symplectic properties of the TDVP also conserve other symmetries. Assume that the Hamiltonian is invariant under the action of a unitary symmetry operator $\operator{U}$, such that $[\ham,\operator{U}]=0$. In order to be able to transfer this symmetry to the uMPS manifold $\varM$, we need to assume that for any state $\ket{\Psi(A)}\in\varM$, the action of $\operator{U}$ is mapped to a new state $\ket{\Psi(A_U(A))}=\operator{U}\ket{\Psi(A)}\in\varM$. One particular class of symmetries that fulfill this condition are those for which the symmetry operators $\hat{U}$ decompose into a product of 1-site operators $\hat{U}=\prod_{n\in\mathbb{Z}} \hat{u}_n$ with $\hat{u}$ site independent. We then obtain $A_U^s(A)=\braket{s|\hat{u}|t}A^t$. Because of the unitarity of $\operator{U}$, we have $N(\overline{A}_{U},A_U)= N(A,A)$, where we omit the explicit dependence of $A_U$ on $A$ for the sake of brevity. By taking the logarithm, followed by differentiating with respect to $\overline{A}^{\overline{\imath}}$ and $A^{l}$, we obtain
\begin{equation}
\partial_{\overline{\imath}}\overline{A}^{\overline{j}}_{U} \widetilde{N}_{\overline{\jmath},k}(\overline{A}_U,A_U) \partial_l A_U^k= \widetilde{N}_{\overline{\imath},l}(\overline{A},A),
\end{equation}
The condition $[\ham,\operator{U}]=0$ also allows us to conclude that $H(\overline{A}_{U},A_U)=H(\overline{A},A)$, from which we find
\begin{equation}
\partial_{\overline{\imath}}\overline{A}_U^{\overline{\jmath}} H_{\overline{\jmath}}(\overline{A}_U,A_U)=H_{\overline{\imath}}(\overline{A},A),
\end{equation}
The (modified) metric and the gradient thus transform covariantly under the symmetry transformation and can be used to transform the TDVP equation [Eq.~\eqref{eq:tdvpeq}] into
\begin{multline*}
+\ic\partial_{\overline{\imath}}\overline{A}_U^{\overline{\jmath}}(A(t)) \widetilde{N}_{\overline{\jmath},k}\big(\overline{A}_U(\overline{A}(t)),A_U(A(t))\big)\frac{\d\ }{\d t}A_U^k(A(t))=\\
\partial_{\overline{\imath}}\overline{A}_U^{\overline{\jmath}}\big(A(t))H_{\overline{\jmath}}(\overline{A}_{U}(\overline{A}(t)),A_{U}(A(t))\big)
\end{multline*}
By using the injectivity of the map $A_U(A)$, we can eliminate the Jacobian $\partial_{\overline{\imath}}\overline{A}_U^{\overline{\jmath}}(A)$ in order to obtain the correct TDVP equation in terms of the new coordinates $A_U(t)$. Hence, it should be straightforward to implement symmetry-adapted version of the TDVP with increased computational efficiency.

Finally, we study the case where the symmetry operator $\operator{U}$ corresponds to a continuous symmetry generated by the Hermitian generator $\operator{K}\in\End(\hilbert)$, with $[\operator{K},\operator{H}]=0$. Thus, the expectation value of the generator $\operator{K}$ is conserved under exact evolution. We define a one-parameter family of transformations $\operator{U}(\epsilon)=\exp(\ic \epsilon \operator{K})$. Since we require that for every uMPS $\ket{\Psi(A)}\in\varM$, $\operator{U}(\epsilon)\ket{\Psi(A)}=\ket{\Psi(A_U(A,\epsilon))}\in\varM$, we can differentiate this defining relation with respect to $\epsilon$ and set $\epsilon=0$ in order to learn
\begin{equation}
\ic \operator{K} \ket{\Psi(A)}= \frac{\partial A_U^i}{\partial \epsilon}(A,0)\ket{\partial_{i}\Psi(A)}.
\end{equation}
The action of $\operator{K}$ on a uMPS $\ket{\Psi(A)}$ thus has to be exactly captured in $\Tplane_{0}^{\ket{\Psi(A)}}$, so that we can write
\begin{displaymath}
\operator{P}_{\Tplane_{0}^{\perp}}\operator{K}'\ket{\Psi(A)}=\operator{K}'\ket{\Psi(A)},
\end{displaymath}
where we have used the definition $\hat{K}'=\hat{K}-K(\overline{A},A)$. We then obtain
\begin{multline}
\{H,K\}(\overline{A},A)=\frac{\braket{\Psi(\overline{A})|\ham'\operator{K}'-\hat{K}'\hat{H}'|\Psi(A)}}{\braket{\Psi(\overline{A})|\Psi(A)}}\\
=\frac{\braket{\Psi(\overline{A})|[\operator{H}-H(\overline{A},A),\operator{K}-K(\overline{A},A)]|\Psi(A)}}{\braket{\Psi(\overline{A})|\Psi(A)}}=0.
\end{multline}
Generators of continuous symmetry transformations are thus constants of motion of the evolution according to the time-dependent variational principle, provided that the symmetry transformation can be captured exactly in the uMPS manifold $\varM$.

\subsection{Properties of imaginary time evolution}
\label{ss:imaginary}
The TDVP equation [Eq.~\eqref{eq:tdvpeq}] can also be used to simulate imaginary time evolution by setting $t=-\ic \tau$. Stationary solutions of the TDVP equation satisfy $H_{\overline{\jmath}}(\overline{A},A)=0$, where $H$ is the energy functional that is the central quantity of the time-independent variational principle (TIVP). Hence, stationary solutions of both real or imaginary time evolution governed by the TDVP equation correspond to extremal solutions of the TIVP. But whereas real time evolutions do typically not converge and are thus not required to end up in such extremal solutions, imaginary time evolution necessarily has to converge for $\tau\to\infty$ to a solution with $H_{\overline{\jmath}}(\overline{A},A)=0$. The reason for this is the rate of change of the energy expectation value is given by
\begin{equation}
\frac{\d }{\d \tau} H=-2 H_i \widetilde{N}^{i,\overline{\jmath}} H_{\overline{\jmath}} \leq 0.\label{eq:energyrate}
\end{equation}
where we have omitted the arguments $(\overline{A}(\tau),A(\tau))$ for the sake of brevity. Hence, the energy expectation value decreases monotonically under imaginary time evolution. Note that even for our infinite size system, the previous equation makes sense, since $H$ is proportional to the number of sites $\lvert \mathbb{Z}\rvert$, and the pseudo-inverse metric contains a factor $\lvert\mathbb{Z}\rvert^{-1}$. 

In the full Hilbert space $\hilbert$, imaginary time evolution will converge any random initial state to the exact ground state, provided that the initial state is not orthogonal to this ground state. Note that imaginary time evolution in combination with the modified Schr\"odinger equation [Eq.~\eqref{eq:schrodingermod}] does not change the norm of the state. Imaginary time evolution then describes a continuous version of steepest descent for a convex energy function $H(\overline{\Psi},\Psi)$ in the convex subspace of constant norm $\braket{\Psi|\Psi}$ and thus converges monotonically to the unique minimum\footnote{The full energy function $H(\overline{\Psi},\Psi)$ in the restriction of $\hilbert$ to a convex subspace of constant norm and phase can have saddle points corresponding to higher eigenstates, but has a single minimum corresponding to the ground state.}.

In the restricted manifold of uMPS, the energy functional $H(\overline{A},A)$ might have many local minima, and there is no guarantee that the flow of the TDVP converges towards the global minimum (which is assumed to provide the best approximation of the exact ground state). However, if the uMPS manifold $\varM$ is able to accurately approximate the exact imaginary time evolution, one can hope that the flow inherits the global minimizing character of the exact imaginary time flow and does indeed converge to the global optimum for most random initial states. Note that imaginary time evolution according to the TDVP equation [Eq.~\eqref{eq:tdvpeq}]  does not resemble a simple steepest descent in parameter space, since the (pseudo-inverse) metric explicitly takes the geometry of the manifold into account. It would thus be worthwhile to investigate geometrically covariant formulations of more advanced optimization methods as discussed in Ref.~\onlinecite{Absil:2009uy} for finding a uMPS ground state approximation.

\subsection{A simple implementation}
\label{ss:tdvp:implementation}
We now discuss a simple first order implementation, which can be considered as an improved version of the algorithm originally introduced in \cite{2011PhRvL.107g0601H}. More advanced numerical integration schemes will be discussed elsewhere\cite{inprep}. 

Let us first discuss how to compute $\tilde{N}^{i,\overline{\jmath}}H_{\overline{\jmath}}(\overline{A},A)$. For either choice of the gauge fixing conditions [Eq.~\eqref{eq:leftgaugeuB} or Eq.~\eqref{eq:rightgaugeuB}], one non-local term survives in the expression for $\braket{\Phi_{p}(\overline{B};\overline{A})|\ham|\Psi(A)}$ [Eq.~\eqref{eq:grad}], which requires the computation of the pseudo-inverse of $\voperator{\one}-\voperator{E}$. An exact computation of $(\voperator{\one}-\voperator{E})^{\mathsf{P}}$ would be an operation that scales as $\order(D^{6})$, but an iterative strategy is also possible. If we represent $B$ as $\mathscr{B}(X)$ [Eq.~\eqref{eq:defBrepresentation}], so that the left gauge fixing conditions are fulfilled, we have to compute
\begin{equation}
\rbra{K}=\rbra{l}\voperator{H}^{AA}_{AA}(\voperator{\one}-\voperator{E})^{\mathsf{P}}=\rbra{l}\voperator{H}^{AA}_{AA}\voperator{Q}(\voperator{\one}-\voperator{Q}\voperator{E}\voperator{Q})^{-1}\voperator{Q}
\end{equation}
where the last $\voperator{Q}$ is obsolete and where $\rbra{l}\voperator{H}^{AA}_{AA}\voperator{Q}$ can be computed efficiently. Since the action of $(\voperator{\one}-\voperator{Q}\voperator{E}\voperator{Q})$ on a vector $\rbra{K}$ can also be implemented as an operation with computational efficiency $\order(D^{3})$ using the maps $\mathscr{E}$ and $\widetilde{\mathscr{E}}$ [Eq.~\eqref{eq:defEpsilon} and Eq.~\eqref{eq:defEpsilontilde}], and since $(\voperator{\one}-\voperator{Q}\voperator{E}\voperator{Q})$ itself is non-singular, an iterative solver such as the biconjugate gradient stabilized method can be used to compute $\rbra{K}$ with a computational cost that scales as $\order(D^{3})$. Setting $C^{s,t}=\sum_{u,v=1}^{d}\braket{s,t|\hat{h}|u,v} A^{u}A^{v}$, we then also define
\begin{multline}
F=\sum_{s,t=1}^{d} {V^{s}_{L}}^{\dagger} l^{1/2} C^{s,t} r {A^{t}}^{\dagger} r^{-1/2}\\
+\sum_{s,t=1}^{d}{V^{s}_{L}}^{\dagger} l^{-1/2} {A^{t}}^{\dagger} l C^{t,s} r^{1/2}+ \sum_{s=1}^{d}  {V^{s}_{L}}^{\dagger} l^{-1/2} KA^{s} r^{1/2},\label{eq:defF}
\end{multline}
in order to obtain
\begin{equation}
\braket{\Phi_{p}(\overline{\mathscr{B}}(\overline{X}))|\ham|\Psi(A)}=2\pi\delta(p)\tr\left[X^{\dagger}F\right].
\end{equation}
The solution to the minimization of $\big\lVert \ket{\Phi_{0}(B)}-\operator{H}\ket{\Psi(A)}\big\rVert^{2}$ is then given by $B=\mathscr{B}(F)$.

A simple first order update scheme is obtained by using the Euler update rule $A(t+\d t) = A(t) -\ic \d t \mathscr{B}(F)$ in case of real time evolution or $A(\tau+\d \tau) = A(t) -\d \tau \mathscr{B}(F)$. $F$ is computed using the current state $A(t)$ or $A(\tau)$ and $\d t$ or $\d \tau$ is the chosen real or imaginary time step. This implementation for uniform matrix product states requires $\order(N_{\text{iter}} D^{3})$ operations, with $N_{\text{iter}}$ the number of iterations necessary in the iterative eigensolvers for $l$, $r$ and in the iterative linear solver for $K$.

In the case where $A(t)$ also satisfies the left orthonormalization condition $\sum_{s=1}^{d} A^{s}(t)^{\dagger} A^s(t)=\one_{D}$, the left gauge fixing condition for the tangent vector $B$ imposes that this orthonormalization is preserved up to first order: $\sum_{s=1}^{d} A^{s}(t+\d t)^{\dagger} A^{s}(t+\d t)= \one_{D}+(\d t)^2 \sum_{s=1}^{d} B^{s\dagger} B^{s}$.  The terms of order $\d t$ are zero because they correspond exactly to the left hand side of the left gauge fixing condition [Eq.~\eqref{eq:leftgaugeuB}] and its Hermitian conjugate. We can then use an improved update rule such that $A(t+dt)$ satisfies the left orthonormalization exactly. We therefore define $[A]$ as a matrix representation of $A$ with dimension $Dd \times D$, where the physical index $s=1,\ldots,d$ and the row index of the matrices $A^s$ is combined: $[A]_{(\alpha s),\beta}=A^{s}_{\alpha,\beta}$. The left orthonormalization expresses the isometric character of $[A]$ (\textit{i.e.}~$[A]^{\dagger} [A]=\one_{D}$). We now state the alternative update rule for real-time evolution
\begin{equation}
[A(t+\d t)]=\exp(-\ic\d t \{ [B][A(t)]^{\dagger}+[A(t)][B]^{\dagger}\})[A(t)].
\end{equation}
or for imaginary time evolution
\begin{equation}
[A(t+\d t)]=\exp(-\d \tau\{ [B][A(t)]^{\dagger}-[A(t)][B]^{\dagger}\})[A(t)].
\end{equation}
Expanding the exponential to first order and using $[B]^{\dagger}[A(t)]=0$ shows that this update rule is equivalent at first order in $\d t$. Since the argument of the matrix exponential is antihermitian, the isometric character of $A$ is exactly preserved. In case of large bond dimension $d$, it might be disadvantageous to compute a matrix exponential of a $dD\times dD$ matrix. We can expand the exponential into its Taylor series and, using the properties of $A$ and $B$, resum the series to obtain
\begin{equation}
A^{s}(t+\d t)= A^s(t)\cos\left(\d t \lvert B\rvert\right)  + \ic  B^s \lvert B\rvert^{-1} \sin\left(\d t\lvert B\rvert\right)
\end{equation}
in case of real time evolution, or
\begin{equation}
A^{s}(\tau+\d \tau)= A^s(\tau)\cos\left(\d \tau \lvert B\rvert\right)  -  B^s \lvert B\rvert^{-1} \sin\left(\d \tau\lvert B\rvert\right)
\end{equation}
in case of imaginary time evolution. In these equations, the $D\times D$ matrix $\lvert B\rvert$ is given by
\begin{equation}
\lvert B\rvert=\left([B]^{\dagger}[B]\right)^{1/2}=\left(\sum_{s=1}^{d} B^{s\dagger} B^{s}\right)^{1/2}.
\end{equation}
Hence, instead of one matrix exponential of a $dD\times dD$ matrix, we have to compute a matrix sine and cosine of a $D\times D$ matrix, which boils down to two matrix exponentials of this $D\times D$ matrix.

\subsection{Error and convergence measures}
\label{ss:error}
According to the geometric formulation of the TDVP [Eq.~\eqref{eq:TDVPinH}], we can assess the difference between the TDVP evolution and the exact evolution as
\begin{equation}
\begin{split}
\epsilon(\overline{A},A)&=\lVert [\hat{\one}-\hat{P}_{\mathbb{T}_0^{\perp}}] [\hat{H}-H(\overline{A},A)]\ket{\Psi(A)}\rVert\\
&=\sqrt{\Delta H(\overline{A},A)^2-\eta(\overline{A},A)^2}
\end{split}
\end{equation}
where we have defined the norm of the TDVP evolution vector as
\begin{equation}
\eta(\overline{A},A)=\lVert\hat{P}_{\mathbb{T}_0^{\perp}} \hat{H}\ket{\Psi(A)}\rVert
\end{equation}
and the energy deviation as
\begin{equation}
\Delta H(\overline{A},A)=\braket{\Psi(\overline{A})|(\hat{H}-H(\overline{A},A))^2|\Psi(A)}^{1/2}.
\end{equation}
For an infinite system, each of these quantities diverges as $\lvert\mathbb{Z}\rvert^{1/2}$. This is an infinitesimal manifestation of the orthogonality catastrophe, which for the case of uMPS expresses the fact that any two injective uMPS that are not gauge-equivalent, are necessary orthogonal. Consequently, even the infinitesimal variation from $\ket{\Psi(A)}$ to $\ket{\Psi(A+\d A)}$ corresponds to a state $\ket{\Phi_0^{(A)}(\d A)}$ with diverging norm.
\begin{widetext}
However, we are mostly interested in local properties of systems and should thus use error measures based on regions of finite size. A suitable length scale for such a region which guarantees good global properties of the state is given by the correlation length $\xi$. For example, for a Hamiltonian $\ham=\sum_{n\in\mathbb{Z}} \hat{T}^n \hat{h} \hat{T}^{-n}$ where $\hat{T}$ is the shift operator and the local interaction term $\hat{h}$ only acts on nearest neigbor sites, we obtain
\begin{equation}
\begin{split}
\Delta H(\overline{A},A)^{2}&=\braket{\Psi(\overline{A})|\left[\sum_{n\in\mathbb{Z}}\operator{T}^{n}(\operator{h}-h(\overline{A},A)\operator{T}^{-n}\right]^{2}|\Psi(A)}\\
&=\lvert\mathbb{Z}\rvert\bigg(\sum_{n=-1}^{1}\braket{\Psi(\overline{A})|(\operator{h}-h(\overline{A},A))\operator{T}^{n}(\operator{h}-h(\overline{A},A))|\Psi(A)}+2\rbraket{l|\voperator{E}^{C}_{AA}(\voperator{\one}-\voperator{E})^{\mathsf{P}}\voperator{E}^{AA}_{C}|r}\bigg).\label{eq:mps:defdh}
\end{split}
\end{equation}
The first term results in
\begin{displaymath}
\braket{\Psi|(\operator{h}-h)(\operator{h}-h)|\Psi}= \sum_{s,t,u,v=1}^{d}\braket{u,v|(\operator{h}-h)^{2}|s,t}\rbraket{l|A^{s}A^{t}\otimes \overline{A}^{u} \overline{A}^{v}|r}=\Delta h(\overline{A},A)^{2}
\end{displaymath}
and
\begin{displaymath}
\braket{\Psi|(\operator{h}-h)\operator{T}(\operator{h}-h)|\Psi}=\sum_{r,s,t,u,v,w=1}^{d}\braket{u,v,w|(\operator{h}-h)\operator{T}(\operator{h}-h) \operator{T}^{-1}|r,s,t}\rbraket{l|A^{r}A^{s}A^{t}\otimes \overline{A}^{u} \overline{A}^{v}\overline{A}^{w}|r},
\end{displaymath}
for $n=0$ and $n=1$ respectively, and in the complex conjugate of the last expression for $n=-1$. By taking out the diverging factor $\lvert Z\rvert^{1/2}$ from $\Delta H(\overline{A},A)$, we obtain a local and finite measure for the energy deviation which is based on the correlation between any two Hamiltonian terms up to the correlation length. We similarly define a local measure
\begin{displaymath}
\widetilde{\eta}(\overline{A},A)=\lvert Z\rvert^{-1/2} \eta(\overline{A},A)=\left[\rbraket{l|\voperator{E}^{B'}_{B}|r}+\rbraket{l|\voperator{E}^{A}_{B} \voperator{Q} (\voperator{\one}-\ec^{\ic p}\voperator{Q}\voperator{E}\voperator{Q})^{-1}\voperator{Q}\voperator{E}^{B'}_{A}|r}+\rbraket{l|\voperator{E}^{B'}_{A} \voperator{Q}(\voperator{\one}-\ec^{-\ic p}\voperator{Q}\voperator{E}\voperator{Q})^{-1}\voperator{Q}\voperator{E}^{A}_{B}|r}\right]
\end{displaymath}
where $\rbraket{l|\voperator{E}^{B}_{A}|r}$ was assumed. Using the representation $B=\mathscr{B}(x)$ with $x=F$ as constructed in Eq.~\eqref{eq:defF}, we obtain
\begin{equation}
\widetilde{\eta}(\overline{A},A)=\sqrt{\tr[F^{\dagger} F]}=\lVert F\rVert_{\mathrm{F}},
\end{equation}
where $\lVert\cdot \rVert_{\text{F}}$ denotes the Frobenius norm. A local measure for the error made by the TDVP evolution with respect to the full evolution is then given by
\begin{equation}
\tilde{\varepsilon}(\overline{A},A)= \sqrt{\Delta H(\overline{A},A)^{2}/\lvert\mathbb{Z}\rvert - \widetilde{\eta}(\overline{A},A)^{2}}.\label{eq:mps:deflocalepsilon}
\end{equation}
When the exact time evolution is accurately captured in the manifold of (uniform) matrix product states, $\tilde{\varepsilon}$ contains the difference of two terms which are of comparable size. In addition, the computation of $\Delta H(\overline{A},A)^{2}/\lvert\mathbb{Z}\rvert$ contains four terms that can be both positive and negative and can neutralize each other. This can result in large numerical errors in the computation of these quantities. A better strategy for evaluating $\widetilde{\epsilon}(\overline{A},A)$ as a sum of strictly positive terms is constructed in Subsection~\ref{ss:extensions:dynamic}.
\end{widetext}

For imaginary-time evolution, we expect the evolution to converge to a point $\ket{\Psi(A^{\ast})}$ where $\widetilde{\eta}(\overline{A}^{\ast},A^{\ast})=0$. We can easily motivate that at any point $\ket{\Psi(A(\tau))}$ in the evolution, $\widetilde{\eta}(\overline{A}(\tau),A(\tau))$ can be used as a local measure for the difference between the current uMPS and the final state $\ket{\Psi(A^{\ast})}$. Indeed, we can show that the change of the expectation value of an operator $\hat{O}$ with support on $N$ sites satisfies
\begin{displaymath}
\left\lvert\frac{\d }{\d \tau} O\big(\overline{A}(\tau),A(\tau)\big)\right\rvert\leq c \lVert \hat{O}\rVert (2\xi+N+2) \widetilde{\eta}\big(\overline{A}(\tau),A(\tau)\big)
\end{displaymath}
where $c$ is some constant and $\xi$ is the correlation length, set by the second largest eigenvalue of the transfer matrix as in Eq.~\eqref{eq:defcorr}. One special example is when we look at the expectation value of the Hamiltonian density $\hat{h}$, for which we obtain
\begin{displaymath}
\frac{\d }{\d \tau} h\big(\overline{A}(\tau),A(\tau)\big)=-\widetilde{\eta}(\overline{A}(\tau),A(\tau))^2.
\end{displaymath}
Thus, the energy density $h(\overline{A}(\tau),A(\tau))$ converges quadratically as fast as the local measure for the state error $\widetilde{\eta}$, a result that parallels the quadratic convergence of the total energy in the global state error. 

Having reached a minimum $\ket{\Psi(A^{\ast})}$ of the uMPS manifold [$\widetilde{\eta}(\overline{A},A)=0$], we can use the local error measure $\widetilde{\epsilon}(\overline{A}^{\ast},A^{\ast})=\Delta H(\overline{A}^{\ast},A^{\ast})/\lvert Z\rvert^{1/2}$ to assess the difference between the variational optimum and the exact ground state.
\section{Variational ansatz for excitations}
\label{s:excitations}
The DMRG was originally developed for finding ground states of strongly correlated quantum lattice systems in one spatial dimension. By applying the variational principle to a state that is enforced to be orthogonal to previously found states, low-lying excited states on finite lattices can be found. Typically, these are not the states that one is interested in. On the finite lattice with open boundary conditions, the momentum quantum number does not exist. Low-lying excited states can easily be related to boundary effects and have no relation to the momentum eigenstates in the bulk of a macroscopic system. In the thermodynamic limit, the suggested approach fails, since any two states are likely to be orthogonal due to the orthogonality catastrophe. Even if we were able to construct a uMPS approximation for the lowest lying excited state with momentum zero, the finite excitation energy would spread out over an infinite lattice and is undetectable from computing the expectation value of the energy density. States with a different energy density as the ground state contain a finite density (and thus an infinite number) of elementary excitations. On a more mathematical level, we do not expect the class of matrix product states in the thermodynamic limit to have the correct properties for describing elementary excited states, since MPS are normalizable, and excited states with definite momentum are not. 

Two different strategies for solving this problem emerge. Information about the spectrum of excited states can be obtained from the pole structure of the Fourier transform of dynamic correlation functions. Initially, algorithms for directly evaluating these correlation functions in frequency domain were developed\cite{1995PhRvB..52.9827H,1999PhRvB..60..335K,2002PhRvB..66d5114J,2008arXiv0808.2620J}. But since the development of the TEBD, time evolution can be approximated so well that modern state-of-the-art algorithms first compute the time-dependent correlation function for some finite interval $t\in [0,T]$, and then compute the Fourier transform \cite{Pereira:2008aa,White:2008aa}. Starting from a disturbance in the ground state of a large but finite lattice, the time evolution can be computed for any time $T$ below which the information of the disturbance has not yet reached the edges of the lattice. The finite time $T$ results in a broadening of the spectral function, but by combining advanced linear prediction techniques to extend $T$ beyond the computable range with complex statistical machinery for isolating the location of the poles, a fairly accurate determination of the dispersion relation of the elementary excitation in the Heisenberg model was obtained \cite{White:2008aa}. Because of the (approximately) linear increase of entropy under time evolution\cite{2006PhRvL..97e0401B,2006PhRvL..97o0404E}, very large bond values are required in order to accurately approximate the time evolution all the way up to time $T$. This is in sharp contrast with the observation that low-lying excited states also satisfy an approximate area law for the scaling of entanglement entropy. For free field theories, power-law corrections to the area law were found when the field is in a superposition of its ground state and low-lying excited states \cite{2006PhRvD..73l1701D,2007CQGra..24.5299D,2008PhRvD..77f4013D}. For large areas, these are negligible and the area law still holds. An area law was also found for low-lying excited states in an integrable one-dimensional lattice model \cite{2009JSMTE..10..020A}. In Ref.~\onlinecite{2009PhRvA..80e2104M} an area law is proven for all low-energy states ---not restricted to eigenstates--- of short-range interacting lattice models under some technical conditions, including a sufficiently rapid decay of connected correlation functions and an upper bound on the number of low-lying excitations in a subsystem corresponding to a compact spatial region.

\begin{widetext}
Given these considerations on the entanglement entropy of excited states, an alternative strategy is thus to construct a variational ansatz that is suited to directly probe the spectrum of excited states. Nevertheless, variational ansatzes for excited states based on the matrix product concept have not been very common. Studying energy-momentum dispersion relations seems to automatically redirect us to a lattice with periodic boundary conditions, for which the MPS algorithms are less efficient, unless one can work in the thermodynamic limit. However, given the remarks above, a direct construction in the thermodynamic limit seems far from trivial. The first proposal for a variational ansatz for excitations was made by Rommer and \"Ostlund \cite{1995PhRvL..75.3537O,1997PhRvB..55.2164R}. In their seminal work on MPS, they also suggested to study excitations with momentum $p=2\pi n/N$ on a ring of length $N$ using the ansatz
\begin{equation}
\ket{\widetilde{\Phi}_{p}(x)}=\sum_{n=1}^{N}\ec^{\ic p n}\operator{T}^{n}\sum_{\{s_n\}=1}^{d}\tr\left[xA^{s_{1}}A^{s_{2}} \cdots A^{s_{N}} \right] \ket{s_{1}s_{2}\cdots s_{N}},\label{eq:mps:rommerostlund}
\end{equation}
which allowed them to get an early estimate of the Haldane gap in the spin $1$ Heisenberg chain. The matrices $A^{s}$ are fixed to the value for which the uniform matrix product state $\ket{\Psi(A)}$ (of finite size $N$) best approximates the ground state, and one can hope that several branches of the energy-momentum spectrum can be captured by different values of $x$. The rationale of this ansatz is that low-lying excited states can be described as a momentum superposition of a local disturbance, which is encoded in the virtual system using the virtual operator $x\in\End(\mathbb{C}^{D})$. Using an analytic series expansion in the system size $N$, Rommer and \"{O}stlund were even able to extrapolate their results to the thermodynamic limit. A different type of variational class corresponds to the so-called ``projected entangled multipartite states'', given by the ansatz  \cite{2006PhRvB..73a4410P}
\begin{equation}
\ket{\Upsilon_{p}[A]}=\frac{1}{\sqrt{N}}\sum_{n=1}^{N}\ec^{\ic p n}\operator{T}^{n}\sum_{\{s_n\}=1}^{d}\tr\left[A^{s_{1}}(1) A^{s_{2}}(2) \cdots A^{s_{N}}(N) \right] \ket{s_{1}s_{2}\cdots s_{N}},\label{eq:mps:pems}
\end{equation}
which contains a momentum superposition of the non-translation invariant MPS. Here all matrices $A^{s}(n)$ are variational parameters, and different branches of the spectrum are obtained by creating mutually orthogonal states at a fixed momentum $p$. This specific superposition is expected to be able to introduce long-range information: writing $\ket{\Upsilon_{0}[A]}$ as a uMPS $\ket{\tilde{\Psi}(\tilde{A})}$ requires a bond dimension $\tilde{D}=ND$ if $D$ represents the bond dimension of the matrices $A^s(n)$. The computational complexity of this algorithm scales as $\order(N^{2}D^{5})$, and it is thus restricted to lattices of moderate size and small values of the bond dimension $D$. This last aspect is partially compensated by the higher entanglement that is allowed in this state. 
\end{widetext}

The idea that low-lying excited states can be regarded as (momentum superpositions of) local disturbances on the ground state is of course inspired by the case of quadratic theories, where creation operators $\operator{a}^{\dagger}(p)$ can be defined that create elementary excitations when acting on the ground state. For elementary excitations, this pointlike structure is often a good assumption. Bijl, Feynman and Cohen generalized this concept by acting on the ground state with general operators $\operator{O}(p)$, which represent the Fourier transform of some local operator $\operator{O}$ with compact support, for studying excitations in liquid Helium \cite{Bijl:1941aa,Feynman:1954aa,Feynman:1956aa}. If $\{\operator{O}^{\alpha}\}$ represents a complete set of local observables, then the Feynman-Bijl operator $\operator{O}$ can be expanded as $\operator{O}=c_{\alpha}\operator{O}^{\alpha}$ and $\{c_{\alpha}\}$ can be treated as the set of variational parameters. This ansatz was first used in the context of spin systems by Arovas, Auerbach and Haldane \cite{Arovas:1988aa} and is then referred to as the single-mode approximation. The single-mode approximation was first combined with matrix product states in Ref.~\onlinecite{2003EPJB...31..209B}, and generalized to local operators acting on up to $4$ sites\cite{2009PhLA..373.2277C}. 

It is now clear that the tangent vectors $\ket{\Phi_{p}(B)}$ defined in Eq.~\eqref{eq:defumpstangentp} generalize both the construction of \"{O}stlund and Rommer, where the excitation is represented as an operator $x$ in the virtual space (choose $B^s=x A^s$), and the single-mode approximation, where the excitation is represented as an operator $\hat{O}$ in the physical space (choose $B^s=\braket{s|\hat{O}|t}A^t$). Feynman-Bijl operators with a larger support of $n>1$ sites are not strictly included in this variational class, but by transferring information along the virtual space all operators acting on $n\approx 2\log_{q} D$ sites are effectively included. We can even hope that the $D$ left and $D$ right Schmidt vectors throw away irrelevant information on the nearest sites in favor of keeping relevant information on sites that are further away. In the final section of this paper, we will also discuss a generalized excitation ansatz where we replace the ground state matrices $A$ on a contiguous block of several sites. The ansatz $\ket{\Phi_{p}(B)}$ was used on a ring of $N$ sites in Ref.~\onlinecite{2012PhRvB..85c5130P}. However, the real power of this ansatz is unleashed by using it in the thermodynamic limit, where we can hopefully reproduce the $\order(D^{3})$ scaling that we have grown accustomed to from DMRG and that was also reproduced in the TDVP calculations. This of course requires that we can consistently compute the variational excitation energy directly in the thermodynamic limit. In Subsection~\ref{ss:excitations:toptriv} we will illustrate how to compute the expectation value of the Hamiltonian $\ham$ in full detail, using the same techniques as in the previous section. Being able to formulate our techniques in the thermodynamic limit also allows us to describe topologically non-trivial excitations, as is sketched in Subsection~\ref{ss:excitations:topnontriv}. In one spatial dimension, topologically non-trivial excitations commonly appear in systems with discrete symmetry breaking as kinks or domain walls that interpolate between two ground states with a different value of the order parameter. This ansatz includes the topologically non-trivial analogue of the Feynman-Bijl operators, which are the Mandelstam operators \cite{Mandelstam:1975aa}. Subsection~\ref{ss:excitations:reltdvp} also relates our excitation ansatz to the TDVP from the previous section. Finally, by noting that we have a variational estimate not only for the excitation energies but also for the corresponding wave functions, we look at dynamical correlation functions in Subsection~\ref{ss:excitations:spectral}.

\subsection{Topologically trivial states}
\label{ss:excitations:toptriv}
Let $\ham$ be a given translation invariant Hamiltonian on an infinite lattice, which we assume to contain only nearest neighbor interactions for reasons of notational simplicity: $\ham=\sum_{n\in\mathbb{Z}} \operator{T}^{n}\operator{h}\operator{T}^{-n}$ with $\hat{T}$ the shift operator. We assume that the ground sate is well approximated by a uMPS $\ket{\Psi(A)}\in \varM$. We henceforth assume that $A$ is the value of at least a local ---and hopefully the global--- minimum. We again assume that the uMPS $\ket{\Psi(A)}$ is pure ($A\in\manifold{A}$) and normalized to unity, so that $\voperator{E}$ has a unique eigenvalue $1$ and all the other eigenvalues lie strictly within the unit circle. We now apply the time-independent variational principle to the set of states $\ket{\Phi_{p}^{(A)}(B)}\in\Tplane_{p}^{\ket{\Psi(A)}}$. Since we are interested in excited states, we need to impose orthogonality to the ground state approximation $\ket{\Psi(A)}$. We can thus restrict to $\Tplane^{\ket{\Psi(A)}\perp}$. As in the case of TDVP, this restriction enters naturally. We can thus recycle the parameterization $B=\mathscr{B}^{(A)}(X)$ from Eq.~\eqref{eq:defBrepresentation} in terms of the $D(d-1)\times D$ matrix $X$. With $A$ assumed to be fixed throughout this section, we omit the explicit reference to $A$ and $\ket{\Psi(A)}$ in the notation of the states $\ket{\Phi_{p}(B)}$, the spaces $\Tplane_{p}^{\perp}$ and the representation $\mathscr{B}(X)$. Since our variational manifold for excitations corresponds to a linear subspace $\Tplane_{p}^{\perp}$ of Hilbert space $\hilbert$, for which we have a linear representation through the series of linear maps $X\mapsto \mathscr{B}(X)\mapsto \ket{\Phi(\mathscr{B}(X))}$, the variational optimization problem reduces to a Rayleigh-Ritz problem and we will have to solve a generalized eigenvalue equation in $X$. In fact, because of the way the representation $\mathscr{B}(X)$ was constructed, the effective norm matrix constructed in Eq.~\eqref{eq:effnormrepresentation} is proportional to the unit matrix and we end up with normal eigenvalue problem. 

Two remarks are in order. Firstly, the ansatz states $\ket{\Phi_{p}(B)}$ are momentum eigenstates in an infinite volume and can thus not be normalized to unity. Secondly, unlike for the ground state, we cannot restrict to an evaluation of the energy density expectation value. As explained in the introduction, the finite excitation energy in a momentum eigenstate is spread out over the complete lattice, and the energy density
\begin{displaymath}
\frac{\braket{\Phi_{p}(\overline{B})|\operator{h}|\Phi_{p'}(B')}}{\braket{\Phi_{p}(\overline{B})|\Phi_{p'}(B')}}
\end{displaymath}
is indistinguishable from its ground state value $h(\overline{A},A)=\braket{\Psi(\overline{A})|\operator{h}|\Psi(A)}$. We thus have to evaluate the expectation value of the full Hamiltonian $\braket{\Phi_{p}(\overline{B})|\ham|\Phi_{p'}(B')}$, where the excitation energy is present as a finite shift (times the infinite normalization $\braket{\Phi_{p}(\overline{B})|\operator{h}|\Phi_{p'}(B')})$ above a divergent contribution from the extensive ground state energy $H(\overline{A},A)=\lvert\mathbb{Z}\rvert h(\overline{A},A)$ (times the infinite normalization $\braket{\Phi_{p}(\overline{B})|\operator{h}|\Phi_{p'}(B')}$). Subtracting this ground state energy can quickly become a source of errors, as we have to subtract precisely $\lvert\mathbb{Z}\rvert$ times the ground state energy density, and counting errors are easily made. The safest strategy is to subtract $H(\overline{A},A)$ from $\ham$ from the beginning. Note that, unlike in the evaluation of $\braket{\Phi_{p}(\overline{B})|\ham|\Psi(A)}$ that was required for the TDVP, the ground state energy contribution is not automatically subtracted by restricting to tangent vectors $\ket{\Phi_{p}(B)}$ that are orthogonal to $\ket{\Psi(A)}$. We thus redefine $\operator{h}\leftarrow \operator{h}-h(\overline{A},A)$, where $h(\overline{A},A)=\rbraket{l|\voperator{H}^{AA}_{AA}|r}$ [see Eq.~\eqref{eq:defsupopH}]. With this newly defined $\operator{h}$, we obtain $\rbraket{l|\voperator{H}^{AA}_{AA}|r}=0$. 

\begin{widetext}
We are now ready to evaluate the effective Hamiltonian appearing in the Rayleigh-Ritz equation. It corresponds to the restriction of the full Hamiltonian to the subspace $\Tplane_{p}^{\perp}$. Expanding $\braket{\Phi_{p}(\overline{B})|\ham|\Phi_{p'}(B')}$ is a lot more involved then either the norm $\braket{\Phi_{p}(\overline{B})|\Phi_{p'}(B')}$ or the TDVP gradient $\braket{\Phi_{p}(\overline{B})|\ham|\Psi(A)}$, as we now have to deal with three infinite sums. The three summation indices indicate the position of $B$, $B'$ and the first site acted upon by $\operator{h}$. In between these three positions are transfer matrices $\voperator{E}$, which can be decomposed into connected contributions coming from $\voperator{Q}\voperator{E}\voperator{Q}$ and disconnected contributions coming from $\voperator{S}=\rket{r}\rbra{l}$. Thanks to the redefinition of the hamiltonian terms $\operator{h}$, we obtain $\rbra{l}\voperator{H}^{AA}_{AA}\voperator{S}=0$ and $\voperator{S}\voperator{H}^{AA}_{AA}\rket{r}=0$ and no disconnected contributions coming from $\voperator{H}^{AA}_{AA}$  can arise. The connected contributions yield finite results, and we are free to introduce substitutions of the summation indices. Disconnected contributions coming from $\voperator{E}^{A}_{B}$ and $\voperator{E}^{B'}_{A}$ might give rise to additional divergences and should be treated carefully. The total expression is of the general format
\begin{multline*}
\braket{\Phi_{p}(\overline{B})|\operator{H}|\Phi_{p'}(B')}=\sum_{n=-\infty}^{+\infty}\sum_{n'=-\infty}^{+\infty}\sum_{n_{0}=-\infty}^{+\infty}\ec^{\ic p' n' - \ic p n}\left[\text{$B$ at site $n$, $B'$ at site $n'$ and $\operator{h}$ on sites $n_{0}$ and $n_{0}+1$}\right]
\end{multline*}
We first focus on the terms where everything is connected, thus where all transfer operators have been replaced by their corresponding regularized version $\voperator{Q}\voperator{E}\voperator{Q}$. We can now safely introduce the substitution $n'\leftarrow n_{\text{c}}$, $n\leftarrow n_{\text{c}}+\Delta n$ and $n_{0}\leftarrow n_{\text{c}}+\Delta n_{0}$. The summation over $n_{\text{c}}$ immediately yields the momentum conserving factor $2\pi\delta(p'-p)$, since the terms within the summation are independent of the global position $n_{\text{c}}$. If we change $\Delta n$ to $n$ and $\Delta n_{0}$ to $n_{0}$ for notational simplicity and omit the overall factor $2\pi\delta(p'-p)$, we are left with
\begin{align*}
&\rbraket{l|\voperator{H}^{B'A}_{BA}|r}+(l| \voperator{H}^{AB'}_{AB}| r)
+ \sum_{ n_{0}=1}^{+\infty} ( l|\voperator{E}^{B'}_{B}\voperator{Q}\voperator{E}^{n_{0}-1}\voperator{Q}\voperator{H}^{AA}_{AA}| r)+ \sum_{ n_{0}=-\infty}^{-2} \rbraket{l|\voperator{H}^{AA}_{AA}\voperator{Q}\voperator{E}^{-n_{0}-2}\voperator{Q}\voperator{E}^{B'}_{B}|r}\\
&+\sum_{ n=-\infty}^{-1}\ec^{-\ic p  n}\Big[\theta(n=-1)\rbraket{l|\voperator{H}^{AB'}_{BA}|r}
+\rbraket{l|\voperator{E}^{A}_{B}\voperator{Q}\voperator{E}^{ n-1}\voperator{Q} \voperator{H}^{B'A}_{AA}|r}+\rbraket{l|\voperator{H}^{AA}_{AB}\voperator{Q}\voperator{E}^{-n-1}\voperator{Q} \voperator{E}^{B'}_{A}|r}\\
&\qquad\qquad\qquad\quad+\theta( n < -1) \rbraket{l|\voperator{E}^{A}_{B}\voperator{Q}\voperator{E}^{-n-2}\voperator{Q} \voperator{H}^{AB'}_{AA}|r}+\theta(n < -1) \rbraket{l|\voperator{H}^{AA}_{BA}\voperator{Q}\voperator{E}^{-n-2}\voperator{Q} \voperator{E}^{B'}_{A}|r}\\
&\qquad\qquad\qquad\quad+\sum_{n_{0}= 1}^{+\infty}\rbraket{l|\voperator{E}^{A}_{B}\voperator{Q}\voperator{E}^{-n-1}\voperator{Q}\voperator{E}^{B'}_{A} \voperator{Q}\voperator{E}^{n_{0}-1}\voperator{Q} \voperator{H}^{AA}_{AA}|r}+\sum_{n_{0}=-\infty}^{n-2} \rbraket{l|\voperator{H}^{AA}_{AA}\voperator{Q}\voperator{E}^{-n_{0}+n-2}\voperator{Q}\voperator{E}^{A}_{B}\voperator{Q}\voperator{E}^{-n-1}\voperator{Q}\voperator{E}^{B'}_{A}|r}\\
&\qquad\qquad\qquad\quad+\theta(n<-2) \sum_{n_{0}=n+1}^{-2}\rbraket{l|\voperator{E}^{A}_{B}\voperator{Q}\voperator{E}^{-n+n_{0}-1}\voperator{Q}\voperator{H}^{AA}_{AA}\voperator{Q}\voperator{E}^{-n_{0}-2}\voperator{Q}\voperator{E}^{B'}_{A}|r}\Big]\\
&+\sum_{ n=1}^{+\infty}\ec^{-\ic p  n}\Big[\theta(n=1) \rbraket{l|\voperator{H}^{B'A}_{AB}|r}+\rbraket{l|\voperator{E}^{B'}_{A}\voperator{Q}\voperator{E}^{n-1}\voperator{Q} \voperator{H}^{AA}_{BA}|r}+\rbraket{l|\voperator{H}^{AB'}_{AA}\voperator{Q}\voperator{E}^{n-1}\voperator{Q} \voperator{E}_{B}^{A}|r}\\
&\qquad\qquad\qquad\quad+\theta(n>1) \rbraket{l|\voperator{E}_{A}^{B'}\voperator{Q}\voperator{E}^{n-2}\voperator{Q} \voperator{H}_{AB}^{AA}|r}+\theta(n>1) \rbraket{l|\voperator{H}_{AA}^{B'A}\voperator{Q}\voperator{E}^{n-2}\voperator{Q} \voperator{E}_{B}^{A}|r}\\
&\qquad\qquad\qquad\quad+\sum_{n_{0}=n+1}^{+\infty}\rbraket{l|\voperator{E}_{A}^{B'}\voperator{Q}\voperator{E}^{n-1}\voperator{Q}\voperator{E}_{B}^{A} \voperator{Q}\voperator{E}^{n_{0}-n-1}\voperator{Q} \voperator{H}^{AA}_{AA}|r}+\sum_{n_{0}=-\infty}^{-2} \rbraket{l|\voperator{H}^{AA}_{AA}\voperator{Q}\voperator{E}^{-n_{0}-2}\voperator{Q}\voperator{E}_{A}^{B'}\voperator{Q}\voperator{E}^{-n-1}\voperator{Q}\voperator{E}_{B}^{A}|r}\\
&\qquad\qquad\qquad\quad+\theta(n>2) \sum_{n_{0}=1}^{n-2}\rbraket{l|\voperator{E}_{A}^{B'}\voperator{Q}\voperator{E}^{n_{0}-1}\voperator{Q}\voperator{H}_{AA}^{AA}\voperator{Q}\voperator{E}^{n-n_{0}-2}\voperator{Q}\voperator{E}_{B}^{A}|r}\Big].
\end{align*}
The terms on the first line correspond to $n=0$, \textit{i.e.}~where $B$ and $B'$ are on the same site. Then we have all the terms corresponding to $n<0$ and all the terms corresponding to $n>0$. For most terms, we can immediately evaluate the geometric series for $n_{0}$, followed by an evaluation of the additional geometric series in $n$ for some terms. The only exception are the terms with $\theta(n<-2)$ and $\theta(n>2)$, where it is better to first switch the two sums and express the summation bounds of $n$ in terms of $n_{0}$. Then we first evaluate the geometric series in $n$, followed by the one in $n_{0}$. We obtain
\begin{multline*}
\rbraket{l|\voperator{H}^{B'A}_{B A}|r}+\rbraket{l|\voperator{H}^{AB'}_{AB}|r}+\ec^{+\ic p}\rbraket{l|\voperator{H}^{AB'}_{BA}|r}+\ec^{-\ic p}\rbraket{l|\voperator{H}^{B'A}_{AB}|r}+\rbraket{l|\voperator{E}^{B'}_{B}(\voperator{\one}-\voperator{E})^{\mathsf{P}}\voperator{H}^{AA}_{AA}|r}+\rbraket{l|\voperator{H}^{AA}_{AA}(\voperator{\one}-\voperator{E})^{\mathsf{P}}\voperator{E}^{B'}_{B}|r}\\
+\ec^{+\ic p}\rbraket{l|\voperator{E}^{A}_{B}(\voperator{\one}-\ec^{+\ic p}\voperator{E})^{\mathsf{P}}\voperator{E}^{B'}_{A}(\voperator{\one}-\voperator{E})^{\mathsf{P}}\voperator{H}^{AA}_{AA}|r}+\ec^{-\ic p}\rbraket{l|\voperator{E}^{B'}_{A}(\voperator{\one}-\ec^{-\ic p}\voperator{E})^{\mathsf{P}}\voperator{E}^{A}_{B}(\voperator{\one}-\voperator{E})^{\mathsf{P}}\voperator{H}^{AA}_{AA}|r}\\
+\ec^{+\ic p}\rbraket{l|\voperator{H}^{AA}_{AA}(\voperator{\one}-\voperator{E})^{\mathsf{P}}\voperator{E}^{A}_{B}(\voperator{\one}-\ec^{+\ic p}\voperator{E})^{\mathsf{P}}\voperator{E}^{B'}_{A}|r}+\ec^{-\ic p}\rbraket{l|\voperator{H}^{AA}_{AA}(\voperator{\one}-\voperator{E})^{\mathsf{P}}\voperator{E}^{B'}_{A}(\voperator{\one}-\ec^{-\ic p}\voperator{E})^{\mathsf{P}}\voperator{E}^{A}_{B}|r}\\
+\ec^{+\ic p}\rbraket{l|\voperator{E}^{A}_{B}(\voperator{\one}-\ec^{+\ic p}\voperator{E})^{\mathsf{P}}\voperator{H}^{B'A}_{AA}|r}+\ec^{-\ic p}\rbraket{l|\voperator{E}^{B'}_{A}(\voperator{\one}-\ec^{-\ic p}\voperator{E})^{\mathsf{P}}\voperator{H}^{AA}_{BA}|r}\\
+\ec^{+\ic p}\rbraket{l|\voperator{H}^{AA}_{AB}(\voperator{\one}-\ec^{+\ic p}\voperator{E})^{\mathsf{P}}\voperator{E}^{B'}_{A}|r}+\ec^{-\ic p}\rbraket{l|\voperator{H}^{AB'}_{AA}(\voperator{\one}-\ec^{-\ic p}\voperator{E})^{\mathsf{P}}\voperator{E}^{A}_{B}|r}\\
+\ec^{+2\ic p}\rbraket{l|\voperator{E}^{A}_{B}(\voperator{\one}-\ec^{+\ic p}\voperator{E})^{\mathsf{P}}\voperator{H}^{AB'}_{AA}|r}+\ec^{-2\ic p}\rbraket{l|\voperator{E}^{B'}_{A}(\voperator{\one}-\ec^{-\ic p}\voperator{E})^{\mathsf{P}}\voperator{H}^{AA}_{AB}|r}\\
+\ec^{+2\ic p}\rbraket{l|\voperator{H}^{AA}_{BA}(\voperator{\one}-\ec^{+\ic p}\voperator{E})^{\mathsf{P}}\voperator{E}^{B'}_{A}|r}+\ec^{-2\ic p}\rbraket{l|\voperator{H}^{B'A}_{AA}(\voperator{\one}-\ec^{-\ic p}\voperator{E})^{\mathsf{P}}\voperator{E}^{A}_{B}|r}\\
+\ec^{+3\ic p}\rbraket{l|\voperator{E}^{A}_{B}(\voperator{\one}-\ec^{+\ic p}\voperator{E})^{\mathsf{P}}\voperator{H}^{AA}_{AA}(\voperator{\one}-\ec^{+\ic p}\voperator{E})^{\mathsf{P}}\voperator{E}^{B'}_{A}|r}+\ec^{-3\ic p}\rbraket{l|\voperator{E}^{B'}_{A}(\voperator{\one}-\ec^{-\ic p}\voperator{E})^{\mathsf{P}}\voperator{H}^{AA}_{AA}(\voperator{\one}-\ec^{-\ic p}\voperator{E})^{\mathsf{P}}\voperator{E}^{A}_{B}|r}.
\end{multline*} 
The symbolic notation $(\operator{\one}-\ec^{\pm \ic p}\voperator{E})^{\mathsf{P}}=\voperator{Q}(\voperator{\one}-\ec^{\pm \ic p}\voperator{Q}\voperator{E}\voperator{Q})^{-1}\voperator{Q}$ was introduced in the previous section. Only for $p=0$ is this truly a pseudo-inverse. For $p\neq 0$, the $\operator{\one}-\ec^{\pm \ic p}\voperator{E}$ is not really singular. Nevertheless, we had to separate the eigenvalue $\ec^{\pm \ic p}$ with modulus $1$ from the operator $\ec^{\pm \ic p}\voperator{E}$ in order to use the formula for the geometric series. 

We now consider the contributions resulting from disconnecting either $B$ or $B'$. They cannot be disconnected both, since this would also imply that $\operator{h}$ is disconnected, which we've excluded above. Whenever $B'$ appears on the complete left (right) side of a term, and is separated from the rest by a transfer operator $\voperator{E}$, there is such a disconnected contribution. We assume that we can still make the substitution to the global position $n_{\text{c}}$ and the relative positions $n$ and $n_{0}$. Only making substitutions that changes the value of finite bounds in the sum result in a possibility of miscounting contributions and making errors. The summation over the global position again yields the momentum conservation. The total (left and right) contribution from disconnecting $B$ is given by (omitting the momentum conserving factor $2\pi\delta(p'-p)$)
\begin{align*}
\rbraket{l|E^{A}_{B}|r}\bigg[&\rbraket{l| \voperator{H}^{B'A}_{AA}|r}\Big(\sum_{n=-\infty}^{-1} \ec^{-\ic p n} + \sum_{n=2}^{+\infty} \ec^{-\ic p n} \Big) + \rbraket{l| \voperator{H}^{AB'}_{AA}|r} \Big(\sum_{n=-\infty}^{-2} \ec^{-\ic p n} + \sum_{n=1}^{+\infty} \ec^{-\ic p n} \Big)\\
&+\sum_{n=-\infty}^{-1}\ec^{-\ic p n}\sum_{n_{0}=1}^{+\infty}\rbraket{l|\voperator{E}^{B'}_{A} \voperator{Q}\voperator{E}^{n_{0}-1}\voperator{Q} \voperator{H}^{AA}_{AA}|r}+\sum_{n=3}^{+\infty}\ec^{-\ic p n}\sum_{n_{0}= 1}^{n-2}\rbraket{l|\voperator{E}^{B'}_{A} \voperator{Q}\voperator{E}^{n_{0}-1}\voperator{Q} \voperator{H}^{AA}_{AA}|r}\\
& +\sum_{n=-\infty}^{-3}\ec^{-\ic p n} \sum_{n_{0}=n+1}^{-2}\rbraket{l|\voperator{H}^{AA}_{AA}\voperator{Q}\voperator{E}^{-n_{0}-2}\voperator{Q}\voperator{E}^{B'}_{A}|r}+\sum_{n=1}^{+\infty}\ec^{-\ic p n} \sum_{n_{0}=-\infty}^{-2}\rbraket{l|\voperator{H}^{AA}_{AA}\voperator{Q}\voperator{E}^{-n_{0}-2}\voperator{Q}\voperator{E}^{B'}_{A}|r}\bigg].
\end{align*}
These terms should be treated carefully. We expect them to generate a divergence at $p=0$ through a $2\pi \delta(p)$, since we have not expressed orthogonality with respect to the ground state yet. But for any other $p\neq 0$, they should be finite, as orthogonality to the ground state is automatic. By inserting the result for the finite geometric sums in $n_{0}$, we obtain for the terms between the square brackets
\begin{align*}
&\rbraket{l| \voperator{H}^{B'A}_{AA}|r}\Big(2\pi\delta(p)-1-\ec^{-\ic p} \Big) + \rbraket{l| \voperator{H}^{AB'}_{AA}|r} \Big(2\pi\delta(p)-1- \ec^{+\ic p} \Big)+\sum_{n=-\infty}^{-1}\ec^{-\ic p n}\rbraket{l|\voperator{E}^{B'}_{A} (\voperator{\one}-\voperator{E})^{\mathsf{P}} \voperator{H}^{AA}_{AA}|r}\\
&+\sum_{n=3}^{+\infty}\ec^{-\ic p n}\rbraket{l|\voperator{E}^{B'}_{A} (\voperator{\one}-\voperator{E})^{\mathsf{P}}(\voperator{Q}-\voperator{Q}\voperator{E}^{n-2}\voperator{Q})\voperator{H}^{AA}_{AA}|r} +\sum_{n=-\infty}^{-3}\ec^{-\ic p n} \rbraket{l|\voperator{H}^{AA}_{AA}(\voperator{\one}-\voperator{E})^{\mathsf{P}}(\voperator{Q}-\voperator{Q}\voperator{E}^{-n-2}\voperator{Q})\voperator{E}^{B'}_{A}|r}\\
&+\sum_{n=1}^{+\infty}\ec^{-\ic p n} \rbraket{l|\voperator{H}^{AA}_{AA}(\voperator{\one}-\voperator{E})^{\mathsf{P}}\voperator{E}^{B'}_{A}|r}.
\end{align*}
Since $(\voperator{\one}-\voperator{E})^{\mathsf{P}}\voperator{Q}=(\voperator{\one}-\voperator{E})^{\mathsf{P}}$, we can now group the third and fourth term, as well as the fifth and sixth term, and complete the sums in $n$ to $\sum_{n=-\infty}^{+\infty} \ec^{\pm \ic p}=2\pi\delta(p)$, in order to obtain
\begin{align*}
&\rbraket{l| \voperator{H}^{B'A}_{AA}|r}\Big(2\pi\delta(p)-1-\ec^{-\ic p} \Big) + \rbraket{l| \voperator{H}^{AB'}_{AA}|r} \Big(2\pi\delta(p)-1- \ec^{+\ic p} \Big)+\big(2\pi\delta(p)-1-\ec^{-\ic p}-\ec^{-2\ic p}\big)\rbraket{l|\voperator{E}^{B'}_{A} (\voperator{\one}-\voperator{E})^{\mathsf{P}} \voperator{H}^{AA}_{AA}|r}\\
&-\sum_{n=3}^{+\infty}\ec^{-\ic p n}\rbraket{l|\voperator{E}^{B'}_{A} (\voperator{\one}-\voperator{E})^{\mathsf{P}}\voperator{Q}\voperator{E}^{n-2}\voperator{Q}\voperator{H}^{AA}_{AA}|r}+\big(2\pi\delta(p)-1-\ec^{+\ic p}-\ec^{+2\ic p}\big) \rbraket{l|\voperator{H}^{AA}_{AA}(\voperator{\one}-\voperator{E})^{\mathsf{P}}\voperator{E}^{B'}_{A}|r}\\
&-\sum_{n=-\infty}^{-3}\ec^{-\ic p n} \rbraket{l|\voperator{H}^{AA}_{AA}(\voperator{\one}-\voperator{E})^{\mathsf{P}}\voperator{Q}\voperator{E}^{-n-2}\voperator{Q}\voperator{E}^{B'}_{A}|r}.
\end{align*}
Finally, we have to compute two converging geometric series in $n$. Note that the power of $(\voperator{Q}\voperator{E}\voperator{Q})$ starts at one instead of zero (for $n=3$ on line 2 and for $n=-3$ on line 3). We can absorb the part with factor $\ec^{-2\ic p}$ from from the last term of the first line, and the part with factor $\ec^{+2\ic p}$ from the last term of the second line respectively, in order to have a geometric series in $(\voperator{Q}\voperator{E}\voperator{Q})$ starting at power zero. We hence obtain for the total disconnected contribution of $B$
\begin{align*}
\rbraket{l|E^{A}_{B}|r}\bigg[&\rbraket{l| \voperator{H}^{B'A}_{AA}|r}\Big(2\pi\delta(p)-1-\ec^{-\ic p} \Big) + \rbraket{l| \voperator{H}^{AB'}_{AA}|r} \Big(2\pi\delta(p)-1- \ec^{+\ic p} \Big)+\big(2\pi\delta(p)-1-\ec^{-\ic p}\big)\rbraket{l|\voperator{E}^{B'}_{A} (\voperator{\one}-\voperator{E})^{\mathsf{P}} \voperator{H}^{AA}_{AA}|r}\\
&-\ec^{-\ic 2 p}\rbraket{l|\voperator{E}^{B'}_{A} (\voperator{\one}-\voperator{E})^{\mathsf{P}}(\voperator{\one}-\ec^{-\ic p}\voperator{E})^{\mathsf{P}}\voperator{H}^{AA}_{AA}|r}+\big(2\pi\delta(p)-1-\ec^{+\ic p}\big) \rbraket{l|\voperator{H}^{AA}_{AA}(\voperator{\one}-\voperator{E})^{\mathsf{P}}\voperator{E}^{B'}_{A}|r}\\
&-\ec^{+\ic 2 p} \rbraket{l|\voperator{H}^{AA}_{AA}(\voperator{\one}-\voperator{E})^{\mathsf{P}}(\voperator{\one}-\ec^{+\ic p}\voperator{E})^{\mathsf{P}}\voperator{E}^{B'}_{A}|r}\bigg].
\end{align*}
By adding a similar result from disconnecting $B'$, we obtain the final result
\begin{equation}
\begin{split}
\langle\Phi_{p}(\overline{B})|\operator{H}|\Phi_{p'}&(B')\rangle=\overline{B}^{\overline{\imath}} H_{\overline{i},j}(p,p') B^{j} =2\pi\delta(p'-p) H_{\overline{\imath},j}(p)\\
2\pi\delta(p-p')\Bigg\{&\rbraket{l|\voperator{H}^{B'A}_{B A}|r}+\rbraket{l|\voperator{H}^{AB'}_{AB}|r}+\ec^{+\ic p}\rbraket{l|\voperator{H}^{AB'}_{BA}|r}+\ec^{-\ic p}\rbraket{l|\voperator{H}^{B'A}_{AB}|r}\\
&+\rbraket{l|\voperator{E}^{B'}_{B}(\voperator{\one}-\voperator{E})^{\mathsf{P}}\voperator{H}^{AA}_{AA}|r}+\rbraket{l|\voperator{H}^{AA}_{AA}(\voperator{\one}-\voperator{E})^{\mathsf{P}}\voperator{E}^{B'}_{B}|r}\\
&+\ec^{+\ic p}\rbraket{l|\voperator{E}^{A}_{B}(\voperator{\one}-\ec^{+\ic p}\voperator{E})^{\mathsf{P}}\voperator{E}^{B'}_{A}(\voperator{\one}-\voperator{E})^{\mathsf{P}}\voperator{H}^{AA}_{AA}|r}+\ec^{-\ic p}\rbraket{l|\voperator{E}^{B'}_{A}(\voperator{\one}-\ec^{-\ic p}\voperator{E})^{\mathsf{P}}\voperator{E}^{A}_{B}(\voperator{\one}-\voperator{E})^{\mathsf{P}}\voperator{H}^{AA}_{AA}|r}\\
&+\ec^{+\ic p}\rbraket{l|\voperator{H}^{AA}_{AA}(\voperator{\one}-\voperator{E})^{\mathsf{P}}\voperator{E}^{A}_{B}(\voperator{\one}-\ec^{+\ic p}\voperator{E})^{\mathsf{P}}\voperator{E}^{B'}_{A}|r}+\ec^{-\ic p}\rbraket{l|\voperator{H}^{AA}_{AA}(\voperator{\one}-\voperator{E})^{\mathsf{P}}\voperator{E}^{B'}_{A}(\voperator{\one}-\ec^{-\ic p}\voperator{E})^{\mathsf{P}}\voperator{E}^{A}_{B}|r}\\
&+\ec^{+\ic p}\rbraket{l|\voperator{E}^{A}_{B}(\voperator{\one}-\ec^{+\ic p}\voperator{E})^{\mathsf{P}}\voperator{H}^{B'A}_{AA}|r}+\ec^{-\ic p}\rbraket{l|\voperator{E}^{B'}_{A}(\voperator{\one}-\ec^{-\ic p}\voperator{E})^{\mathsf{P}}\voperator{H}^{AA}_{BA}|r}\\
&+\ec^{+2\ic p}\rbraket{l|\voperator{E}^{A}_{B}(\voperator{\one}-\ec^{+\ic p}\voperator{E})^{\mathsf{P}}\voperator{H}^{AB'}_{AA}|r}+\ec^{-2\ic p}\rbraket{l|\voperator{E}^{B'}_{A}(\voperator{\one}-\ec^{-\ic p}\voperator{E})^{\mathsf{P}}\voperator{H}^{AA}_{AB}|r}\\
&+\ec^{+\ic p}\rbraket{l|\voperator{H}^{AA}_{AB}(\voperator{\one}-\ec^{+\ic p}\voperator{E})^{\mathsf{P}}\voperator{E}^{B'}_{A}|r}+\ec^{-\ic p}\rbraket{l|\voperator{H}^{AB'}_{AA}(\voperator{\one}-\ec^{-\ic p}\voperator{E})^{\mathsf{P}}\voperator{E}^{A}_{B}|r}\\
&+\ec^{+2\ic p}\rbraket{l|\voperator{H}^{AA}_{BA}(\voperator{\one}-\ec^{+\ic p}\voperator{E})^{\mathsf{P}}\voperator{E}^{B'}_{A}|r}+\ec^{-2\ic p}\rbraket{l|\voperator{H}^{B'A}_{AA}(\voperator{\one}-\ec^{-\ic p}\voperator{E})^{\mathsf{P}}\voperator{E}^{A}_{B}|r}\\
&+\ec^{+3\ic p}\rbraket{l|\voperator{E}^{A}_{B}(\voperator{\one}-\ec^{+\ic p}\voperator{E})^{\mathsf{P}}\voperator{H}^{AA}_{AA}(\voperator{\one}-\ec^{+\ic p}\voperator{E})^{\mathsf{P}}\voperator{E}^{B'}_{A}|r}+\ec^{-3\ic p}\rbraket{l|\voperator{E}^{B'}_{A}(\voperator{\one}-\ec^{-\ic p}\voperator{E})^{\mathsf{P}}\voperator{H}^{AA}_{AA}(\voperator{\one}-\ec^{-\ic p}\voperator{E})^{\mathsf{P}}\voperator{E}^{A}_{B}|r}\\
&\rbraket{l|\voperator{E}^{A}_{B}|r}\bigg[\Big(2\pi\delta(p)-1-\ec^{-\ic p} \Big)\big(\rbraket{l| \voperator{H}^{B'A}_{AA}|r}+\rbraket{l|\voperator{E}^{B'}_{A} (\voperator{\one}-\voperator{E})^{\mathsf{P}} \voperator{H}^{AA}_{AA}|r}\big)\\
&\qquad\qquad+ \Big(2\pi\delta(p)-1- \ec^{+\ic p} \Big)\Big( \rbraket{l| \voperator{H}^{AB'}_{AA}|r}+\rbraket{l|\voperator{H}^{AA}_{AA}(\voperator{\one}-\voperator{E})^{\mathsf{P}}\voperator{E}^{B'}_{A}|r}\Big)\\
&\qquad\qquad-\ec^{-\ic 2 p}\rbraket{l|\voperator{E}^{B'}_{A} (\voperator{\one}-\voperator{E})^{\mathsf{P}}(\voperator{\one}-\ec^{-\ic p}\voperator{E})^{\mathsf{P}}\voperator{H}^{AA}_{AA}|r}-\ec^{+\ic 2 p} \rbraket{l|\voperator{H}^{AA}_{AA}(\voperator{\one}-\voperator{E})^{\mathsf{P}}(\voperator{\one}-\ec^{+\ic p}\voperator{E})^{\mathsf{P}}\voperator{E}^{B'}_{A}|r}\bigg]\\
&\rbraket{l|\voperator{E}^{B'}_{A}|r}\bigg[\Big(2\pi\delta(p)-1-\ec^{+\ic p} \Big)\big(\rbraket{l| \voperator{H}^{AA}_{BA}|r}+\rbraket{l|\voperator{E}^{A}_{B} (\voperator{\one}-\voperator{E})^{\mathsf{P}} \voperator{H}^{AA}_{AA}|r}\big)\\
&\qquad\qquad+ \Big(2\pi\delta(p)-1- \ec^{-\ic p} \Big)\Big( \rbraket{l| \voperator{H}^{AA}_{AB}|r}+\rbraket{l|\voperator{H}^{AA}_{AA}(\voperator{\one}-\voperator{E})^{\mathsf{P}}\voperator{E}^{A}_{B}|r}\Big)\\
&\qquad\qquad-\ec^{+\ic 2 p}\rbraket{l|\voperator{E}^{A}_{B} (\voperator{\one}-\voperator{E})^{\mathsf{P}}(\voperator{\one}-\ec^{+\ic p}\voperator{E})^{\mathsf{P}}\voperator{H}^{AA}_{AA}|r}-\ec^{-\ic 2 p} \rbraket{l|\voperator{H}^{AA}_{AA}(\voperator{\one}-\voperator{E})^{\mathsf{P}}(\voperator{\one}-\ec^{-\ic p}\voperator{E})^{\mathsf{P}}\voperator{E}^{A}_{B}|r}\bigg]\Bigg\}.\label{eq:mps:hamoverlappsip}
\end{split}
\end{equation}
\end{widetext}
This equation defines the effective Hamiltonian $H_{\overline{\imath},j}(p,p')$. As expected, because of the translation invariance of $\ham$, the $\delta$ normalizing factor is obtained. The remaining part of the Hamiltonian has been defined as $H_{\overline{\imath},j}(p)$, similarly to how the normalization matrices $N_{\overline{\imath},j}(p,p')=2\pi\delta(p-p')N_{\overline{\imath},j}(p)$ were defined in Eq.~\eqref{eq:defnphip}.

For momentum zero, the additional divergences $\delta(p)$ in $H_{\overline{\imath},j}(p)$ signal the need for imposing $\rbraket{l|\voperator{E}^{A}_{B}|r}=0$ and $\rbraket{l|\voperator{E}^{B'}_{A}|r}=0$, which boils down to restricting to tangent vectors $\ket{\Phi_{0}(B)},\ket{\Phi_{0}(B')}\in\mathbb{T}_{0}^{\perp}$. For other momenta, there is no need to impose these conditions and these terms (\textit{i.e.}~the terms in the square brackets) are finite. However, thanks to the gauge freedom we can still impose this condition, and even the more general right or left gauge fixing conditions in Eq.~\eqref{eq:leftgaugeuB} or \eqref{eq:rightgaugeuB} respectively. The terms in the square brackets then all disappear, together with some additional terms (among which the terms with prefactor $\ec^{\pm \ic 3p}$) on the upper lines. Eq.~\eqref{eq:mps:hamoverlappsip} can thus be simplified by a proper choice of the gauge fixing conditions on $B$.

In particular, by using the representation $B=\mathscr{B}(X)$, the norm matrix for the new variables becomes the identity [Eq.~\eqref{eq:effnormrepresentation}] and we simply have to diagonalize the sum of the remaining terms in the effective Hamiltonian. Since the effective parameterization $X$ has $(d-1)D^2$ components, a complete diagonalization has a computational cost $\order(D^{6})$ and is only feasible for low values of $D$. Since we are mostly interested in the lowest eigenvalues, which correspond to elementary excitations, we can apply an iterative eigensolver. We have already explained how to efficiently compute the gradient $H_{\overline{\imath}}$ in Subsection~\ref{ss:tdvp:implementation}. An efficient implementation of the matrix vector product $H_{\overline{\imath},j}B^{j}$ is a bit more complicated but also possible. Note that we need to iteratively determine the action of the $D^{2}\times D^{2}$ operators $(\voperator{\one}-\voperator{E})^{\mathsf{P}}$ and $(\voperator{\one}-\ec^{\pm\ic p}\voperator{E})^{\mathsf{P}}$ on a $D^{2}$-dimensional vector. An algorithm for efficiently determining the pseudo-inverse $(\voperator{\one}-\voperator{E})^{\mathsf{P}}$ was also sketched in the previous section and is equally applicable to $(\voperator{\one}-\ec^{\pm\ic p}\voperator{E})^{\mathsf{P}}$.

\begin{widetext}
\subsection{Topologically non-trivial states}
\label{ss:excitations:topnontriv}
In a system with discrete symmetry breaking, elementary excitations are often of a topologically non-trivial nature. They appear as domain walls or kinks separating two different ground states at $+\infty$ and $-\infty$. These states are protected from decay into the ground state (or into any other topologically trivial state), as there is an infinitely high energy barrier to tunnel from one ground state into another in a half-infinite region of space\cite{Rajaraman:1982aa}. Let $\ket{\Psi(A)}\in\varM$ and $\ket{\Psi(\tilde{A})}\in\varM$ represent two uMPS with bond dimensions $D$ and $\tilde{D}$ that approximate two different instances from the ground state subspace of $\ham$. Note that we still require these uMPS to be pure, which implies that their connected correlation functions are exponentially decaying to zero. This indicates that they are special instances from the ground state subspace. If $\operator{O}$ is the local order parameter corresponding to the broken symmetry, it is possible to create a symmetric ground state which does not have the exponential clustering property, since $\braket{O}=0$ while $\lim_{n\to\infty}\braket{\hat{O} \hat{T}^{n} \hat{O}\hat{T}^{-n}}\neq 0$. Such a state can only be approximated by a linear combination of several pure MPS, which are each individually good ground state approximations\cite{1996CMaPh.175..565N}. The pure MPS $\ket{\Psi(A)}$ and $\ket{\Psi(\tilde{A})}$ will always show the symmetry breaking locally (\textit{i.e.}~$\braket{\Psi(\overline{A})|\hat{O}|\Psi(A)}\neq 0$). 

We now have two transfer operators $\voperator{E}^{A}_{A}$ and $\voperator{E}^{\tilde{A}}_{\tilde{A}}$, each of which has a unique eigenvalue $1$, and we define the corresponding left and right eigenvectors as $\rbra{l}$, $\rket{r}$ and $\rbra{\tilde{l}}$, $\rket{\tilde{r}}$ respectively. If the symmetry transformations $\operator{U}_g$ corresponding to every element $g$ from the symmetry group can be written as a product of one-site transformations $\operator{U}_g=\prod_{n\in\mathbb{Z}} \operator{T}^{n} \hat{u}_g\operator{T}^{-n}$, then we can probably choose $\tilde{A}^{s}=\sum_{t=1}^{d} \braket{s|\operator{u}_{g}|t} A^{t}$ for some $g\in\mathsf{G}$, so that $\tilde{D}=D$ and $\voperator{E}^{\tilde{A}}_{\tilde{A}}=\voperator{E}^{A}_{A}$, and thus also $\tilde{l}=l$ and $\tilde{r}=r$. We allow for the more general case as well. However, the fact that $\ket{\Psi(A)}$ and $\ket{\Psi(A')}$ are inequivalent (\textit{i.e.}~that there really is symmetry breaking and that they are not just the same state) implies that $\rho(\voperator{E}^{A}_{\tilde{A}})<1$.

An ansatz for approximating the topologically non-trivial state with momentum $p$ that asymptotically looks like $\ket{\Psi(A)}$ at $-\infty$ and like $\ket{\Psi(\tilde{A})}$ at $+\infty$ (\textit{i.e.}~a kink or domain wall) is given by
\begin{equation}
\ket{\widetilde{\Phi}_{p}(B;A;\tilde{A})}=\sum_{n\in\mathbb{Z}}\ec^{\ic p n}\sum_{\{s_n\}=1}^{d} \bm{v}_{\mathrm{L}}^{\dagger}\left[\left(\prod_{m<n} A^{s_{m}}\right) B^{s_{n}} \left(\prod_{m'>n} \tilde{A}^{s_{m'}}\right)\right]\bm{v}_{\mathrm{R}} \ket{\bm{s}},\label{eq:mps:defxip}
\end{equation}
with $B^{s}$ a set of $D\times \tilde{D}$ matrices ($\forall s=1,\ldots,d$). We now impose $(\bm{v}_{\mathrm{L}}^{\dagger} r \bm{v}_{\mathrm{L}})(\bm{v}_{\mathrm{R}}^{\dagger} \tilde{l} \bm{v}_{\mathrm{R}})=1$ so as not to be troubled by the boundary vectors when computing expectation values. In order for this state to have a finite excitation energy, we need to impose $h(\overline{A},A)=h(\overline{\tilde{A}},\tilde{A})$, so that both uMPS approximate their respective ground state equally well. As for the ansatz for topologically non-trivial excitations, the rationale behind the ansatz in Eq.~\eqref{eq:mps:defxip} is that the kink itself is a highly localized or point-like object that is in a momentum superposition. It is not completely restricted to live on a single site, since it can spread out along the virtual dimension, and has non-trivial support over at least $\log_{d} D+\log_{d}\tilde{D}$ sites. Creating a kink through the action of a physical operator, analogously to the Feynman-Bijl operator, was first attempted by Mandelstam \cite{Mandelstam:1975aa} in the context of relativistic quantum field theories. Translated to the lattice case, he proposed to use as operator the Fourier transform of
\begin{equation}
\operator{O}(n)=\operator{o}_{n}\prod_{m>n} \operator{u}_{n},
\end{equation}
with $\operator{o}_{n}$ a completely local operator on site $n$ and $\prod_{m>n} \operator{u}_{n}$ a string of operators that has the effect to transform the ground state to another ground state for $m>n$, \textit{i.e.}~in our context $\operator{u}_{n}=\hat{T}^{n}\operator{u}_g\hat{T}^{-n}$.

The states $\ket{\widetilde{\Phi}_{p}(B;A,\tilde{A})}$ share many properties with the tangent vectors $\ket{\Phi_{p}(B,A)}$. Using $\rho(\voperator{E}^{A}_{\tilde{A}})<1$ even allows for some simplifications. Firstly, $\braket{\Psi(\overline{A})|\widetilde{\Phi}_{p}(B;A,\tilde{A})}=\braket{\Psi(\overline{\tilde{A}})|\widetilde{\Phi}_{p}(B;A,\tilde{A})}=0$ for all values of the momentum $p$, including $p=0$, so that orthogonality with respect to the ground state is automatic and no divergent terms are to be expected. The reason is the appearance of factors $\voperator{E}^{A}_{\tilde{A}}$ in a half-infinite space. Secondly, the linear map
\begin{equation}
\widetilde{\Phi}_{p}^{(A,\tilde{A})}:\mathbb{C}^{D\times d\times\tilde{D}}\mapsto \hilbert:B\mapsto \ket{\widetilde{\Phi}_{p}^{(A,\tilde{A})}(B)}=\ket{\widetilde{\Phi}_{p}(B;A,\tilde{A})}
\end{equation}
has a non-trivial null space $\widetilde{\mathbb{N}}_{p}^{(A,\tilde{A})}$. Indeed, it can easily be checked that the map
\begin{equation}
\widetilde{\mathscr{N}}_{p}^{(A,\tilde{A})}:\mathbb{C}^{D\times\tilde{D}}\mapsto \widetilde{\mathbb{N}}^{(A,\tilde{A})}_{p}: x\mapsto  \mathscr{N}_{p}^{(A,\tilde{A})}(x)\quad \text{with}\quad \mathscr{N}^{(A,\tilde{A})s}_{p}(x)=A^{s}x-\mathrm{e}^{-\ic p} x \tilde{A}^{s}, \forall s=1,\ldots,d
\end{equation}
defines a set of choices $B=\widetilde{\mathscr{N}}_{p}(x)$ that produce $\ket{\widetilde{\Phi}_{p}(B)}=0$. We henceforth omit the explicit notation of $A$ and $\tilde{A}$. It is easy to see that the null space of $\widetilde{\mathscr{N}}_{p}$ itself is empty $\forall p$, including $p=0$, since $\widetilde{\mathscr{N}}_{p}^{s}(x)=0$ ($\forall s=1,\ldots,d$) requires that $\voperator{E}^{A}_{\tilde{A}}\rket{x\tilde{r}}=\ec^{-\ic p}\rket{x\tilde{r}}$. But since $\rho(\voperator{E}^{A}_{\tilde{A}})<1$, this equation can have no solution. If we define
\begin{equation}
\widetilde{\mathbb{T}}_{p}=\{\ket{\widetilde{\Phi}_{p}(B)}\mid B\in\mathbb{C}^{D\times d \times{\tilde{D}}}\}
\end{equation}
then we obtain $\dim \widetilde{\mathbb{T}}_{p}=\dim \mathbb{C}^{D\times d \times{\tilde{D}}}-\dim \widetilde{\mathbb{N}}_{p}= (d-1)D\tilde{D}$, $\forall p\in[-\pi,+\pi)$. To fix the additive gauge freedom in the parameterization $B$, we restrict to parameterizations $B$ in the a horizontal subspace $\widetilde{\mathbb{B}}^{(A,\tilde{A})}$, which is defined as the solution space of either of the following conditions:
\begin{itemize}
\item \emph{left gauge fixing condition:}
\begin{equation}
\rbra{l}\voperator{E}^{B}_{A}=0 \qquad \Leftrightarrow \qquad\sum_{s=1}^{d} {A^{s}}^{\dagger} l B^{s}=0,\label{eq:leftgaugeuBxi}
\end{equation}
\item \emph{right gauge fixing condition:}
\begin{equation}
\voperator{E}^{B}_{\tilde{A}}\rket{\tilde{r}}=0 \qquad \Leftrightarrow \qquad\sum_{s=1}^{d} B^{s}\tilde{r}\tilde{A}^{s\dagger}=0.\label{eq:rightgaugeuBxi}
\end{equation}
\end{itemize}
Either of these two conditions impose in total $D\tilde{D}$ linearly independent equations that completely fix the gauge freedom in $x$. A linear parameterization satisfying \textit{e.g.}~the left gauge fixing condition can easily be obtained as $\widetilde{\mathscr{B}}^s(\tilde{X})=l^{-1/2} V_L^s \tilde{X} \tilde{r}^{-1/2}$, where $\tilde{X}\in\mathbb{C}^{(d-1)D\times \tilde{D}}$ and $V_L$ was constructed in the context of Eq.~\eqref{eq:defBrepresentation}.

Applying the time-independent variational principle to $\widetilde{\mathbb{T}}_{p}$ also boils down to solving a Rayleigh-Ritz problem. We first compute the overlap $\braket{\widetilde{\Phi}_{p}(\overline{B})|\widetilde{\Phi}_{p'}(B')}$. Since the superoperators $\voperator{E}^{A}_{\tilde{A}}$ and $\voperator{E}^{\tilde{A}}_{A}$ that appear between $B$ and $B'$ have a spectral radius smaller than one, they do not need `regularizing' and we cannot have disconnected contributions at all. There is then also no need to introduce pseudo-inverses. We simply obtain
\begin{equation}
\braket{\widetilde{\Phi}_{p}(\overline{B})|\widetilde{\Phi}_{p'}(B')}=2\pi\delta(p'-p)\bigg[\rbraket{l|\voperator{E}^{B'}_{B}|\tilde{r}}+\rbraket{l|\voperator{E}^{A}_{B} (\voperator{\one}-\ec^{+\ic p}\voperator{E}^{A}_{\tilde{A}})^{-1}\voperator{E}^{B'}_{A}|\tilde{r}}+\rbraket{l|\voperator{E}^{B'}_{A} (\voperator{\one}-\ec^{-\ic p}\voperator{E}^{\tilde{A}}_{A})^{-1}\voperator{E}^{A}_{B}|\tilde{r}}\bigg].\label{eq:mps:xioverlap}
\end{equation}
Similarly, all disconnected contributions that were present in $\braket{\Phi_{p}(\overline{B})|\ham|\Phi_{p'}(B')}$ [terms in square brackets in Eq.~\eqref{eq:mps:hamoverlappsip}] disappear in the evaluation of $\braket{\widetilde{\Phi}_{p}(\overline{B})|\ham|\widetilde{\Phi}_{p'}(B')}$. If we also redefine $\operator{h}\leftarrow \operator{h}-h(\overline{A},A)=\operator{h}-h(\overline{\tilde{A}},\tilde{A})$, so that the correct ground state energy is subtracted and there are no disconnected contributions from $\operator{h}$ either, we immediately obtain
\begin{align}
\langle\widetilde{\Phi}_{p}(\overline{B}&)|\operator{H}|\widetilde{\Phi}_{p'}(B')\rangle=2\pi\delta(p'-p)\times\nonumber\\
\bigg[&\rbraket{l|\voperator{H}^{B'\tilde{A}}_{B \tilde{A}}|\tilde{r}}+\rbraket{l|\voperator{H}^{AB'}_{AB}|\tilde{r}}+\ec^{+\ic p}\rbraket{l|\voperator{H}^{AB'}_{B\tilde{A}}|\tilde{r}}+\ec^{-\ic p}\rbraket{l|\voperator{H}^{B'\tilde{A}}_{AB}|\tilde{r}}+\rbraket{l|\voperator{E}^{B'}_{B}(\voperator{\one}-\voperator{E}^{\tilde{A}}_{\tilde{A}})^{\mathsf{P}}\voperator{H}^{\tilde{A}\tilde{A}}_{\tilde{A}\tilde{A}}|\tilde{r}}+\rbraket{l|\voperator{H}^{AA}_{AA}(\voperator{\one}-\voperator{E}^{A}_{A})^{\mathsf{P}}\voperator{E}^{B'}_{B}|\tilde{r}}\nonumber\\
&+\ec^{+\ic p}\rbraket{l|\voperator{E}^{A}_{B}(\voperator{\one}-\ec^{+\ic p}\voperator{E}^{A}_{\tilde{A}})^{-1}\voperator{E}^{B'}_{\tilde{A}}(\voperator{\one}-\voperator{E}^{\tilde{A}}_{\tilde{A}})^{\mathsf{P}}\voperator{H}^{\tilde{A}\tilde{A}}_{\tilde{A}\tilde{A}}|\tilde{r}}+\ec^{-\ic p}\rbraket{l|\voperator{E}^{B'}_{A}(\voperator{\one}-\ec^{-\ic p}\voperator{E}^{\tilde{A}}_{A})^{-1}\voperator{E}^{\tilde{A}}_{B}(\voperator{\one}-\voperator{E}^{\tilde{A}}_{\tilde{A}})^{\mathsf{P}}\voperator{H}^{\tilde{A}\tilde{A}}_{\tilde{A}\tilde{A}}|\tilde{r}}\nonumber\\
&+\ec^{+\ic p}\rbraket{l|\voperator{H}^{AA}_{AA}(\voperator{\one}-\voperator{E}^{A}_{A})^{\mathsf{P}}\voperator{E}^{A}_{B}(\voperator{\one}-\ec^{+\ic p}\voperator{E}^{A}_{\tilde{A}})^{-1}\voperator{E}^{B'}_{\tilde{A}}|\tilde{r}}+\ec^{-\ic p}\rbraket{l|\voperator{H}^{AA}_{AA}(\voperator{\one}-\voperator{E}^{A}_{A})^{\mathsf{P}}\voperator{E}^{B'}_{A}(\voperator{\one}-\ec^{-\ic p}\voperator{E}^{\tilde{A}}_{A})^{-1}\voperator{E}^{\tilde{A}}_{B}|\tilde{r}}\nonumber\\
&+\ec^{+\ic p}\rbraket{l|\voperator{E}^{A}_{B}(\voperator{\one}-\ec^{+\ic p}\voperator{E}^{A}_{\tilde{A}})^{-1}\voperator{H}^{B'\tilde{A}}_{\tilde{A}\tilde{A}}|\tilde{r}}+\ec^{-\ic p}\rbraket{l|\voperator{E}^{B'}_{A}(\voperator{\one}-\ec^{-\ic p}\voperator{E}^{\tilde{A}}_{A})^{-1}\voperator{H}^{\tilde{A}\tilde{A}}_{B\tilde{A}}|\tilde{r}}\nonumber\\
&+\ec^{+\ic p}\rbraket{l|\voperator{H}^{AA}_{AB}(\voperator{\one}-\ec^{+\ic p}\voperator{E}^{A}_{\tilde{A}})^{-1}\voperator{E}^{B'}_{\tilde{A}}|\tilde{r}}+\ec^{-\ic p}\rbraket{l|\voperator{H}^{AB'}_{AA}(\voperator{\one}-\ec^{-\ic p}\voperator{E}^{\tilde{A}}_{A})^{-1}\voperator{E}^{\tilde{A}}_{B}|\tilde{r}}\nonumber\\
&+\ec^{+2\ic p}\rbraket{l|\voperator{E}^{A}_{B}(\voperator{\one}-\ec^{+\ic p}\voperator{E}^{A}_{\tilde{A}})^{-1}\voperator{H}^{AB'}_{\tilde{A}\tilde{A}}|\tilde{r}}+\ec^{-2\ic p}\rbraket{l|\voperator{E}^{B'}_{A}(\voperator{\one}-\ec^{-\ic p}\voperator{E}^{\tilde{A}}_{A})^{-1}\voperator{H}^{\tilde{A}\tilde{A}}_{AB}|\tilde{r}}\nonumber\\
&+\ec^{+2\ic p}\rbraket{l|\voperator{H}^{AA}_{B\tilde{A}}(\voperator{\one}-\ec^{+\ic p}\voperator{E}^{A}_{\tilde{A}})^{-1}\voperator{E}^{B'}_{\tilde{A}}|\tilde{r}}+\ec^{-2\ic p}\rbraket{l|\voperator{H}^{B'\tilde{A}}_{AA}(\voperator{\one}-\ec^{-\ic p}\voperator{E}^{\tilde{A}}_{A})^{-1}\voperator{E}^{\tilde{A}}_{B}|\tilde{r}}\nonumber\\
&+\ec^{+3\ic p}\rbraket{l|\voperator{E}^{A}_{B}(\voperator{\one}-\ec^{+\ic p}\voperator{E}^{A}_{\tilde{A}})^{-1}\voperator{H}^{AA}_{\tilde{A}\tilde{A}}(\voperator{\one}-\ec^{+\ic p}\voperator{E}^{A}_{\tilde{A}})^{-1}\voperator{E}^{B'}_{\tilde{A}}|\tilde{r}}+\ec^{-3\ic p}\rbraket{l|\voperator{E}^{B'}_{A}(\voperator{\one}-\ec^{-\ic p}\voperator{E}^{\tilde{A}}_{A})^{-1}\voperator{H}^{\tilde{A}\tilde{A}}_{AA}(\voperator{\one}-\ec^{-\ic p}\voperator{E}^{\tilde{A}}_{A})^{-1}\voperator{E}^{\tilde{A}}_{B}|\tilde{r}}\bigg].\label{eq:mps:hamoverlapxip}
\end{align}
By imposing either the left or the right gauge fixing conditions of Eq.~\eqref{eq:leftgaugeuBxi} or \eqref{eq:rightgaugeuBxi}, many terms in Eq.~\eqref{eq:mps:xioverlap} and in Eq.~\eqref{eq:mps:hamoverlapxip} cancel. In particular, using the representation $B=\widetilde{\mathscr{B}}(\tilde{X})$, the effective norm matrix becomes
\begin{displaymath}
\braket{\widetilde{\Phi}_{p}(\overline{\widetilde{\mathscr{B}}}(\overline{\tilde{X}}))|\widetilde{\Phi}_{p'}(\widetilde{\mathscr{B}}(\tilde{X}'))}=2\pi\delta(p'-p)\tr[\tilde{X}^{\dagger}\tilde{X}']
\end{displaymath}
As for the topologically trivial excitation ansatz, the final problem boils down to finding eigenvalues of a normal eigenvalue problem and can be solved using in iterative eigensolver with a computation complexity that scales as $\order(D^3)$.

One final remark is in order. With two matrices $A$ and $\tilde{A}$ present in the ansatz for topologically non-trivial states $\ket{\widetilde{\Phi}_{p}(B;A;\tilde{A})}$, both of which can be defined independently from each other, there is some ambiguity present in the definition of the momentum. Suppose we have $A'=\ec^{\ic \varphi} A$. We obtain $\ket{\widetilde{\Phi}_{p}(B;A';\tilde{A})}=\ket{\widetilde{\Phi}_{p}(B;\ec^{\ic \varphi}A;\tilde{A})}\sim \ket{\widetilde{\Phi}_{p+\varphi}(B;A;\tilde{A})}$, where the similarity sign means the two states are equal up to a global phase. This follows simply from inserting $A'$ in \textit{e.g.}~Eq.~\eqref{eq:mps:hamoverlapxip}. It is ultimately related to the fact that topologically non-trivial excitations can only be defined on systems with open boundary conditions, where momentum is not a good quantum number to start with. This problem did not occur in the ansatz for topologically trivial excitations, because $\ket{\Psi_{p}(B)}$ can be defined starting from a finite lattice with periodic boundary conditions, where $p$ is a good quantum number. In systems with periodic boundary conditions, only pairs of kinks ($\ket{\Psi_{p_1}(B_{1};A,\tilde{A})}$) and antikinks ($\ket{\Psi_{p_2}(B_{2};\tilde{A},A)}$) with total momentum $p_1+p_2$can exist, and the momentum ambiguity is resolved since for $A\leftarrow \ec^{\ic\phi} A$ we obtain $p_1\leftarrow p_1-\phi$ and $p_2\leftarrow p_2+\phi$. Nevertheless, it is often useful to study topologically non-trivial excitations as isolated elementary excitations. Since the momentum is tightly connected to $\voperator{E}^{A}_{\tilde{A}}$ in the sense that they appear together as $\ec^{\ic p}\voperator{E}^{A}_{\tilde{A}}$ in Eq.~\eqref{eq:mps:hamoverlapxip}, it makes sense to fix the freedom in the definition of the momentum quantum number by requiring that the relative phase of $A$ and $\tilde{A}$ is chosen so that the eigenvalue with largest modulus of $\voperator{E}^{A}_{\tilde{A}}$ is positive.

\subsection{Relation to the time-dependent variational principle}
\label{ss:excitations:reltdvp}
The ansatz for studying excitations starts from an optimal uMPS approximation $\ket{\Psi(A)}$ of the ground state. Thus, $A$ is an optimum of the time-independent variational principle for the ground state problem in $\varM$. Consequently, is also a stationary solution of the TDVP equations for the uMPS manifold $\varM$, since $H_{\overline{\jmath}}(\overline{A},A)=0$, or thus, for any $\ket{\Phi_{0}(B)}\perp\ket{\Psi(A)}$, $\braket{\Phi_0(\overline{B})|\ham|\Psi(A)}=0$. Translation invariance also dictates that $\braket{\Phi_{p}(\overline{B})|\ham|\Psi(A)}=0$, so that $\ket{\Psi(A)}$ is also stable against all non-translation invariant first order variations.

We can now investigate small perturbations around the stationary and uniform solution $A$. Note that we have only derived the TDVP equation for the manifold of translation invariant MPS. The generalization to MPS with site-dependent tensors $A(n)$ is straightforward. We introduce the notation $\ket{\Psi[\bm{A}]}$, where the square brackets denote the dependence on an infinite set of tensors $\bm{A}=\{A(n)\}_{n\in\mathbb{Z}}$. The corresponding energy is given by $H[\overline{\bm{A}},\bm{A}]=\braket{\Psi[\overline{\bm{A}}]|\ham|\Psi[\bm{A}]}$, where the state is assumed to be normalized to $\braket{\Psi[\overline{\bm{A}}]|\Psi[\bm{A}]}=1$. We also use a big index $I=(n,i)=(n,(\alpha,\beta,s))$ a a collective index containing also the site index, such that $\bm{A}^I=A^{s}_{\alpha,\beta}(n)$, and let $\partial_{I}$ denote the partial derivative with respect to $\bm{A}^I$. The generalized TDVP equation then reads
\begin{equation}
\braket{\partial_{\overline{I}}\Psi[\overline{\bm{A}}(t)]|\partial_J \Psi[\bm{A}(t)]} \dot{\bm{A}}^{J}(t)=\braket{\partial_{\overline{I}}\Psi[\overline{\bm{A}}(t)]|\ham-H[\overline{\bm{A}}(t),\bm{A}(t)]| \Psi[\bm{A}(t)]}
\end{equation}
For small perturbations $\bm{A}(t)=\bm{A}_0+\epsilon \bm{B}(t)$, where $\bm{A}_0$ denotes the uniform solution $\{ A\}_{n\in\mathbb{Z}}$ and $\bm{B}(t)=\{B(n,t)\}_{n\in\mathbb{Z}}$, we can linearize the TDVP equation as
\begin{multline}
\braket{\partial_{\overline{I}}\Psi[\overline{\bm{A}}_0]|\partial_J \Psi[\bm{A}_0]} \dot{\bm{B}}^{J}(t)=\braket{\partial_{\overline{I}}\Psi[\overline{\bm{A}}_0]|\ham-H[\overline{\bm{A}}_0,\bm{A}_0]| \partial_{J}\Psi[\bm{A}_0]}\bm{B}^{J}(t)\\+\braket{\partial_{\overline{I}}\partial_{\overline{J}}\Psi[\overline{\bm{A}}_0]|\ham-H[\overline{\bm{A}}_0,\bm{A}_0]| \Psi[\bm{A}_0]}\overline{\bm{B}}^{\overline{J}}(t).\label{eq:linearTDVP}
\end{multline}
More generally, the right hand side should read $\partial_{\overline{I}}\partial_{J} H[\overline{\bm{A}}_0,\bm{A}_{0}]\bm{B}^{J}+\partial_{\overline{I}}\partial_{\overline{J}} H[\overline{\bm{A}}_0,\bm{A}_{0}]\overline{\bm{B}}^{\overline{J}}$. However, using the fact that first order derivatives of $H$ at $\bm{A}_{0}$ are zero due to stationarity, and that we are only considering variations orthogonal to the original state, only the terms above survive. Note that we had to introduce a second order derivative $\ket{\partial_{I}\partial_{J}\Psi[\bm{A}_0]}$ in the right hand side. For $\bm{B}_1=\{B_1(n)\}_{n\in\mathbb{Z}}$ and $\bm{B}_2=\{B_2(n)\}_{n\in\mathbb{Z}}$, we now introduce the general notation
\begin{align}
\ket{\Upsilon^{(A)}[\bm{B}_1,\bm{B}_2]}=&\bm{B}_1^{I}\bm{B}_{2}^{J}\ket{\partial_{I}\partial_{J}\Psi[\bm{A}_0]}=\sum_{m,n\in\mathbb{Z}} B^{i}_{1}(m) B_{2}^{j}(n) \left.\frac{\partial^{2}\ }{\partial A^{i}(m)\partial A^{j}(n)} \ket{\Psi[\bm{A}]}\right|_{\bm{A}=\bm{A}_0}\nonumber\\
=&\sum_{m<n\in\mathbb{Z}}\sum_{\{s_n\}=1}^{d} \bm{v}_{\mathrm{L}}^{\dagger}\Bigg[\bigg(\prod_{k<m} A^{s_{k}}\bigg)B_{1}^{s_{m}}(m) \bigg(\prod_{m<k'<n} A^{s_{k'}}\bigg) B_{2}^{s_{n}}(n)\bigg(\prod_{n<k''}A^{s_{k''}}\bigg)\Bigg]\bm{v}_{\mathrm{R}} \ket{\bm{s}}\nonumber\\
&+\sum_{m>n\in\mathbb{Z}}\sum_{\{s_n\}=1}^{d} \bm{v}_{\mathrm{L}}^{\dagger}\Bigg[\bigg(\prod_{k<w} A^{s_{k}}\bigg) B_{2}^{s_{n}}(n) \bigg(\prod_{n<k'<m} A^{s_{k'}}\bigg) B_{1}^{s_{m}}(m)\bigg(\prod_{n<k''}A^{s_{k''}}\bigg)\Bigg]\bm{v}_{\mathrm{R}} \ket{\bm{s}},\label{eq:defdoubletangentvec}
\end{align}
These states span the double tangent space $\mathbb{T}^{(2)}$ of the MPS manifold. They were already introduced in Ref.~\onlinecite{2012arXiv1210.7710H} for calculating the Levi-Civita connection.

Returning to the linearized TDVP equation [Eq.~\eqref{eq:linearTDVP}], we can now look for specific solutions 
\begin{equation}
B(n,t)=B_{+}\ec^{\ic p n - \ic \omega t} + B_{-} \ec^{-\ic p n + \ic \omega t}
\end{equation}
and contract the free index $I$ of this equation with a test vector $\bm{B}'=\{B'(n)\}_{n\in\mathbb{Z}}$ with $B'(n)=B\ec^{\ic p'n}$. This essentially boils down to a Fourier transform of the linearized TDVP equation with respect to both space and time, and results in
\begin{equation}
\begin{split}
\omega \braket{\Phi_{p'}(\overline{B}')|\Phi_{p}(B_{+})}&=\braket{\Phi_{p'}(\overline{B}')|\ham-H|\Phi_{p}(B_+)}+\braket{\Upsilon_{p',-p}(\overline{B}',\overline{B}_{-})|\ham-H|\Psi(A)}\\
-\omega \braket{\Phi_{p'}(\overline{B}')|\Phi_{-p}(B_{-})}&=\braket{\Phi_{p'}(\overline{B}')|\ham-H|\Phi_{-p}(B_{-})}+\braket{\Upsilon_{p',p}(\overline{B}',\overline{B}_{+})|\ham-H|\Psi(A)}
\end{split}\label{eq:linearizedtdvpfourier}
\end{equation}
where $H=H(\overline{A},A)=H[\overline{\bm{A}}_0,\bm{A}_0]$ and we have introduced a new basis for the double tangent space as
\begin{align}
\ket{\Upsilon_{p_1,p_2}(B_1,B_2)}=&\sum_{m<n\in\mathbb{Z}}\sum_{\{s_n\}=1}^{d}\ec^{\ic p_1 m+\ic p_2 n} \bm{v}_{\mathrm{L}}^{\dagger}\Bigg[\bigg(\prod_{k<m} A^{s_{k}}\bigg)B_{1}^{s_{m}} \bigg(\prod_{m<k'<n} A^{s_{k'}}\bigg) B_{2}^{s_{n}}\bigg(\prod_{n<k''}A^{s_{k''}}\bigg)\Bigg]\bm{v}_{\mathrm{R}} \ket{\bm{s}}\nonumber\\
&+\sum_{m>n\in\mathbb{Z}}\sum_{\{s_n\}=1}^{d}\ec^{\ic p_1 m+\ic p_2 n} \bm{v}_{\mathrm{L}}^{\dagger}\Bigg[\bigg(\prod_{k<w} A^{s_{k}}\bigg) B_{2}^{s_{n}} \bigg(\prod_{n<k'<m} A^{s_{k'}}\bigg) B_{1}^{s_{m}}\bigg(\prod_{n<k''}A^{s_{k''}}\bigg)\Bigg]\bm{v}_{\mathrm{R}} \ket{\bm{s}}.\label{eq:defdoubletangentvecp}
\end{align}
We then also define
\begin{equation}
\braket{\Upsilon_{p_1,p_2}(\overline{B}_1,\overline{B}_2)|\hat{H}-H|\Psi(A)}= K_{\overline{\imath},\overline{\jmath}}(p_1,p_2)\overline{B}_1^{\overline{\imath}}\overline{B}_{2}^{\overline{\jmath}}=2\pi\delta[(p_1+p_2)\bmod 2\pi] K_{\overline{\imath},\overline{\jmath}}(p_1) \overline{B}_1^{\overline{\imath}}\overline{B}_{2}^{\overline{\jmath}}
\end{equation}
where the expression for $K_{\overline{\imath},\overline{\jmath}}(p_1)$ can easily be derived using the techniques that were used for the calculation of $H_{\overline{\imath},j}(p)$. By taking the complex conjugate of the second line in Eq.~\eqref{eq:linearizedtdvpfourier}, we can now write the Fourier transformed linearized TDVP equation as a generalized eigenvalue equation
\begin{equation}
\omega \begin{bmatrix} N(p) & 0 \\0 & -\overline{N}(-p)\end{bmatrix}\begin{bmatrix}B_{+}\\ \overline{B_{-}}\end{bmatrix} = \begin{bmatrix}H(p) & K(p) \\\overline{K}(-p) & \overline{H}(-p)\end{bmatrix}\begin{bmatrix}B_{+}\\ \overline{B_{-}}\end{bmatrix}.\label{eq:linearizedtdvpfourier2}
\end{equation}
Note that the matrix appearing in the right hand side of this equation is Hermitian, since $H(p)$ and $N(p)$ are Hermitian, and $K(p)$ satisfies $K(p)=K(-p)^{\mathrm{T}}$. If we would have linearized the original TDVP equation for uMPS, we would only have recovered the $p=0$ case of the above equation. For $p=0$, the matrix in the right hand side can easily be recognized as the Hessian of the energy functional $H(\overline{A},A)$ at the point $A$. We could thus also have denoted $K_{\overline{\imath},\overline{\jmath}}(0,0)$ as $H_{\overline{\imath},\overline{\jmath}}=\partial_{\overline{\imath}}\partial_{\overline{\jmath}}H$. By considering all momenta, we have actually constructed a momentum decomposition of the full Hessian of the energy functional in the manifold of all MPS with site-dependent matrices. Since $A$ is a minimum of the energy functional in the class of uMPS, this Hessian is positive (semi)-definite at momentum zero. If the uniform solution is also a minimum in the full class of generic MPS, the corresponding Hessian is positive (semi)-definite at any momentum $p$. In that case, all eigenvalues of the generalized eigenvalue equation in Eq.~\eqref{eq:linearizedtdvpfourier2} are real. In addition, for any eigenvector $\omega^{(k)}(p)$ with corresponding eigenvector $\big(B_{+}^{(k)}(p),\overline{B}_{-}^{(k)}(p)\big)$ at momentum $p$, there exists a dual eigenvalue $\omega^{(-k)}(-p)=-\omega^{(k)}(p)$ with eigenvector $\big(B_{+}^{(-k)}(-p)=B_{-}^{(k)}(p),\overline{B}_{-}^{(-k)}(-p)=\overline{B}_{+}^{(k)}(p)\big)$. In particular, at momentum zero, the spectrum is even around zero. Since many systems are invariant under spatial reflections, we expect that $N(p)$, $H(p)$ and $K(p)$ will be gauge equivalent to $N(-p)$, $H(-p)$ and $K(-p)$. In that case, the spectrum of eigenvalues of Eq.~\eqref{eq:linearizedtdvpfourier2} is even around zero at all momenta, since we then have $\omega^{(-k)}(p)=\omega^{(-k)}(-p)=-\omega^{(k)}(p)$. 

From linear response theory, it can be argued that the eigenvalues $\omega$ of Eq.~\eqref{eq:linearizedtdvpfourier2} correspond to resonances in the system that should be related to the excitation energies of the system\cite{2009JChPh.130r4111D}. Clearly, the Rayleigh-Ritz problem obtained for our variational ansatz for topologically trivial excitations is equivalent to the linearized TDVP problem of Eq.~\eqref{eq:linearizedtdvpfourier2} if the off-diagonal blocks $K(p)=0$. We now discuss the effect of having $K(p)\neq 0$. The matrix $K(p)$ is obtained as (a momentum decomposition of) the projection of $\operator{H}-H\ket{\Psi}$ to the double tangent space $\mathbb{T}^{(2)}$, which is spanned by the second order derivatives $\ket{\partial_{I}\partial_{J}\Psi[A_0]}$. We have introduced the notation $\ket{\Upsilon_{p_1,p_2}(B_1,B_2)}$ for momentum states in $\mathbb{T}^{(2)}$ in Eq.~\eqref{eq:defdoubletangentvecp}. We will elaborate on the physical interpretation of this set of states in Section~\ref{s:extensions}. For now, the structure of Eq.~\eqref{eq:linearizedtdvpfourier2} seems to predict corrections to the variational excitation energies (obtained from the diagonal part) by taking into account a larger part of Hilbert space, namely the part that is captured by $\mathbb{T}^{(2)}$. 
The problem with these corrections is that they are not variational in nature, and can therefore be unbounded and unphysical. We illustrate in the next section how Eq.~\eqref{eq:linearizedtdvpfourier2} can give rise to spurious excitation energies which are not physical and which do not show up when using the variational ansatz $\ket{\Phi_{p}(B)}$. Since $K(p)$ is obtained from projecting $\operator{H}-H\ket{\Psi}$ onto the double tangent space, one can show that $\lVert K(p)\rVert < \widetilde{\epsilon}(\overline{A},A)$ (some care has to be taken to deal with the diverging prefactors), where $\widetilde{\epsilon}(\overline{A},A)=\Delta H(\overline{A},A) \lvert \mathbb{Z}\rvert^{-1/2}$ is the local measure for the state error. In the limit where the MPS approximation is becoming very accurate, $K(p)\to 0$ and the variational energies will match the eigenvalues of Eq.~\eqref{eq:linearizedtdvpfourier2}.

Nevertheless, recognizing the right hand of Eq.~\eqref{eq:linearizedtdvpfourier2} as the Hessian of the energy functional is important for interpreting the variational excitation energies, which correspond to the eigenvalues of the diagonal block $H(p)$. We have already mentioned that $\ket{\Psi(A)}$ is a stationary solution, both in the manifold of uMPS as well as in the manifold of generic MPS. Hence, there are no corrections of first order in $\epsilon$ to the energy $H[\overline{\bm{A}},\bm{A}]$ when replacing $\bm{A}\leftarrow \bm{A}_0+\epsilon\bm{B}$. If for some momentum $p$, the variational energies obtained from $H(p)$ contain negative values, this indicates that condensation of this excitation is possible so as to create a lower-energy MPS approximation of the ground state that is not uniform. Clearly, this should not happen at momentum $p=0$, as this would indicate that the stationary solution $A$ is not a minimum but rather a saddle point or local maximum. For example, let $H(p)$ have a negative energy $\omega<0$ with corresponding eigenvector $B_{+}$, and let there be reflection invariance such that $H(-p)$ also has eigenvalue $\omega$ with corresponding eigenvector $B_{-}$ (related to $B_{+}$ by a gauge transform). For $\bm{A}\leftarrow \bm{A}_0+\epsilon\bm{B}$, with $\bm{B}=\{B(n)\}_{n\in\mathbb{Z}}$ and $B(n)=\exp(\ic p n) B_{+}+\exp(-\ic p n) B_{-}$, the second order correction to the energy density will be
\begin{equation}
H[\overline{\bm{A}},\bm{A}]=H[\overline{\bm{A}}_0,\bm{A}_0]+\frac{\epsilon^2}{2} \mathbb{Z}\left( \overline{B}_{+}^{\overline{\imath}} H_{\overline{\imath},j}(p)B_{+}^{j}+\overline{B}_{-}^{\overline{\imath}} H_{\overline{\imath},j}(-p)B_{-}^{j} +\overline{B}_{+}^{\overline{\imath}} \overline{B}_{-}^{\overline{\jmath}} K_{\overline{\imath},\overline{\jmath}}(p)+\overline{K}_{i,j}(p) B_{+}^{i}B_{-}^{j} \right)
\end{equation}
The first two terms in the brackets reduce to $\omega<0$, whereas we can certainly choose the phase of $B_{+}$ and $B_{-}$ in such a way that the last two terms are zero or negative as well. If this happens for $p\neq 0$, it indicates that the uniform solution $\ket{\Psi(A)}$ is only a minimum in the restricted manifold of uniform MPS, but not in the manifold of generic MPS.
\end{widetext} 

Finally, we can also discuss the ansatz $\ket{\widetilde{\Phi}_{p}(B)}$ for topologically non-trivial excitations. The states $\ket{\widetilde{\Phi}_{\kappa}(B)}$ do not live in the tangent space $\mathbb{T}$ of $\mathcal{M}$. However, if we construct a larger manifold $\{\ket{\Psi'[\bm{A}']}\}$ of MPS of bond dimension $D'=D+\tilde{D}$ and define the uniform solution $\bm{A}'_{0}=\{A'\}_{n\in\mathbb{Z}}$ with
\begin{equation}
A^{'s}=\begin{bmatrix} A^{s} & 0 \\0 & \widetilde{A}^{s}\end{bmatrix},
\end{equation}
then the topologically non-trivial excitation $\ket{\widetilde{\Phi}_{p}(B)}$ is obtained as the tangent vector $\ket{\Phi'[\bm{C}]}=C^{I}\ket{\partial_{I} \Psi'[\bm{A}_0']}$ with 
\begin{equation}
C^s(n)=\begin{bmatrix} 0 & B \mathrm{e}^{\ic p n}\\0 & 0\end{bmatrix}.
\end{equation}
Because of the special structure of $A'$, the point $\bm{A}'=\bm{A}'_{0}$ is a stationary point of the energy functional $H[\overline{\bm{A}'},\bm{A}]$. However, $\bm{A}'_{0}$ is not expected to be a true minimum in the variational manifold of MPS with bond dimension $2D$, and the effective Hamiltonian for this variational ansatz can have negative eigenvalues. This would indicate that the uMPS $\ket{\Psi'(A')}$ is not stable at second order against fluctuations that mix the two ground state approximations $\ket{\psi(A)}$ and $\ket{\psi(\widetilde{A})}$ with maximal symmetry breaking. Quantum fluctuations can thus have the tendency to decrease the expectation value of the order parameter with respect to the value obtained with mean field theory, or with an MPS approximation of finite $D$. As $D$ grows larger and the approximation improves, the negative eigenvalues should eventually disappear.

\subsection{Dynamical correlation functions}
\label{ss:excitations:spectral}
We conclude this section by discussing how the tangent space can assist in the computation of dynamical correlation functions. Let the ground state of a system be well approximated by the uMPS $\ket{\Psi(A)}$. As discussed in the introduction, dynamical correlation functions can either be computed in the time domain, or directly in the frequency domain. Traditional MPS algorithms for computing correlation functions in the time domain had to put the system on a finite lattice, although recently a number of competing algorithms have been developed to deal with localized perturbations in an otherwise translation invariant background \cite{Phien:2012dr,2012arXiv1207.0862Z,2012arXiv1207.0678P,2012arXiv1207.0691M}.

Dynamical correlation functions can be obtained from Fourier transforming the Green's function $G^{(\alpha,\beta)}(p,p',\omega)$ of two operators $\hat{O}^{(\alpha)}$ and $\hat{O}^{(\beta)}$, which is defined as
\begin{multline}
G^{(\alpha,\beta)}(p,p',\omega)=\\
\braket{\Psi(A)| \hat{O}^{(\alpha)\dagger}_{p}\frac{1}{\omega-[\hat{H}-H(\overline{A},A)]+\ic \eta}\hat{O}^{(\beta)}_{p'}|\Psi(A)}\label{eq:defgreen}
\end{multline}
where the Fourier transform of the operators $\hat{O}$ is defined as
\begin{equation}
\hat{O}_{p}=\sum_{n\in\mathbb{Z}} \ec^{\ic p n} \hat{T}^{n} \hat{O} \hat{T}^{-n}.\label{eq:fourieroperator}
\end{equation}
We have subtracted the approximate ground state energy $H(\overline{A},A)$ so as to get poles at finite excitation energies. Due to the translation invariance, $G^{(\alpha,\beta)}(p,p',\omega)$ is of the form $2\pi \delta(p-p') G^{(\alpha,\beta)}(p,\omega)$ where $G^{(\alpha,\beta)}(p,\omega)$ is finite (aside from its poles due to the excitation energies).

Clearly, for $\hat{O}^{(\alpha)}$ being one-site operators, the state $\hat{O}^{(\alpha)}_{m}\ket{\Psi(A)}$ exactly corresponds to a tangent vector $\ket{\Phi(B^{(\alpha)})}$ with $B^{(\alpha)s}=\braket{s|\hat{O}^{(\alpha)}|t}A^{t}$. We will generalize this to the case where $\hat{O}^{\alpha}$ acts on $K$ consecutive sites in Section~\ref{s:extensions}. For now, we can, without approximation, write 
\begin{multline*}
G^{(\alpha,\beta)}(p,p',\omega)=\\
\braket{\Phi(B^{(\beta)})|\hat{P}_{\mathbb{T}}\frac{1}{\omega-[\hat{H}-H(\overline{A},A)]+\ic \eta}\hat{P}_{\mathbb{T}}|\Psi(B^{(\alpha)})}\end{multline*}
Note that existing methods for directly computing the Green's function\cite{1995PhRvB..52.9827H,1999PhRvB..60..335K,2002PhRvB..66d5114J,2008arXiv0808.2620J,2011PhRvB..83p1104D,2011PhRvB..83s5115H,2012PhRvB..85t5119D}, which are almost always formulated for finite systems, try to approximate $\hat{O}^{\alpha}\ket{\Psi(A)}$, and sometimes even $({\omega-[\hat{H}-H(\overline{A},A)]+\ic \eta})^{-1}\hat{O}^{\alpha}\ket{\Psi(A)}$ as matrix product states. While this has led to successful methods on the finite lattice, the thermodynamic limit makes it clear that these states do not have the normalizable structure of MPS, but rather the structure of tangent vectors to the MPS manifold.

As we illustrate in the next section, the tangent space is typically suited for capturing elementary excitations with a single-particle-like or point-like structure. If we expect that these excitations are responsible  for the dominant contribution to the Green's function $G^{(\alpha,\beta)}(p,\omega)$, than we can replace Eq.~\eqref{eq:defgreen} with
\begin{multline*}
G^{(\alpha,\beta)}(p,p',\omega)\approx\\
\braket{\Phi(B^{(\beta)})|\hat{P}_{\mathbb{T}}\frac{1}{\omega-[\hat{P}_{\mathbb{T}}\hat{H}\hat{P}_{\mathbb{T}}-H(\overline{A},A)]+\ic \eta}\hat{P}_{\mathbb{T}}|\Psi(B^{(\alpha)})}
\end{multline*}
which leads to
\begin{multline}
G^{(\alpha,\beta)}(p,p',\omega)\approx\\
2\pi\delta(p-p') \overline{B}^{(\beta)\overline{\imath}} \left[\frac{1}{\omega-H(p)+\ic \eta}\right]_{\overline{\imath},j} B^{(\alpha)j},
\end{multline}
where the part multiplying the $2\pi\delta(p-p')$ has been defined as $G^{(\alpha,\beta)}(p,\omega)$ above. In practice we use of course the representation of $H(p)$ and $B$ in terms of the parameterization $X$, in order to eliminate the null modes. As will be illustrated in the next section, this approach is not able to accurately describe the contribution of the multi-particle continuum. Nevertheless, since $\hat{O}^{(\alpha)}_p\ket{\Psi(A)}$ is exactly in the tangent space, we do not lose any spectral weight.

Finally, we note that we do not need a full inverse of $\omega-H(p)$. If for example, we are interested in the imaginary part of $G^{(\alpha,\beta)}(p,\omega)$, which is often known as a spectral function or a structure factor, then we can write
\begin{equation}
\Im\left[ G^{(\alpha,\beta)}(p,\omega)\right]\approx\overline{B}^{(\beta)}\left[\delta\{\omega-H(p)\}\right]_{\overline{\imath},j} B^{(\alpha)j}\label{eq:defgreen2}
\end{equation}
The point is that we do not need to know the full eigenvalue decomposition of $H(p)$ (which would require an $O(D^6)$ calculation) to compute the $\delta$ function in the previous equation. Instead, we can decompose the $\delta$-function into Chebychev polynomials\cite{2011PhRvB..83s5115H}. Indeed, we can easily and efficiently compute arbitrary polynomials of the effective Hamiltonian $H(p)$ acting on $\ket{\Phi(B^{(\alpha)})}$ without making any further approximation. This allows to compute the right hand side of Eq.~\eqref{eq:defgreen2} at a fixed value of $p$, but for an arbitrary range of $\omega$ values, with a computational efficiency that scales as $\order(N D^3)$, where $N$ is the number of Chebychev moments that is used in the decomposition of the $\delta$-function.

\section{Illustrative examples}
\label{s:examples}
We now provide a few selected examples that can be used to illustrate some of the general statements of the previous section regarding the tangent space framework for excitations and dynamical correlation functions. 
\subsection{Variational ansatz for excitations}
\begin{figure}[h]
\centering
\includegraphics[width=\columnwidth]{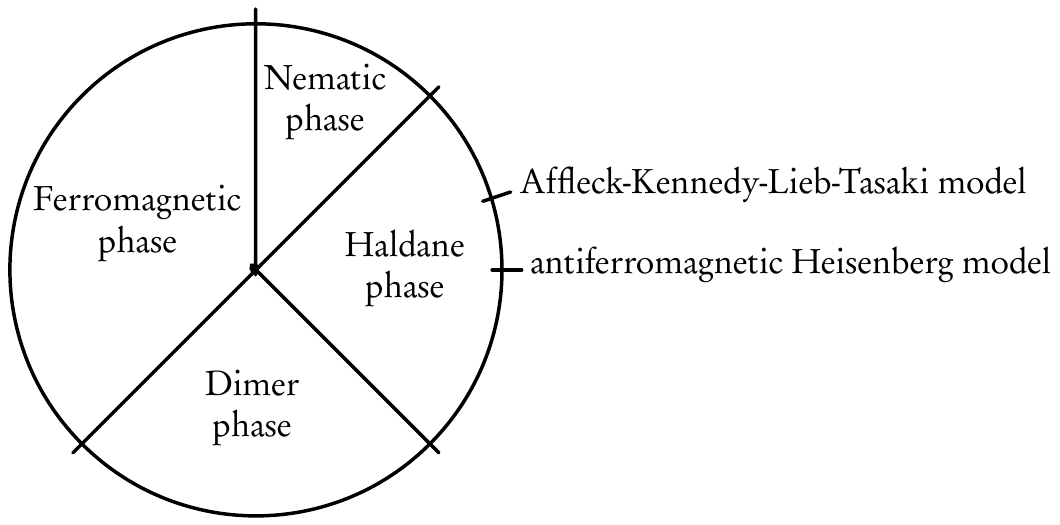}
\caption{Phase diagram of the bilinear-biquadratic Heisenberg model (taken from Ref.~\onlinecite{Rachel:2008aa}).}
\label{fig:mps:bbphase}
\end{figure}

An interesting model to illustrate the behavior of the variational ansatz for excitations is the bilinear-biquadratic antiferromagnetic Heisenberg model with $S=1$ spins, which is described by the $\mathsf{SO}(3)$-invariant Hamiltonian
\begin{equation}
\ham_{\text{BB}}=J\sum_{n\in\mathbb{Z}}\cos \theta \left(\operator{\bm{S}}_{n}\cdot \operator{\bm{S}}_{n+1}\right)+\sin \theta  \left(\operator{\bm{S}}_{n}\cdot \operator{\bm{S}}_{n+1}\right)^{2},\label{eq:bbham}
\end{equation}
with an energy scale $J>0$ and an angle $\theta\in [-3\pi/4,5\pi/4)$. This model has many interesting phases and phase transitions as a function of $\theta$. There cannot be antiferromagnetic order due to the Mermin-Wagner theorem\cite{Mermin:1966aa}. Ferromagnetic order can exist, since the ferromagnetic order parameter commutes with the Hamiltonian, and is indeed present for $\theta\in (\pi/2,5\pi/4)$. At $\theta=0$, this Hamiltonian describes the antiferromagnetic Heisenberg model. The ground state is then in a symmetry protected topologically ordered phase, the Haldane phase, and the lowest lying excitation is an $S=1$ triplet with finite mass $\Delta_{\text{Haldane}}$, referred to as the Haldane gap \cite{Haldane:1983aa,Haldane:1983ab,2010PhRvB..81f4439P}. The same phase exists throughout $\theta\in (-\pi/4,\pi/4)$. In particular, for $\tan \theta=1/3$ this is the model studied by Affleck, Kennedy, Lieb and Tasaki\cite{1987PhRvL..59..799A,1988CMaPh.115..477A}, which has an exact matrix product state representation with bond dimension $D=2$. At $\theta=\pm\pi/4$, an exact solution in terms of the Bethe ansatz is possible and the model is critical. The dispersion relation shows nodes for $p=0$ and $p=\pi$ for $\theta=-\pi/4$ (the Takhtajan-Babudjan point), and the model undergoes a second order phase transition to a dimer phase (only  invariant under $\operator{T}^{2}$), which exists for $\theta\in (-3\pi/4,-\pi/4)$. The existence of a small nematic phase between the dimer phase and the ferromagnetic phase has recently been ruled out \cite{Grover:2007aa}. For $\theta=\pi/4$ , the dispersion relation of the elementary excitation has nodes at $p=0$ and $p=\pm 2\pi/3$ for $\theta=\pi/4$. This critical behavior exists throughout the range $\theta\in[\pi/4,\pi/2)$, no trimerization occurs and the system is a nematic phase. The transition at $\theta=\pi/4$ is supposedly of the Kosterlitz-Thouless type. The whole phase diagram is summarized in FIG.~\ref{fig:mps:bbphase} (see \cite{Lauchli:2006aa,Rachel:2008aa} and references therein). 

\begin{figure}[h]
\centering
\includegraphics[height=0.5\textheight]{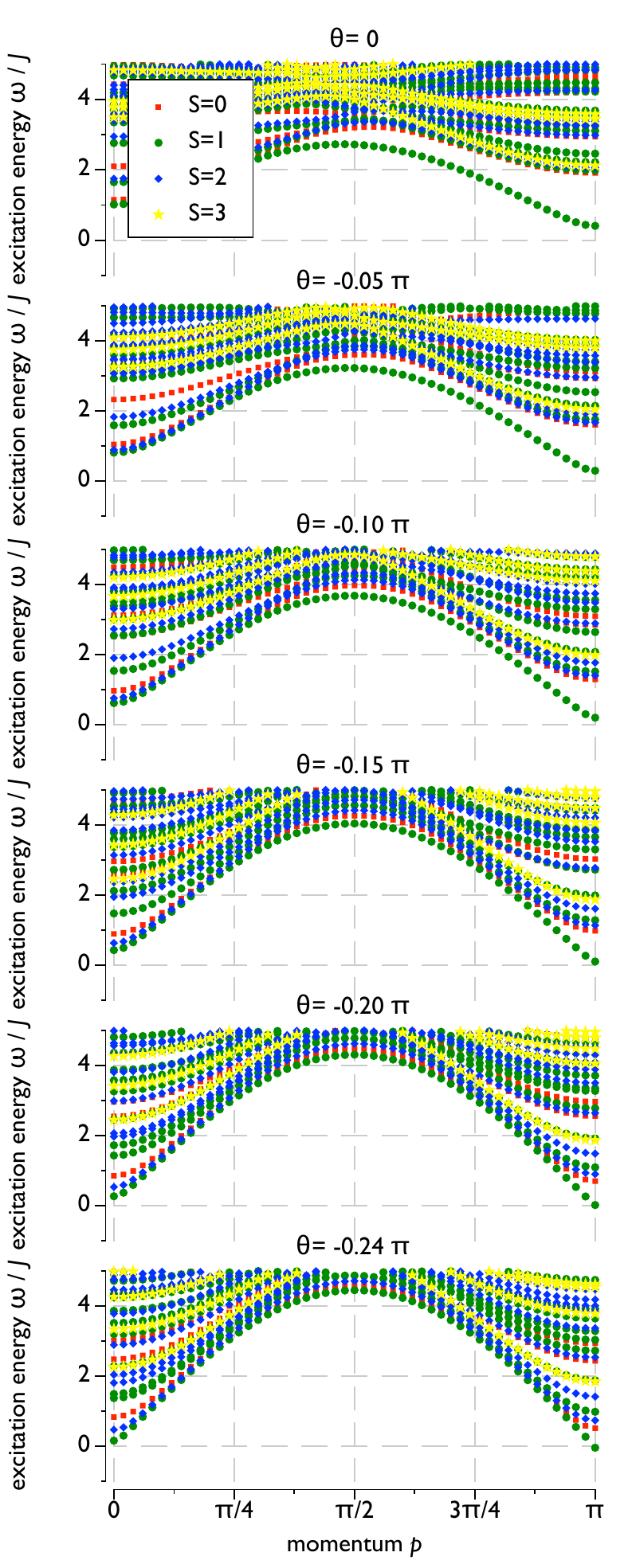}\includegraphics[height=0.5\textheight]{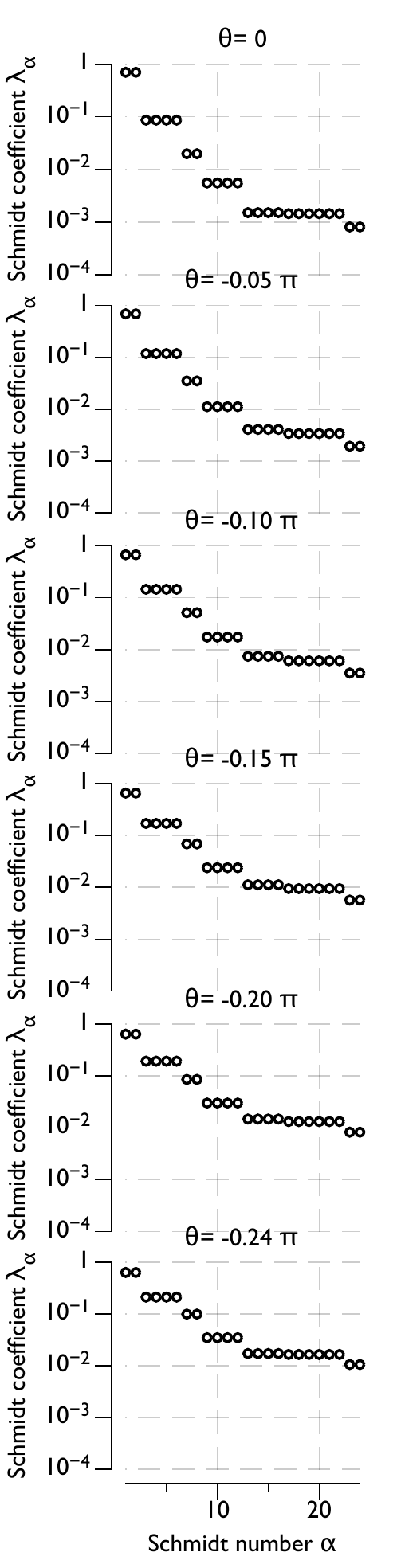}
\caption{Excitation spectrum (left) and Schmidt spectrum (right) for the bilinear-biquadratic antiferromagnetic Heisenberg model in the region $\theta\in (-\pi/4,0]$ obtained with a uMPS ansatz with bond dimension $D=24$.}
\label{fig:mps:bbspectrum1}
\end{figure}

In Ref.~\onlinecite{2012PhRvB..85j0408H}, the variational ansatz was used at the Heisenberg point $\theta=0$, in order to compute the value of the Haldane gap $\Delta_{\text{Haldane}}/J$ up to 12 digits of precision. We now turn to the more general bilinear-biquadratic Heisenberg model in the Haldane region $-\pi/4<\theta <+\pi/4$. As explained in the beginning of this paragraph, the model becomes critical at the end points. Figure~\ref{fig:mps:bbspectrum1} depicts the full set of eigenvalues of $H(p)$ for $\theta$ decreasing from $0$ to $-\pi/4$. Let us first discuss some general properties of the spectra obtained with our variational strategy. Firstly, one can observe that we have labeled the different eigenvalues according to their $\mathsf{SU}(2)$ spin quantum number $S$. We did not impose this symmetry explicitly, but were able to read of the irreducible representation to which each eigenvalue belonged simply from its degeneracy, which was almost perfect up to machine precision. The reason for this is that we started from a ground state approximation $\ket{\Psi(A)}$ where the bond dimension $D=24$ was chosen so as to contain a complete number of irreducible representations, and which was then converged ---also without imposing the $\mathsf{SU}(2)$ symmetry explicitly--- up to a state error $\widetilde{\epsilon}(\overline{A},A)\leq 10^{-12}$.

Around momentum $p=\pi$, the lowest excitation energy is separated from the rest of the spectrum and can be identified with the elementary spin 1 magnon excitation, which we can approximate well with our ansatz, as can be inferred from the numerical precision with which the Haldane gap can be estimated\cite{2012PhRvB..85j0408H}. Other excitations all fall within the multi-magnon continuum, starting with the 2-magnon ($S=0,1,2$) or 3-magnon ($S=0,1,2,3$) bands. Because we are diagonalizing a finite matrix $H(p)$ (dimensions $(d-1)D^2 \times (d-1)D^2$) for each momentum $p$, we obtain only a finite number of excitation energies. We thus need to interpret the discrete eigenvalues which are contained within the continuous spectrum of the full Hamiltonian. Clearly, the tangent space is not suited as a variational subspace for describing multiparticle excitations, as this would require an ansatz containing several perturbations of the ground state, which move independently with different momenta. Since only the total momentum of a state is a well-defined quantum number, we can however create superpositions of these multiparticle states in order to obtain a kind of artificial bound state, where all particles are staying together in a small spatial region. Such states can be described in the uMPS tangent space $\mathbb{T}_{p}$ and are thus obtained from diagonalizing $H(p)$. They are, however, no good approximations for eigenstates of the full Hamiltonian $\ham$, since they contain a superposition of many exact eigenstates with different energies.  The only exception is when the exact dispersion relation of the elementary excitation(s) of the full Hamiltonian $H$ is flat. In that case, all multiparticle excitations should also have a flat dispersion relation.  In general, we thus need a different strategy to study multiparticle excitations, on which we elaborate in Section~\ref{s:extensions}.

For larger bond dimensions, a full diagonalization of $H(p)$ becomes infeasible. However, we now know that we should only be interested in the lowest excitation energies and can thus resort to an iterative eigensolver, which can be implemented with computational complexity scaling as $\order(D^3)$, as discussed in the previous section. We can also make predictions about the convergence behavior of the variational excitation energies as a function of increasing bond dimension. Note that the excitation energies are no longer truly variational in the most strict sense, since they can have negative errors. Indeed, there are two competing sources of errors within our variational strategy for studying excitations. Firstly, the variational energy, which is always positive, comes from trying to represent the exact excited state as a uMPS tangent vector. However, since we compute the energy from a redefined Hamiltonian $\ham\leftarrow \ham-H(\overline{A},A)$, we are also subtracting an approximation to the ground state energy which is too large. The result of this second effect is a negative error. In fact, both sources produce errors which are infinitely large, unless the uMPS $\ket{\Psi(A)}$ captures the ground state exactly. However, these errors cancel each other, because they have opposite sign and are both related to an imperfect representation of the ground state. The result is a finite error, containing a positive contribution from the variational assumption that the excited state can be obtained from a local perturbation of the ground state, and a negative contribution from subtracting a ground state energy density which is too large in the region in which this local perturbation has support. For multiparticle excitations, the first error is clearly dominant and the resulting energies decrease as the bond dimension is increased, since this has the effect of enlarging the spatial region on which the excitation has support. In addition, new eigenvalues can appear within the multiparticle continuum when $D$ is increased. For elementary excitations or bound states on the other hand, the locality assumption is often appropriate, and the second error typically dominates. We then observe a variational excitation energy that increases as $D\to \infty$. It can however happen that the spatial region on which the excitation has support is too small for small values of $D$, in which case we also observe a decreasing energy at low bond dimensions. Such effects were indeed observed in Ref.~\onlinecite{2012arXiv1212.1114D}, to which we refer for a more elaborate discussion on this subject.

We now return to discuss the elementary magnon excitation in the bilinear-biquadratic antiferromagnetic Heisenberg model. For $\theta$ decreasing from $0$ to $-\pi/4$, the excitation gap at $p=\pi$ decreases and eventually vanishes (see Figure~\ref{fig:mps:bbspectrum1}). Simultaneously, the entanglement increases as can be noticed by the Schmidt coefficients shifting upwards. At $\theta=-0.24\pi$, the elementary magnon has in fact a slightly negative excitation energy at $p=\pi$. While this is of course an artifact of the low value of the bond dimension $D=24$, since the critical point is not until $\theta=-0.25\pi$, it does indicate a tendency of these magnons to condense, resulting in a ground state that breaks translation invariance down to invariance under shifts over two sites. Indeed, for $\theta<-\pi/4$ the ground state of the system has a dimer configuration. While a zero (negative) excitation energy of the elementary magnon at $p=\pi$ results in zero (negative) excitation energies for some two-magnon excitations with total momentum $p=0$, the excitation energies found with our ansatz are positive at $p=0$, as required by the fact that our uMPS approximation is a minimum for the energy functional on our variational manifold. This is a consequence of the aforementioned fact that our ansatz is not suitable for the description of two-particle excitations. The negative energy for the elementary excitation at $p=\pi$ does however indicate that we could have found a lower energy density if we would have used a two-periodic MPS with bond dimension $D=24$, as was explained at the end of Subsection~\ref{ss:excitations:reltdvp}.

\begin{figure}[h]
\centering
\includegraphics[height=0.50\textheight]{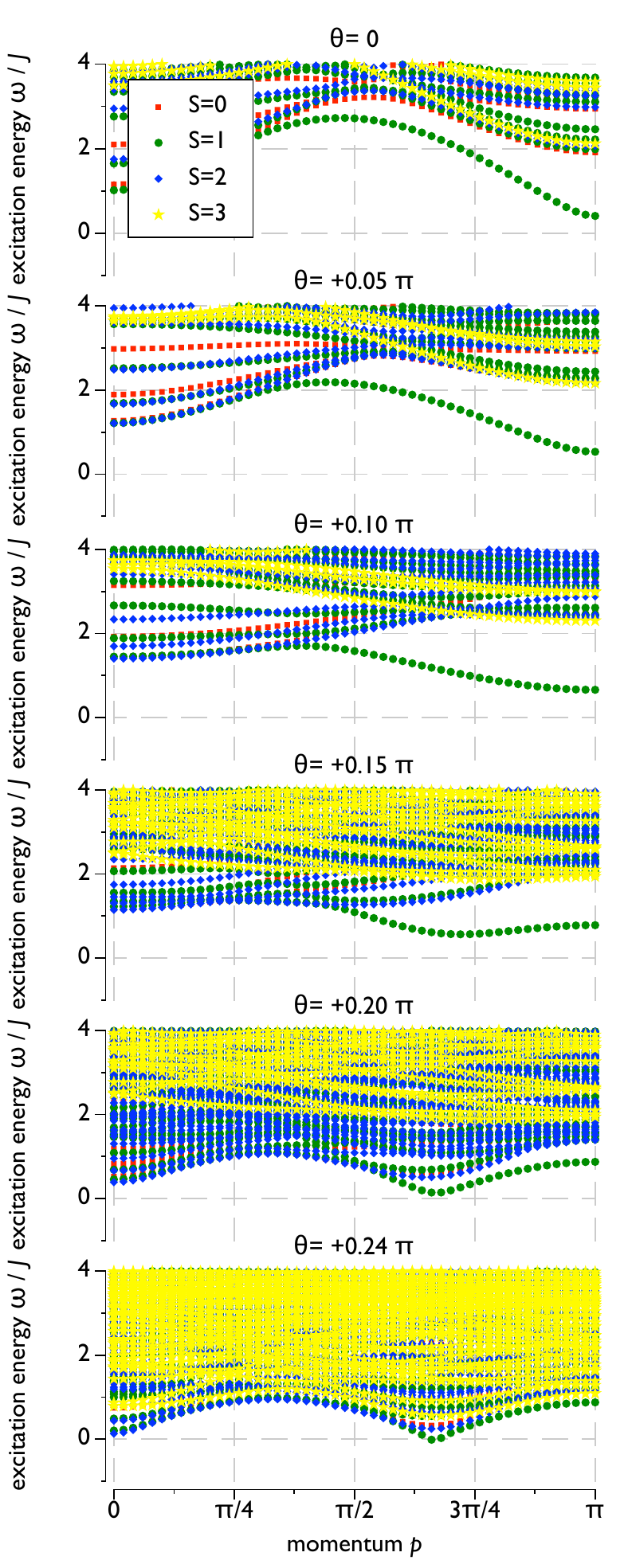}\includegraphics[height=0.50\textheight]{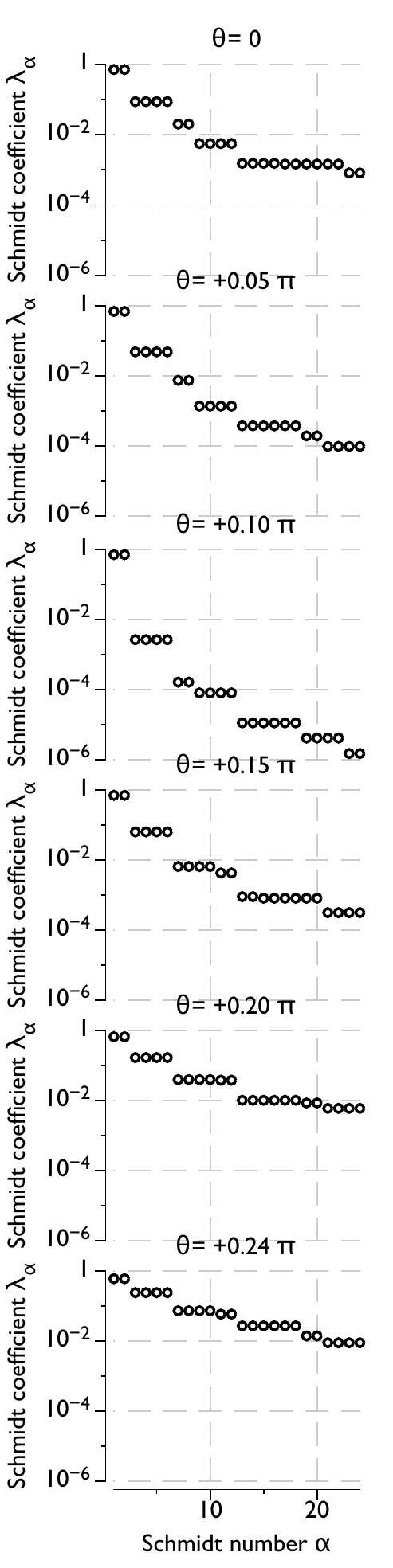}
\caption{Excitation spectrum (left) and Schmidt spectrum (right) for the bilinear-biquadratic antiferromagnetic Heisenberg model in the region $\theta\in [0,\pi/4)$ obtained with a uMPS ansatz with bond dimension $D=24$.}
\label{fig:mps:bbspectrum2}
\end{figure}

The same analysis can be repeated for $\theta$ increasing from $0$ to $\pi/4$, which is sketched in Figure~\ref{fig:mps:bbspectrum2}. For $\theta$ slightly larger than $0$, the excitation energy of the elementary magnon around $p=\pi$ increases, resulting in smaller correlation length and a decrease of the entanglement entropy (as indicated by the Schmidt values shifting downwards). Indeed, at $\theta=\arctan(1/3)\approx 0.1024\pi$ this is the AKLT-model, for which the ground state has an exact matrix product state representation with $D=2$. For $\theta=0.10\pi$, the importance of the Schmidt coefficients $\lambda_{\alpha}$ for $\alpha>2$ has strongly decreased. If $\theta$ is increased further, the excitation energy of the elementary magnon starts to decrease around $p=2\pi/3$ and eventually a null mode develops. Once again, the excitation energy is slightly negative for $\theta=0.24\pi$, which is an artefact of the small bond dimension $D=24$. This could again be interpreted as an indication for condensation of elementary magnons with momentum $2\pi/3$ in the phase transition at $\theta=\pi/4$, which would result in the breaking of translation invariance down to invariance under shifts over three sites. However, in the exact solution no such trimerization occurs and the model remains critical throughout $\theta\in [\pi/4,\pi/2)$. Hence, while the MPS approximation and derived methods for excitations can provide valuable information about the phase of a system and the nature of a phase transition, such information is not always reliable and should be used carefully.

\subsection{Linearized time-dependent variational principle}
\begin{figure}[h]
\centering
\includegraphics[width=\columnwidth]{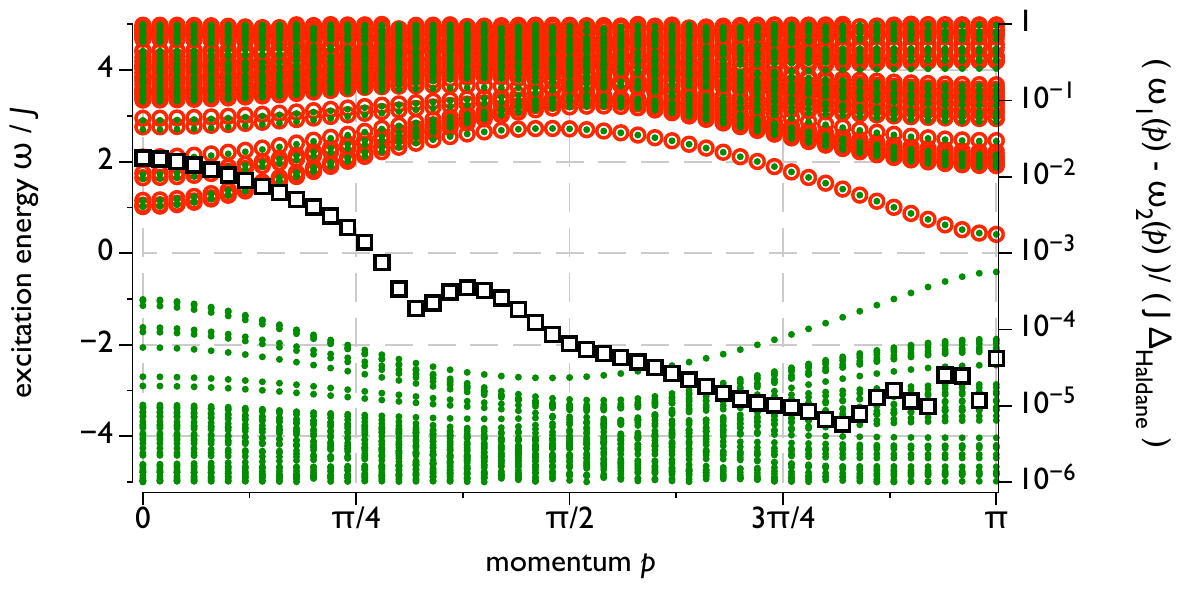}
\caption{Comparison of the excitation spectrum $\omega$ obtained with the linearized TDVP equation [Eq.~\eqref{eq:linearizedtdvpfourier2}] (green dots) with the excitation spectrum obtained with the tangent space variational ansatz for excitations $\ket{\Phi_{p}(B)}$ (red circles), for the Heisenberg model studied at bond dimension $D=24$. Also shown is the absolute value of the difference between the dispersion relations (black squares, to be read on the right axis) for the lowest excitation, which corresponds to the elementary spin-$1$ magnon in the momentum range $\frac{\pi}{4}\lessapprox \lvert p\rvert \leq \pi$.}
\label{fig:heisenbergspectrumextended}
\end{figure}

We now elaborate on the relation between the tangent space as variational ansatz for excitations, and the resonance frequencies that we obtain from linearizing the TDVP equation around the variational optimum $\ket{\Psi(A)}$. For the $S=1$ Heisenberg antiferromagnet [$\theta=0$ in Eq.~\eqref{eq:bbham}], we 
perform an exact diagonalization of the generalized eigenvalue problem of Eq.~\eqref{eq:linearizedtdvpfourier2} and compare the eigenvalues to those from by diagonalizing $H(p)$. Note that, in the latter case, we have a variational ansatz state $\ket{\Phi_{p}(B)}$ for the corresponding excitation, whereas in the former, we do not. For bond dimension $D=24$, a comparative results between both sets of eigenvalues is sketched in Figure~\ref{fig:heisenbergspectrumextended}. The doubled spectrum of eigenvalues of Eq.~\eqref{eq:linearizedtdvpfourier2} is reflection invariant around zero. For the lowest (positive) eigenvalue, we also plot the absolute value of the difference between both approaches. 

For $D=24$, the localized state error $\widetilde{\epsilon}(\overline{A},A)$ is approximately $3\times 10^{-3}$. In the region around $p=\pi$, the lowest eigenvalue obtained with both methods corresponds to the elementary $S=1$ magnon excitation. It can be observed from FIG.~\ref{fig:heisenbergspectrumextended} that the difference between both dispersion relations is much smaller than $\widetilde{\epsilon}(\overline{A},A)$ in that region. The second tangent space correction that is contained in the larger eigenvalue problem of Eq.~\eqref{eq:linearizedtdvpfourier2} contributes little to the elementary excitations. Hence, for larger values of the bond dimension $D$, we expect the difference between the two approaches to be practically inexistent for elementary excitations. However, around $p\approx \pi/4$, the two-magnon continuum becomes lower in energy and the single-magnon excitation ceases to exist\cite{2008PhRvB..77m4437W}. In that region, the difference between both approaches increases, which is to be expected, since the uMPS tangent space $\mathbb{T}$ is no longer a good variational subspace for multiparticle excitations. Corrections obtained from taking the second tangent space into account thus become more substantial.

Note, however, that the larger eigenvalue problem obtained from linearizing the TDVP is not equivalent to a variational ansatz taking the second tangent space into account. This would be a completely different approach, that is far more difficult, as even within a given sector of fixed total momentum $p$, the second tangent space to the manifold of MPS contains infinitely many degrees of freedom. We return to this question in Subsection~\ref{ss:extensions:fock}. There is no variational ansatz and hence no variational principle associated to the eigenvalues obtained from Eq.~\eqref{eq:linearizedtdvpfourier2}. This lack of an underlying variational principle results in uncontrollable errors and the possibility for spurious solutions, as we now demonstrate.

\begin{figure}[h]
\centering
\includegraphics[width=\columnwidth]{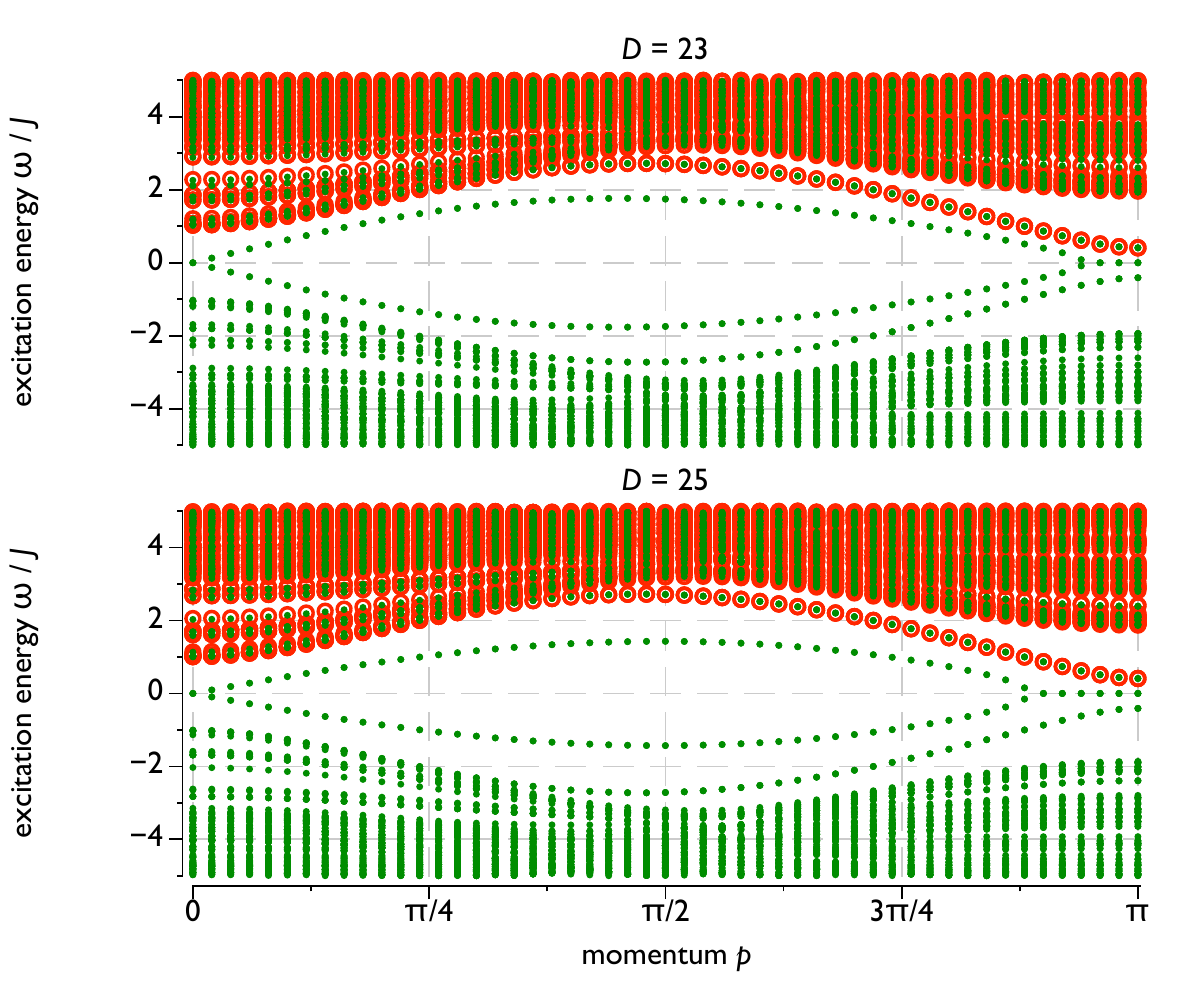}
\caption{Similar comparison of the excitation spectrum $\omega$ obtained with the linearized TDVP equation [Eq.~\eqref{eq:linearizedtdvpfourier2}] (green dots) with the excitation spectrum obtained with the tangent space variational ansatz for excitations $\ket{\Phi_{p}(B)}$ (red circles), for the Heisenberg model at bond dimension $D=23$ and $D=25$.}
\label{fig:heisenbergspectrumextended2}
\end{figure}

All went well for the results in FIG.~\ref{fig:heisenbergspectrumextended}, since we tuned the bond dimension to be commensurable with the $\mathsf{SO}(3)$ degeneracy in the virtual space, \textit{i.e.}~the spectrum of Schmidt coefficients contains an exact number of spin multiplets. As is well known, the Haldane phase is a symmetry protected topological phase, since it only has non-trivial projective representations of $\mathsf{SO}(3)$ in its virtual space, \textit{i.e.}~it only has half integral spin multiplets and every Schmidt value has an even degeneracy\cite{2010PhRvB..81f4439P}. If we choose an odd-valued bond dimension such as $D=23$ or $D=25$, the virtual space cannot entirely consist out of complete half-integral spin representations, and some vectors of an incomplete multiplet will have to be added. The variational optimum is then no longer unique: there will be a flat valley of equally good choices $A$ that produce the same energy expectation values $\braket{\Psi(A)|\ham|\Psi(A)}$. Different values of $A$ correspond to different choices of Schmidt vector associated to the smallest Schmidt coefficient. A first effect of this degenerate valley of energy optima is that the location of the energy optimum with an imaginary time simulation according to the time-dependent variational principle converges much slower. In addition, the corresponding optimum $\ket{\Psi(A)}$ will have a slight breaking of the $\mathsf{SO}(3)$ symmetry of the Hamiltonian, of the same order as the size of of the Schmidt coefficients of the broken multiplet. The effect on the excitation spectrum obtained with the variational ansatz and with the linearized TDVP equation is shown in Fig.~\ref{fig:heisenbergspectrumextended2}. The variational ansatz $\ket{\Phi(B;A)}$ still produces a qualitatively correct excitation spectrum, where the degeneracy of the elementary $S=1$ magnon is slightly lifted. The energy difference between the three states is easily an order of magnitude smaller than the size of the Schmidt coefficients of the broken multiplet, \textit{i.e.}~in the order of $10^{-4}$ to $10^{-5}$. In contrast, when using the linearized TDVP equation around a given point in this optimal value, then it produces a fake null mode. Indeed, let $A(s)$ be a one-parameter group of tensors $A$ that runs through this valley. Then $\braket{\Psi(A(s))|\ham|\Psi(A(s))}$ is $s$-independent. Since all of these are variational optima, we automatically obtain $\braket{\Psi(\overline{A(s)})|\ham|\Phi_{p}(B;A(s))}=0$ for any $\ket{\Phi_{p}(B;A(s))}\in \Tplane_{\text{MPS}}^{\perp}(A(s))$. We also obtain
\begin{multline*}
\left.\frac{\d^{2}\ }{\d s^{2}}\braket{\Psi(A(s))|\ham|\Psi(A(s))}\right\vert_{s=0}=\\
2\braket{\Phi_{0}(\overline{B}|\ham|\Phi_{0}(B)}+\braket{\Psi(A)|\ham|\Upsilon_{0,0}(B,B)}\\
+\braket{\Upsilon_{0,0}(\overline{B},\overline{B})|\ham|\Psi(A)}=\\
\begin{bmatrix} \bm{B}^{\dagger} & \overline{\bm{B}}^{\dagger}\end{bmatrix}\begin{bmatrix}H(0) & K(0) \\ \overline{K(0)} & \overline{H(0)}\end{bmatrix} \begin{bmatrix}\bm{B} \\ \overline{\bm{B}}\end{bmatrix}=0
\end{multline*}
with $A=A(0)$ and $B=\d A/\d s(0)$. Hence, the Hessian of the energy appearing in the left hand side of Eq.~\eqref{eq:linearizedtdvpfourier2} for $p=0$ has a zero eigenvalue, and so does the generalized eigenvalue equation of Eq.~\eqref{eq:linearizedtdvpfourier2} itself. In fact, for the present case, there are two independent null modes. This could be argued to be a good thing, since the above proof essentially shows the existence of Goldstone modes whenever the ground state breaks a continuous symmetry. However, the problem with this null mode is that it is always there, irrespective of how good the ground state approximation $\ket{\Psi(A)}$ is and how small the remaining symmetry breaking is. If the symmetry breaking is only tiny, so is the lift of the degeneracy of the spin multiplets in the eigenvalue spectrum of $H(p)$. In contrast, the spectrum of eigenvalues of Eq.~\eqref{eq:linearizedtdvpfourier2} completely loses its multiplet structure as it starts with a two-fold degenerate Goldstone mode and the energy difference between the different magnon excitations can be as large as $\order(10^{-2})$. This indicates how fragile the eigenvalue structure of Eq.~\eqref{eq:linearizedtdvpfourier2} is. 

\begin{figure}[h]
\centering
\includegraphics[width=\columnwidth]{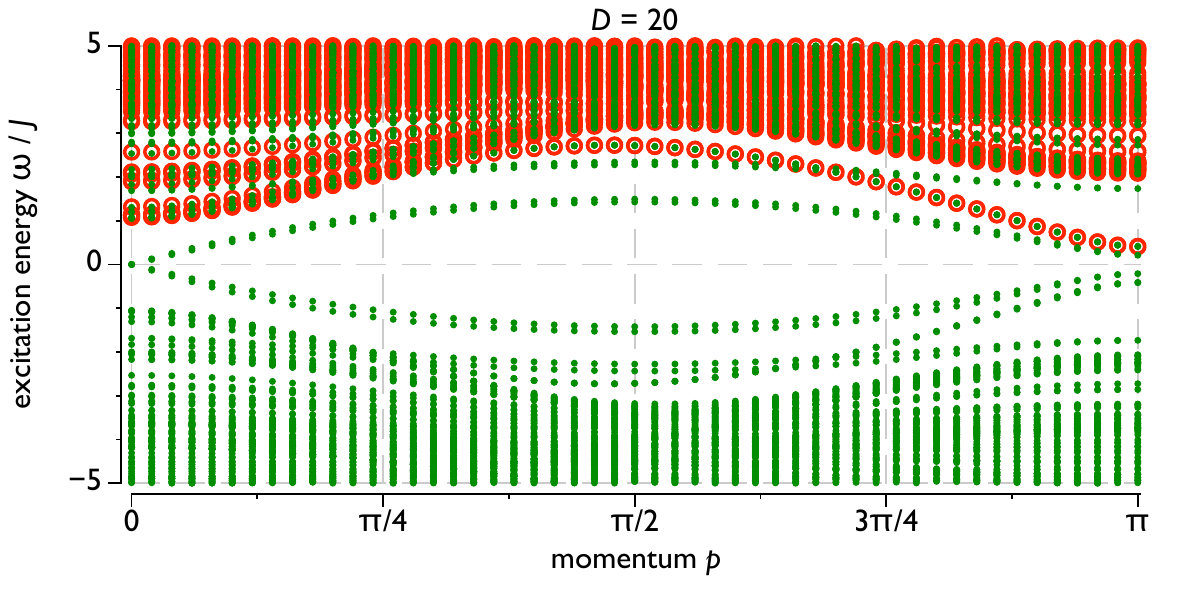}
\caption{Similar comparison of the excitation spectrum $\omega$ obtained with the linearized TDVP equation [Eq.~\eqref{eq:linearizedtdvpfourier2}] (green dots) with the excitation spectrum obtained with the tangent space variational ansatz for excitations $\ket{\Phi_{p}(B)}$ (red circles), for the Heisenberg model at bond dimension $D=20$.}
\label{fig:heisenbergspectrumextended3}
\end{figure}

If we have unwillingly broken an unknown symmetry of the Hamiltonian, no matter how small the symmetry breaking is, the linearized TDVP equation [Eq.~\eqref{eq:linearizedtdvpfourier2}] will always force a zero energy mode upon us. It is unable to detect whether the symmetry breaking is artificial ---\textit{i.e.}~the optimal valley of $A$'s map to states $\ket{\Psi(A)}\in\varM$ which are very close in Hilbert space and are centered around one unique direction $\ket{\Psi_0}$ (the true ground state) that is not exactly contained in the manifold--- or whether the symmetry breaking is physical ---\textit{i.e.}~the optimal valley of $A$'s map to states $\ket{\Psi(A)}$ that are lying in completely different direction of Hilbert space and are close to the different directions in the exact degenerate ground state subspace. In the latter case, if we can well approximate the different ground state by uMPS $\ket{\Psi(A)}$, we expect that $H(p)$ itself contains very small eigenvalues corresponding to Goldstone modes, since the terms $K(0)$ vanish for the exact ground state anyway. As a final example, we plot the same  comparison for bond dimension $D=20$ in FIG.~\ref{fig:heisenbergspectrumextended3}. Even though this value of the bond dimension is even, it still cuts through a spin multiplet in the virtual space, since the Schmidt coefficients with number 17 to 22 all correspond to an $S=5/2$ multiplet. For $D=23$ and $D=25$, we had two Goldstone modes that became massless both at $p=0$ and $p=\pi$, but aside from those, Eq.~\eqref{eq:linearizedtdvpfourier2} did reproduce a nearly degenerate gapped magnon excitation around the same energy as the variational ansatz. For $D=20$, the situation is completely different. The two Goldstone modes are only massless at momentum $p=0$, but mix with other excitations near momentum $\pi$. Indeed, aside from the magnon excitation and the two Goldstone modes, we find other excitations below the multiparticle continuum around momentum $\pi$.

\subsection{Spectral functions}
\begin{figure}[h]
\centering
\includegraphics[width=\columnwidth]{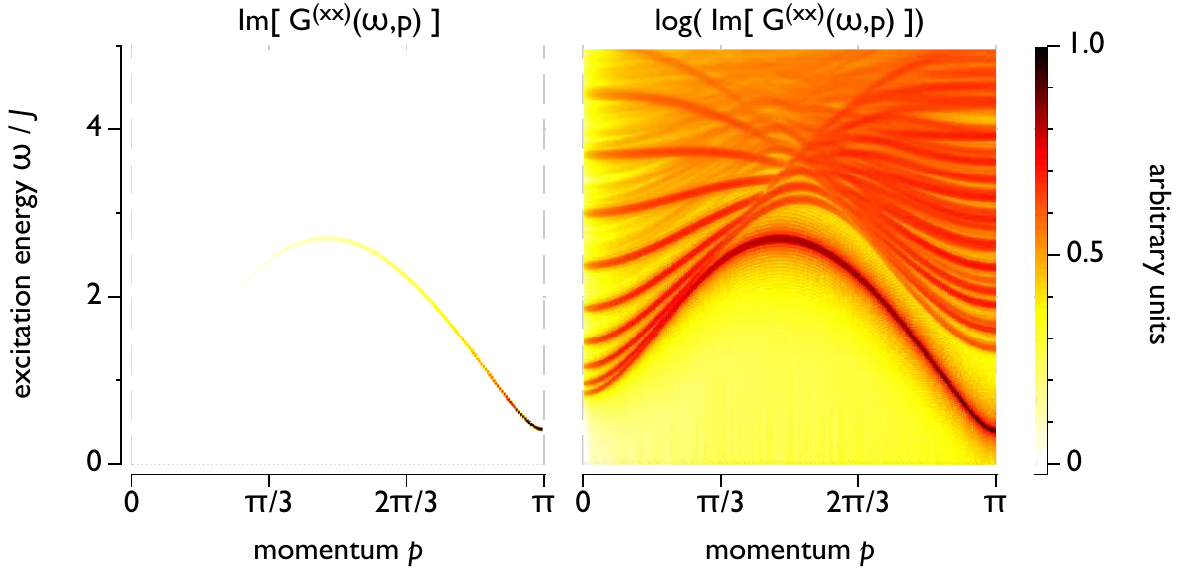}
\caption{Spectral function $\Im[G^{(xx)}(\omega,p)]$ for the $S=1$ Heisenberg antiferromagnet with operators $\hat{O}^{(\alpha)}=\hat{O}^{(\beta)}=\hat{S}^{x}$, as obtained from Eq.~\eqref{eq:defgreen2}. We have applied the Chebychev decomposition of the $\delta$ function using the first $N=1000$ polynomials. The bond dimension was chosen as $D=192$.}
\label{fig:spectral}
\end{figure}

We close this section by showing a simple result for the spectral function function obtained by computing Eq.~\eqref{eq:defgreen2}. For the $S=1$ Heisenberg antiferromagnet, where we use $\hat{O}^{(\alpha)}=\hat{O}^{(\beta)}=\hat{S}^{x}$, we obtain the result in FIG.~\ref{fig:spectral}. If we look at absolute scale, we note that the spin operators have a large overlap with the single magnon excitation around momentum $p=\pi$, but that this overlap goes to zero very quickly as the momentum gets closer to the point where the single magnon excitation vanishes in the two-particle continuum. By looking at the logarithm of the spectral function, we show that part of the spectral weight is also distributed over the discrete eigenvalues in the multi-magnon continuum. Since the tangent space $\mathbb{T}$ captures $\hat{S}^{(x)}_{p}$ acting on the ground state exactly, no spectral weight is lost. Since we accurately reproduce the single magnon dispersion relation\cite{2012PhRvB..85j0408H}, we assume that the corresponding variational estimate for the eigenstate $\ket{\Phi_{p}(B)}$ is sufficiently accurate to correctly capture the spectral overlap with the spin operators. Then, we necessarily also have an accurate estimate for the fraction of the spectral weight that is lost to the 2- and 3-magnon continuum, but instead of this spectral weight being spread out in a continuous distribution over $\omega$, it is localized at the artificial, discrete set of eigenvalues that is supported within the tangent space. Hence, while a tangent-space based computation of spectral functions is acceptable if the only interest is in the contribution of the elementary excitations, it cannot accurately reproduce the contribution of multiparticle excitations. A more advanced strategy is required, the foundations of which are discussed in Subsection~\ref{ss:extensions:fock}.

\begin{figure}[h]
\centering
\includegraphics[width=\columnwidth]{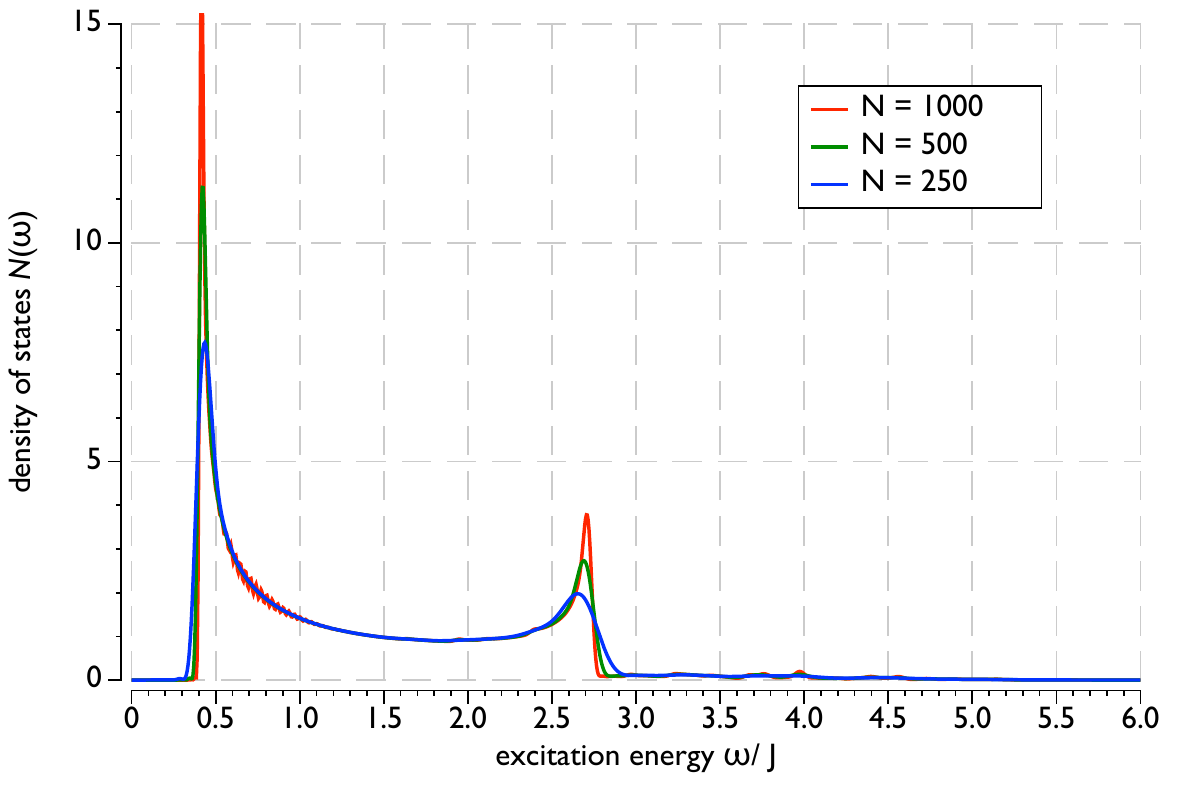}
\caption{Density of states $N(\omega)$ obtained from integrating $\Im[G^{(xx)}(\omega,p)]$ over $p\in[0,2\pi)$ for the $S=1$ Heisenberg antiferromagnet. We show results where we have applied the Chebychev decomposition of the $\delta$ function using the a varying number $N=250$, $500$ or $1000$ polynomials. The bond dimension was chosen as $D=192$.}
\label{fig:density}
\end{figure}

Finally, we can also integrate $\Im[G^{(xx)}(\omega,p)]$ over the momentum to obtain the density of states $N(\omega)$. This is illustrated in FIG.~\ref{fig:density} and compares well to the results in FIG.~5 and FIG.~6 of Ref.~\onlinecite{2008PhRvB..77m4437W}. Clearly, the tangent space offers a very efficient method to directly generate spectral functions or the density of states once we have obtained the uMPS approximation $\ket{\Psi(A)}$ of the ground state. With the tangent space approach, we did not need any kind of statistical procedures to extract or improve the result, nor did we need to take any kind of boundary effects or finite size effects into account. Note that, as the number of terms $N$ in the Chebychev decomposition of the $\delta$-function increases, the peaks become sharper and higher, and will eventually develop into singularities for $N\to\infty$. The fringes that appear for $N=1000$ indicate that the discretization of the momentum $p$ (namely $\d p=\pi/200$) is too rough, which is why the contribution of individual momentum states in the integrated quantity $N(\omega)$ become visible. 

\section{Analogies with other variational methods}
\label{s:analogies}
The TDVP is a general method that can be applied to any variational manifold for any quantum problem, be it a single-particle or a many-particle quantum system. However, it is not guaranteed that for any variational manifold, the tangent space provides a suitable ansatz for elementary excitations of the system. In fact, it is a remarkable property of the manifold of MPS that its tangent space contains states which seem to have the correct structure for well approximating elementary excitations. The best motivation for recognizing that this is indeed the correct structure is not via the linearized TDVP equation, but by observing that the MPS tangent vectors generalize the Feynman-Bijl variational ansatz for excitations. In fact, this observation allows us to simply generalize the excitation ansatz to states which are no longer in the MPS tangent space, as described in the next section.

Nevertheless, we now try to compare the methods constructed in the previous sections to similar developments with other variational methods that are used in the context of quantum many body systems. The most prominent example is without doubt Hartree-Fock theory, \textit{i.e.}~mean-field theory for a system of $N$ fermionic particles\cite{Hartree:1928aa,Hartree:1928ab,Slater:1930aa,Fock:1930aa,Fock:1932aa}. The variational manifold is the set of all Slater determinants, and can be identified with a Grassmann manifold. Applying the TDVP for $D>1$ results in time-dependent Hartree-Fock theory, and by linearizing these equations around the stationary solution, one obtains the so-called random phase approximation\cite{PhysRevA.22.2362,Ring:2004fk}. In this context, the random phase approximation can also be derived using Green's function techniques\cite{Dickhoff:2005uq} or by using the equations-of-motion approach\cite{RevModPhys.40.153}.

In line with what we have illustrated in the previous section, is it well-known in quantum chemistry and nuclear physics that the random phase approximation can be plagued by instabilities. In that case, it is also better to ignore the off-diagonal blocks in the generalized eigenvalue equation, which boils down to a diagonalization of the Hamiltonian within the tangent space\cite{Dickhoff:2005uq}, and which is known as the Tamm-Dancoff approximation\cite{Tamm:1945vn,Dancoff:1950kx}. However, unlike as for MPS, which has a refinement parameter $D$ (the bond dimension), there is no systematic way to improve the Hartree-Fock approximation for the ground state, and the random phase approximation very often provides the only way of taking correlations in the ground state into account.

For Hartree-Fock theory, the tangent space consists of all single particle-hole excitations, and there is one-to-one mapping between derivatives with respect to the variational parameters and physical excitation operators creating the same tangent state. Similarly, the second tangent space of the Hartree-Fock manifold corresponds to all states with two particle-hole excitations on top of the Slater determinant. We discuss the possibility for excitation operators that are in one-to-one correspondence to the tensors $B$ for the case of MPS tangent vectors in the next section, as these are required for further developments that take us beyond the tangent space. For a more detailed comparison between Hartree-Fock theory and matrix product states on finite lattices, we refer the reader to Ref.~\onlinecite{papersebastian}. This paper also describes and implements an MPS analogue of the post-Hartree-Fock method CISD, which is an abbreviation for configuration interaction with singles and doubles, and boils down to an exact diagonalization of the Hamiltonian within the subspace obtained by combining the optimal MPS, its tangent space and its double tangent space. However, such a method breaks down in the thermodynamic limit, as it lacks extensivity. Indeed, the MPS tangent space and double tangent space have a different normalization structure and cannot lower the energy density. Constructing an MPS-analogue of an extensive post-Hartree-Fock methods such as coupled cluster theory, on the other hand, could also be relevant for calculations in the thermodynamic limit.

Mean-field theory for bosons can start from the $N$-particle wave function $\Psi(\vec{r}_{1},\ldots,\vec{r}_N)=\prod_{n=1}^{N} \varphi(\vec{r}_{n})$ or, for systems in second quantization, from the coherent state $\ket{\Psi}=\exp\big[\int \varphi(\vec{r}) \hat{\psi}^{\dagger}(\vec{r}) \,\d \vec{r}\big]\ket{\Omega}$. In both cases, the variational parameters correspond to the choice of single-particle function $\phi(\vec{r})$. Applying the TDVP results in the Gross-Pitaevskii equation for $\varphi(\vec{r})$\cite{Gross:1961aa,Pitaevskii:1961aa}. Linearization around the stationary point results in the Bogoliubov-de Gennes equation\cite{Gennes:1966aa}. As in Hartree-Fock theory, there is a clear link between the tangent vectors and the action of physical operators, since
\begin{align*}
\frac{\delta\ }{\delta \varphi(\vec{r})} \ket{\Psi}&=\frac{\delta\ }{\delta \varphi(\vec{r})}\exp^{\int \varphi(\vec{r}) \hat{\psi}^{\dagger}(\vec{r}) \,\d \vec{r}}\ket{\Omega}\\
&=\hat{\psi}^{\dagger}(\vec{r})\ket{\Psi}
\end{align*}
and the Bogoliubov-de Gennes equation can also be obtained from a bosonic expansion of the Hamiltonian up to second order, which clearly shows its non-variational character (since higher order terms of the expansion are being ignored). The relation between the Gross-Pitaevskii theory and our framework is best studied with continuous MPS\cite{2010PhRvL.104s0405V,2010PhRvL.105z0401O,2012arXiv1211.3935H}. Indeed, a continuous MPS with bond dimension $D=1$ corresponds exactly to the mean field coherent state $\ket{\Psi}$, whereas correlations are systematically included for larger $D$. Applying the TDVP for $D>1$ results in a non-commutative version of the Gross-Pitaevskii equation\footnote{F.~Verstraete \textit{et. al.}, in preparation.}.

\section{Beyond the tangent space}
\label{s:extensions}
Most of this paper so far has been focussed on the MPS tangent space $\mathbb{T}$, although we also briefly encountered states in the double tangent space $\mathbb{T}^{(2)}$. We now discuss several extensions that go beyond the strict MPS tangent space $\mathbb{T}$. As there is no natural reason to stick to the tangent space when constructing a variational ansatz for excitations, we discuss the effect of using tensors $B$ that act on several sites in the first subsection. While the tangent space did appear naturally for time evolution, we show in the second subsection that there can also be a role for momentum zero states with a tensor $B$ acting on several sites. Finally, in the last subsection, we give a general outlook on how successively replacing more and more $A$ matrices corresponds to building multiparticle states, eventually leading to the construction of a complete Fock space in which we can define an effective low-energy theory. Therefore, we need to link the action of replacing tensors $A$ to physical operators.

\begin{widetext}
\subsection{Spreading blocks over several sites}
As a variational ansatz for excitations, the tangent vectors $\ket{\Phi_{p}(B)}$ constructed at the base point $\ket{\Psi(A)}$ were motivated because they capture exactly the Feynman-Bijl ansatz for one-site operators. We expect that these states can accurately capture the effect of operators with bigger support as well, since via the virtual dimension, the tensor $B$ can have have an effect over a distance approximately $\log_{d} D$ away from the site on which $B$ is living. However, there is no intrinsic reason why we should stick to the tangent space, and we can define a generalized excitation ansatz where we replace the tensors $A$ on a block of $K$ consecutive sites by a single tensor $B$ as
\begin{equation}
\ket{\Phi_{p,K}(B;A)}=\sum_{n\in\mathbb{Z}}\ec^{\ic p n}\sum_{\{s_n\}=1}^{d} \bm{v}_{\mathrm{L}}^{\dagger}\left[\left(\prod_{m<n} A^{s_{m}}\right) B^{s_{n}s_{n+1} \ldots s_{n+K-1}} \left(\prod_{m'\geq n+K} A^{s_{m'}}\right)\right]\bm{v}_{\mathrm{R}} \ket{\bm{s}}
\end{equation}
One particular example where this construction is extremely useful, is when the ground state is an exact MPS, as for the AKLT model. In that case, it is impossible to systematically improve the tangent space ansatz $\ket{\Phi_{p}(B)}=\ket{\Phi_{p,1}(B)}$, since there is no point in increasing the bond dimension. Note that in this case the error on the approximate excitation energies is purely variational, \textit{i.e.}~the variational energies are an upper bound for the exact value. By increasing the spatial support of $B$, the variational energies can systematically be lowered and will thus converge to the exact result. We can show that this convergence is exponentially fast\footnote{J.~Haegeman, S.~Michalakis, B.~Nachtergaele, T.~J.~Osborne, N.~Schuch and F.~Verstraete, \textit{in preparation.}}. Table~\ref{tab:aklt} shows the value for the magnon gap at momentum $p=\pi$ in the AKLT model, obtained with this ansatz for $K$ ranging from $1$ to $12$, and it can be observed that the energy converges at a rate of approximately $1$ digit per extra site. The $K=1$ case corresponds to $E_{\text{magnon}}^{(K=1)}=10/27$, as was calculated analytically in Ref.~\onlinecite{1988PhRvL..60..531A}. 

\begin{table}[h]
\caption{Variational excitation energy for the single magnon excitation at momentum $p=\pi$ in the AKLT model, obtained by using the variational ansatz $\ket{\Phi_{p=\pi,K}(B)}$, where the tensor $B$ replaces the matrices $A$ on $K$ consecutive sites.}
\label{tab:aklt}
\begin{tabular}{|c|c|}
\hline
$K$ & $E_{\text{magnon}}$ \\
\hline
1 & 0.370370370370370\\
2 & 0.350634581086138\\
3 & 0.350165202217295\\
4 & 0.350129173076820\\
5 & 0.350124768941854\\
6 & 0.350124225439426\\
7 & 0.350124164567491\\
8 & 0.350124158096952\\
9 & 0.350124157417523\\
10 & 0.350124157346044\\
11 & 0.350124157338490\\
12 & 0.350124157337687\\
\hline
\end{tabular}
\end{table}

For systems where the ground state is not an exact MPS, it can of course also be useful to combine the scaling in $D$ with a scaling in $K$, in particular for approximating excitations of which we expect that they might be spread out over a large number of sites, such as bound states with a very weak binding energy.

For dynamic correlation functions, it can also be useful to work in the larger subspace $\mathbb{T}_{p,K}$, in particular when trying to evaluate the Green's function of operators $\hat{O}^{(\alpha)}$ which have support on more then one site. As long as the size of the support of $\hat{O}^{(\alpha)}$ is smaller than or equal to $K$, the subspace $\mathbb{T}_{p,K}$ can capture $\hat{O}^{(\alpha)}_{p}$ [defined in Eq.~\eqref{eq:fourieroperator}] exactly and no spectral weight is lost if $\hat{H}$ is being replaced by the effective Hamiltonian on $\mathbb{T}_{p,K}$ in Eq.~\eqref{eq:defgreen}. If, however, $K$ is smaller than the number of sites on which $\hat{O}^{(\alpha)}$ acts, then a truncation occurs and part of the spectral weight will be lost in the approximation.

In order to be able to work with the variational subspace $\mathbb{T}_{p,K}$ efficiently, we need to be able to transfer some of the techniques of parameterizing tangent vectors to this larger space. For example, we can now discuss the null space $\mathbb{N}_{p,K}$ of the linear map $\Phi_{p,K}:\mathbb{C}^{D \times d^K \times D} \to \hilbert$. While we cannot infer the form of the null modes from the principal fibre bundle construction of $\varM$, we can easily generalizing the map $\mathscr{N}_{p}^{(A)}$ defined in Eq.~\eqref{eq:defnphip} and just check explicitly that $\ket{\Phi_{p,K}(B)}=0$ for any $B$ which is of the form
\begin{displaymath}
B^{s_1s_2\ldots s_{K-1} s_{K}}=A^{s_1} x^{s_2\ldots s_{K-1} s_{K}} -\ec^{-\ic p K}  x^{s_1s_{2}\ldots s_{K-1}} A^{s_{K}}, \quad \forall X\in\mathbb{C}^{D\times d^{K-1}\times D}.
\end{displaymath}
The dimension of the null space $\mathbb{N}_{p,K}$ of $\Phi_{p,K}$ is thus given by the number of linearly independent tensors $B$ that can be obtained from this construction. Expressing that $B=0$, multiplying this equation with $(A^{s_1}\cdots A^{s_{K}})^{\dagger} l$ and summing over all physical indices results in the condition $\rbra{l} \voperator{E}^{x}_{A\cdots A}(\one-\ec^{-\ic K p} \voperator{E})=0$. We have introduced the shorthand notation
\begin{displaymath}
\voperator{E}^{x}_{A\cdots A}=\sum_{s_1=1}^{d} \cdots \sum_{s_{K-1}=1}^{d} x^{s_1\cdots s_{K-1}}\otimes \overline{A}^{s_1} \cdots \overline{A}^{s_{K-1}}. 
\end{displaymath}
Clearly, this condition cannot be satisfied for $p\neq 0$, while for $p=0$ the one-dimension solution space is given by $x^{s_1\ldots s_{K-1}}\sim A^{s_1} \cdots A^{s_{K-1}}$. However, expressing that $\ket{\Phi_{p,K}(B)}\perp \ket{\Psi(A)}$ imposes an additional condition at momentum zero, so that the subspaces $\mathbb{T}_{p,K}^{\perp}$ which are orthogonal to the ground state have a physical dimension $(d-1)d^{K-1} D^2$. 

One can then also show that the a complement to the null space $\mathbb{N}_{p,K}$ can be constructed as the subspace $\mathbb{B}_{K}$ of solutions to either the left or right gauge fixing condition
\begin{align}
\sum_{s=1}^{d} A^{s\dagger} l B^{s s_{2} \cdots s_{K}}&=0,\quad \forall s_{2},\ldots,s_{K}=1,\ldots,d&&\text{or}& \sum_{s=1}^{d} B^{s_{1} s_{2}\cdots s_{K-1} s} r A^{s\dagger}&=0,\quad \forall s_{1},\ldots,s_{K-1}=1,\ldots,d.\label{eq:generalizedgaugefixing}
\end{align}
We can then build a linear representation $B^{s_1 s_2\ldots s_{K}}=\mathscr{B}_{K}(X)=l^{-1/2} V_{L}^{s_1} X^{s_2\ldots s_{K}} r^{-1/2}$, where the free parameters are represented by a tensor $X\in\mathbb{C}^{(d-1)D \times d^{K-1} \times D}$, which automatically satisfies the left gauge fixing condition and for which 
\begin{equation}
\braket{\Phi_{p,K}(\overline{\mathscr{B}}_{K}(\overline{X}))|\Phi_{p',K}(\mathscr{B}_{K}(X')}=2\pi \delta(p-p') \sum_{s_{2}=1}^{d}\cdots\sum_{s_{K}=1}^{d}\tr\left[X^{s_2\ldots s_{K}\dagger} X^{'s_2\ldots s_{K}}\right].
\end{equation}
Hence, in terms of the degrees of freedom in $X$, the effective normalization matrix is the unit matrix. We can then define the effective Hamiltonian by generalizing the computation from Section~\ref{s:excitations} and diagonalize it iteratively, to obtain a method that scales as $\order(d^{K} D^3)$. Clearly, the exponential scaling in $K$ limits the block sizes that we can achieve with this approach. However, if for describing some excitation we would like to have access to larger block sizes, the tensor $B$ itself can also be written as matrix product decomposition $B^{s_{1}s_{2}\ldots s_{k}}=B^{s_{1}}(1) B^{s_{2}}(2)\cdots B^{s_{K}}(K)$.

Finally, we elaborate on the obvious statement that $\mathbb{T}^{\perp}_{p}=\mathbb{T}^{\perp}_{p,1}\subset \mathbb{T}^{\perp}_{p,2}\subset \cdots \subset \mathbb{T}^{\perp}_{p,K}$. We can represent any $\ket{\Phi_{p,K-1}(\tilde{B})}$ as a vector $\ket{\Phi_{p,K}(B)}\in\mathbb{T}^{\perp}_{p,K}$. Note that we are restricting our discussion to the tangent spaces orthogonal to the uMPS $\ket{\Psi(A)}$, since this is a restriction that we will always want to make in applications. One way to represent this embedding in the parameterization is by setting $B^{s_1\ldots s_{K-1} s_{K}}=\tilde{B}^{s_1\ldots s_{K-1}} A^{s_K}$, which has the favorable effect that if $\tilde{B}$ satisfies the left gauge fixing condition [first equation in Eq.~\eqref{eq:generalizedgaugefixing}], then so will $B$. Now, for any $K\geq 2$, we can also try to define the space $\mathbb{V}_{p,K}$ as the orthogonal complement of $\mathbb{T}^{\perp}_{p,K-1}$ in $\mathbb{T}^{\perp}_{p,K}$, \textit{i.e.}~$\mathbb{V}_{p,K}$ contains all states $\ket{\Phi_{p,K}(C)}$ which are orthogonal to all vectors $\ket{\Phi_{p,K-1}(B)}$. Note that we can easily parameterize $\mathbb{V}_{p,K}$, by setting
\begin{displaymath}
C^{s_1,\ldots,s_K} = \mathscr{V}^{s_1,\ldots,s_K}(Y) =l^{-1/2} V_{L}^{s_1} Y^{s_2,\ldots,s_{K-1}} V_{R}^{s_{K}} r^{-1/2}
\end{displaymath}
where $Y\in\mathbb{C}^{(d-1)D \times d^{K-2} \times (d-1)D}$. Here, we have introduced the set of $(d-1)D\times D$ matrices $V_R^s$ ($s=1,\ldots,d$) which is such that $\sum_{s=1}^{d} V_{R}^{s\dagger} r^{1/2} A^{s} =0$ and $\sum_{s=1}^{d} V_{R}^{s\dagger} V_{R}^{s}=\one_{(d-1)D}$. This matrix can be created analogously to the construction for $V_{L}$ as discussed in the context of Eq.~\eqref{eq:defBrepresentation}. One would need $V_R$ to build a parameterization of tangent vectors that automatically satisfies the right gauge-fixing condition. Note that the number of parameters in $Y$ corresponds to the dimension of $\mathbb{V}_{p,K}$. Indeed, since $\mathbb{T}_{p,K}^{\perp}$ has a dimension $d^{K-1} (d-1)D^2$, the dimension of $\mathbb{V}_{p,K}$ is given by $ d^{K-1} (d-1)D^2 -  d^{K-2} (d-1)D^2= d^{K-2}(d-1)^2 D^2$. It can also by checked that $\braket{\Phi_{p,K-1}(B)|\Phi_{p,K}(C)}=0$ for any $C=\mathscr{V}(Y)$, and for any $\ket{\Phi_{p,K-1}(B)}\in\mathbb{T}^{\perp}_{p,K-1}$. In addition, we also obtain
\begin{equation}
\braket{\Phi_{p,K}(\overline{\mathscr{V}}(\overline{Y}))|\Phi_{p',K}(\mathscr{V}(Y')}=2\pi\delta(p-p') \tr[Y^{\dagger} Y']
\end{equation}
so that this parameterization corresponds to an effective normalization matrix which is the unit matrix. There is at least one simple application of this construction. The error measure $\epsilon(\overline{A},A)$ was defined in Eq.~\eqref{eq:defEpsilon} as the norm of the part of $[\hat{H}-H(\overline{A},A)]\ket{\Psi(A)}$ that was lost in the projection onto the MPS tangent space $\mathbb{T}_{p}$. If $\ham$ is a nearest neighbor Hamiltonian $\ham=\sum_{n\in\mathbb{Z}} \hat{T}^{n} \hat{h} \hat{T}^{-n}$, with $\hat{h}$ acting on two sites, we can exactly represent $[\hat{H}-H(\overline{A},A)]\ket{\Psi(A)}$ in $\mathbb{T}^{\perp}_{p=0,K=2}$. Hence the error measure corresponds to the norm the projection of $[\hat{H}-H(\overline{A},A)]\ket{\Psi(A)}$ onto $\mathbb{V}_{p=0,K=2}$. Using the parameterization $\mathscr{V}$, we can easily obtain that $\hat{P}_{\mathbb{V}_{p=0,K=2}}[\hat{H}-H(\overline{A},A)]\ket{\Psi(A)} =\ket{\Phi_{p=0,K=2}(\mathscr{V}(Y))}$ where $Y$ is given as
\begin{equation}
Y=\sum_{s,t,u,v=1}^{D}\braket{s,t|\hat{h}|u,v} V_{L}^{s\dagger} l^{1/2}  A^{u} A^{v} r^{1/2} V_{R}^{t\dagger}.\label{eq:projH2sitespace}
\end{equation}
We then obtain for the localized error measure $\tilde{\epsilon}(\overline{A},A)=\lVert \mathbb{Z}\rVert^{-1/2}\epsilon(\overline{A},A) = \lVert \mathbb{Z}\rVert^{-1/2} \lVert \ket{\Phi_{p=0,K=2}(\mathscr{V}(Y))}\rVert = \sqrt{\tr[Y^{\dagger} Y]}=\lVert Y\rVert_{\text{F}}$. Hence, we do not have to compute a difference of terms which are almost equal in magnitude. This is a a much quicker and more stable algorithm for computing $\tilde{\epsilon}(\overline{A},A)$. If $\ham$ contains only local interactions up to $K$ sites, we can easily compute the norm of the projection of $[\hat{H}-H(\overline{A},A)]\ket{\Psi(A)}$ onto all spaces $\mathbb{V}_{p=0,k}$ for $k=2,\ldots,K$. The local error measure $\tilde{\epsilon}(\overline{A},A)$ is then obtained as the square root of the sum of the squares of all these norms, and is thus computed as a sum of positive values, without any risk for contributions that neutralize each other. 

\subsection{Dynamic expansion of the manifold}
\label{ss:extensions:dynamic}
For studying time-evolution, the MPS tangent space appeared naturally. If we want to approximate a time-evolving quantum state as a time-evolving MPS with bond dimension $D$, the time derivative of the state has to be an element of the MPS tangent space at that particular point, and the TDVP is prescribing which tangent vector is optimal. 

Nevertheless, it also interesting to ask oneself would it would imply to try to evolve a uMPS $\ket{\Psi(A)}$ in a direction given by a state $\ket{\Phi_{0,K}(B)}\in\mathbb{T}_{0,K}^{\ket{\Psi(A)}}$ for $K>1$. In particular, for nearest neighbor Hamiltonians $\ham=\sum_{n\in\mathbb{Z}} \hat{T}^{n} \hat{h} \hat{T}^{-n}$ with $\hat{h}$ acting on two sites, we can exactly represent $\hat{H}\ket{\Psi(A)}$ in $\mathbb{T}_{p=0,K=2}$. The way to go beyond to tangent space is by noting that for a variational class such as MPS, which has a refinement parameter $D$, it is not necessary to have a fixed value for $D$ throughout the whole evolution. Suppose that at some point during the evolution, the error measure $\tilde{\epsilon}(\overline{A},A)$, which measures the difference between the TDVP evolution and the exact evolution given by the Schr\"{o}dinger equation, exceeds some predefined tolerance value. We then might try to reduce the error by expanding the variational manifold, by increasing the bond dimension from its original value $D$ to some new value $\widetilde{D}$. We can easily embed a uMPS $\ket{\Psi(A)}\in\varM$ with bond dimension $D$ into the uMPS manifold $\widetilde{\varM}$ with bond dimension $\widetilde{D}$ by writing it as $\ket{\widetilde{\Psi}(\widetilde{A})}$ with
\begin{equation}
\widetilde{A}^{s}=\begin{bmatrix} A^{s} &0\\ 0 &0\end{bmatrix}.
\end{equation}
Note, however, that $\widetilde{A}$ is not in the set of injective MPS $\widetilde{\manifold{A}}$, which has some undeniable consequences. For example, if we follow the TDVP spirit in the most strict sense, we set $\widetilde{A}(t=0)=\widetilde{A}$ and
\begin{equation}
\frac{\d\ }{\d t}\widetilde{A}(t) = \widetilde{B} =\begin{bmatrix} B^{s}_{1,1} & B_{1,2}^{s}\\ B_{2,1}^{s} & B_{2,2}^{s}\end{bmatrix},
\end{equation}
with $B^{s}_{1,1}$ a $D\times D$ matrix, $B^{s}_{1,2}$ a $D\times(\widetilde{D}-D)$ matrix, $B^{s}_{2,1}$ a $(\widetilde{D}-D)\times D$ matrix and $B^{s}_{2,2}$ a $(\widetilde{D}-D)\times (\widetilde{D}-D)$ matrix. We then find the $\widetilde{B}$ that minimizes $\lVert \ket{\widetilde{\Phi}(\widetilde{B},\widetilde{A})}+\ic \ham\ket{\widetilde{\Psi}(\widetilde{A})}\rVert$. By construction, $\ket{\widetilde{\Psi}(\widetilde{A})}=\ket{\Psi(A)}$. However, we also find $\ket{\widetilde{\Phi}(\widetilde{B},\widetilde{A})}=\ket{\Phi(B_{1,1},A)}$ and the newly added directions $B_{2,1}$, $B_{1,2}$ and $B_{2,2}$ do not feature to first order. The TDVP is very keen to restricting its evolution to the original manifold $\mathcal{M}$. Put differently, because $\widetilde{A}$ is not in the set of injective MPS $\widetilde{\mathscr{A}}$, it corresponds to a singular region where the tangent space $\widetilde{\mathbb{T}}^{\ket{\widetilde{\Psi}(\widetilde{A})}}_{0}$ does not have dimensions $(d-1)\widetilde{D}^2$ but rather $(d-1)D^2$. 

A solution for this problem is to use a kind of singular perturbation theory. We set $\widetilde{A}(\d t)=\widetilde{A}+\widetilde{\d A}$ with
\begin{equation}
\widetilde{\d A}^{s}=\begin{bmatrix} \d t B^{s}_{1,1} & (\d t)^{1/2} B_{1,2}^{s}\\ (\d t)^{1/2} B_{2,1}^{s} & B_{2,2}^{s}\end{bmatrix},\end{equation}
We then obtain to first order in $\d t$ that 
\begin{equation}
\ket{\widetilde{\Psi}(\widetilde{A}+\widetilde{\d A})}-\ket{\widetilde{\Psi}(\widetilde{A})}=\ket{\widetilde{\d \Psi}_{0}}+\ket{\widetilde{\d \Psi}_{1}}+\order(\d t^2)
\end{equation}
with
\begin{equation}
\ket{\widetilde{\d \Psi}_{0}}=\d t\ket{\Phi_{p=0,K=1}(B_{1,1};A)}\label{eq:dpsi1}
\end{equation}
and
\begin{equation}
\ket{\widetilde{\d \Psi}_{1}}=\d t
\sum_{n\in\mathbb{Z}}\sum_{\{s_{n}\}=1}^{d}\sum_{m=0}^{+\infty}\bm{v}_{\mathrm{L}}^{\dagger}\bigg(\prod_{k<n} A^{s_k}\bigg) B_{1,2}^{s_n} B_{2,2}^{s_{n+1}}\cdots B_{2,2}^{s_{n+m}} B_{2,1}^{s_{n+m+1}} \bigg(\prod_{k>n+m+1} A^{s_k}\bigg)\bm{v}_{\mathrm{R}} \ket{\bm{s}}.\label{eq:dpsi2}
\end{equation}
In particular, if we set $B_{2,2}=0$, we obtain $\ket{\widetilde{\d \Psi}_{1}}=\d t\ket{\Phi_{p=0,K=2}(B;A)}$ with $B^{s_1s_2}=B_{1,2}^{s_1} B_{2,1}^{s_2}$. Since we already have $\ket{\widetilde{\d\Psi}_{1}}\in\mathbb{T}_{p=0,K=1}$, we can restrict to $\ket{\widetilde{\d\Psi}_{2}}\in\mathbb{V}_{p=0,K=2}$, so that we can parameterize
\begin{align}
B_{1,2}^s&=l^{-1/2} V_{L}^{s} Z_{1,2},&&\text{and}&B_{2,1}^{s}&=Z_{2,1} V_{R}^{s} r^{-1/2},
\end{align}
where $Z_{1,2}\in\mathbb{C}^{(d-1)D\times (\widetilde{D}-D)}$ and $Z_{2,1}\in\mathbb{C}^{(\widetilde{D}-D)\times(d-1)D}$. We then automatically have $\braket{\widetilde{\d \Psi}_{1}|\widetilde{\d \Psi}_{2}}=0$. If we now try to imitate the geometric strategy of the time-dependent variational principle, we have to optimize the parameters $B_{1,1}$, $B_{1,2}$ and $B_{2,1}$ so as to minimize
\begin{equation}
\big\lVert \ket{\widetilde{\d \Psi}_{0}}+\ket{\widetilde{\d \Psi}_{1}} +\ic \d t \ham \ket{\widetilde{\Psi}(\widetilde{A})}\big\rVert^{2}.\label{eq:dynexpgeom}
\end{equation}
Because of the orthogonality of $\ket{\widetilde{\d\Psi}_{1}}$ and $\ket{\widetilde{\d \Psi}_{2}}$, the optimization of the parameters $B_{1,1}$ decouples from the optimization of the parameters $B_{1,2}$ and $B_{2,1}$. For $B_{1,1}$, we can use the standard TDVP prescription, \textit{i.e.} if we use the left-gauge fixing condition then $B_{1,1}=\mathscr{B}(F)$ as described in Subsection~\ref{ss:tdvp:implementation}. For the remaining parameters, we need to extremize
\begin{multline*}
\braket{\widetilde{\d \Psi}_{1}|\widetilde{\d \Psi}_{1}}+\ic \d t \braket{\widetilde{\d \Psi}_{1}|\ham|\Psi(A)}-\ic \d t \braket{\Psi(\overline{A})|\ham|\widetilde{\d \Psi}_{1}}=\\
(\d t)^2 \left\{\tr\big[(Z_{1,2}Z_{2,1})^\dagger (Z_{1,2}Z_{2,1})\big]+\tr\big[(Z_{1,2}Z_{2,1})^\dagger Y\big]+\tr\big[Y^\dagger (Z_{1,2}Z_{2,1})\big]\right\}
\end{multline*}
where $Y$ was defined in Eq.~\eqref{eq:projH2sitespace} for the case of a Hamiltonian with nearest neighbor interactions. We are thus looking for the optimal matrix $Z_{1,2}Z_{2,1}$ of rank $\widetilde{D}-D$ such that $\lVert Z_{1,2}Z_{2,1} - Y \rVert_{\text{F}}$ is minimized, with $\lVert \cdot \rVert_{\text{F}}$ the Frobenius norm or Hilbert-Schmidt norm. The best approximation can thus found by performing a singular value decomposition of the $(d-1)D\times (d-1)D$ matrix $Y$ and retaining the largest $\widetilde{D}-D$ singular values. If $\widetilde{D}=d D$, we can exactly capture one time step of the Schr\"{o}dinger evolution for a nearest neighbor Hamiltonian exactly. This compares well to TEBD, where an exact application of the lowest order Trotter decomposition of a nearest neighbor Hamiltonian also increases the bond dimension from $D$ to $\widetilde{D}=dD$.

Clearly, we can also generalize this construction to be able to evolve according to states in $\mathbb{T}_{p=0,K}$, by using
\begin{equation}
\widetilde{\d A}=(\d t)^{1/K} \begin{bmatrix} 0 & B_{1,2} & 0 & \hdots & 0\\
0 & 0 & B_{2,3} & \hdots & 0\\
\vdots & \vdots & \vdots & \ddots & \vdots \\
0 & 0 & 0 & \hdots & B_{K-1,K}\\
B_{K,1} & 0 & 0 & \hdots & 0\end{bmatrix}
\end{equation}
in which case $\ket{\widetilde{\Psi}(\widetilde{A}+\widetilde{\d A})}-\ket{\widetilde{\Psi}(\widetilde{A})}=\d t \ket{\Phi_{p=0,K}(B;A)}+\order(\d t^2)$ with $B^{s_1s_2\ldots s_K}=B_1,2^{s_1} B^{s_2}_{2,3}\cdots B^{s_K}_{K-1,K}$. 
\subsection{Towards a Fock space construction}
\label{ss:extensions:fock}
As a variational ansatz, the generalized subspaces $\mathbb{T}_{p,K}$ are still only useful for studying elementary excitations with a single-particle like nature, which might include bound states of several fundamental particles of the system. If we want to be able to describe true unbound multi-particle excitations, we need a different variational ansatz that supports two independent perturbations of the ground state that can be arbitrarily far apart. There is not much interesting information in the energy of multi-particle states, as we expect that in a system with local interactions, we should be able to add the momentum (modulus $2\pi$) and energy of any two eigenstates and find a new eigenstate at that point in the energy-momentum diagram. For finite systems, there can be finite size corrections to the energy, but this relation should hold true exactly in the thermodynamic limit, as long as the number of excitations remains finite (\textit{i.e.}~the density of excitations is zero). However, being able to describe multi-particle excitations is relevant for a more accurate description of dynamic correlation functions, as well as to understand the scattering properties of elementary particles and thus the interactions that exist between them. Eventually, the goal is to create a new Fock space on top of the interacting MPS vacuum, in which we can build an effective Hamiltonian that describes the low-energy behavior of the system. This effective Hamiltonian can then be used to study the thermodynamics of the system, \textit{i.e.} the behavior of the system under all kinds of perturbations, such as adding a nonzero temperature, applying external fields, quenching some parameters. 

We now assume that we have a complete description of all elementary particles in the system, labeled by some index $\alpha$, in terms of states $\ket{\Phi_{p,K}(B^{(\alpha)})}$, which are normalized as $\braket{\Phi_{p,K}(\overline{B}^{(\alpha)})|\Phi_{p',K}(B^{(\beta)})}=2\pi\delta(p-p')\delta_{\alpha,\beta}$. In general, the tensors $B^{(\alpha)}$ can depend on the momentum $p$. However, we expect the momentum dependence to be weak if we expand around minima in the dispersion relation $\omega^{(\alpha)}(p)$, and we ignore this subtlety in the present discussion. The span of these states thus defines our single particle space. The logical next step is the definition of two-particle space. The double tangent space $\mathbb{T}^{(2)}$ that we encountered in the context of Subsection~\ref{ss:excitations:reltdvp} seemed to have the correct structure to capture two independent disturbances of the ground state. We can now generalize this for tensors $B$ acting on $K$ consecutive sites by first defining for every $n_2>n_1+K$ the states
\begin{equation}
\ket{\Upsilon_{n_1,n_2,K}(B_1,B_2)}= \bm{v}_{\mathrm{L}}^{\dagger} \bigg(\prod_{k<n_1} A^{s_{k}}\bigg) B_{1}^{s_{n_1}\cdots s_{n_1+K-1}}\bigg(\prod_{n_1+K\leq k<n_2} A^{s_{k}}\bigg) B_2^{s_{n_2}\ldots s_{n_2+K-1}}\bigg(\prod_{n_2+K\leq k} A^{s_k}\bigg)\bm{v}_{\mathrm{R}}\ket{\bm{s}}.
\end{equation}
To construct a full basis of states that is complete but not overcomplete, we have started from the set of states where the perturbations encoded by $B_1$ and $B_2$ are localized on sites $n_1$ and $n_2$ respectively. It is clear that we can only define these states for $n_2>n_1+K$, or, alternatively, if we would like to put $B_2$ on a position $n_2<n_1-K$, we should just write it as $\ket{\Upsilon_{n_2,n_1,K}(B_2,B_1)}$. Hence, we do not expect to be able to label the double tangent space by two completely independent momentum numbers without overcounting. However, by setting $n_1=n$ and $n_2=n+ m$, we can take the Fourier transform with respect to $n$ and obtain a new basis of states with a well defined total momentum (for $m\geq K$)
\begin{multline}
\ket{\Upsilon_{p,m,K}(B_1,B_2)}= \sum_{n\in\mathbb{Z}} \ec^{\ic p n} \bm{v}_{\mathrm{L}}^{\dagger} \bigg(\prod_{k<n} A^{s_{k}}\bigg) B_{1}^{s_n\cdots s_{n+K-1}}\bigg(\prod_{n+K<k<m} A^{s_{k}}\bigg) B_2^{s_{n_2}\ldots s_{n_2+K}}\bigg(\prod_{n_2+K<k} A^{s_k}\bigg)\bm{v}_{\mathrm{R}}\ket{\bm{s}}.
\end{multline}
For fixed $p$ and for all $m\geq K$, these states span a space $\mathbb{T}_{p,K}^{(2)}$. They should be considered as the product state basis, because by taking linear combinations of states $\ket{\Upsilon_{p,m,K}(B_1,B_2)}$ with different $B_1$'s and $B_2$'s, we will create `entanglement' between the two excitations.
\end{widetext}

So far, we have only fixed the total momentum, whereas ideally, we would like to put both excitation in an individual momentum superposition by building a state such as
\begin{equation}
\ket{p-q,q,\alpha,\beta}=\sum_{m>K} \ec^{\ic q m}  \ket{\Upsilon_{p,m,K}(B_1^{(\alpha)},B_2^{(\beta)}}.
\end{equation}
The asymptotic part of this state ($m\gg K$) looks like an excitation $B^{(\alpha)}$ with momentum $p-p_2$ and an excitation $B^{(\beta)}$ with momentum $q$. One can indeed show that the energy of this state is dominated by the asymptotic part and is given by $\omega^{(\alpha)}(p-q)+\omega^{(\beta)K}(q)$. However, this is in itself not a good eigenstate, since it is not even allowed to transform the (internal) position coordinate $m$, which is only labeling sites on a half infinite lattice $m>K$, to momentum space. A proper approximation for a stationary scattering state of the full Hamiltonian has an asymptotic contribution of every other pair of particles with total momentum $p$ and the same energy, in particular from the state $\ket{q,p-q,\beta,\alpha}$. The coefficients with respect to the different states appearing in the asymptotic region determine the content of the scattering matrix. In one dimension, the scattering matrix is determined by the simultaneous effect of interactions and statistics and there is no unambiguous way to distinguish between both. Free particles with one type of statistics have an equivalent description as interacting particles with a different type of statistics. In any way, to get an accurate approximation of the scattering coefficients, we also need an accurate description of the scattering state  in the regime where the two excitations are close to each other and are interacting with each other. Given that the $\mathbb{T}_{p,K}^{(2)}$ corresponds to a half infinite lattice, where every site contains the internal degrees of freedom of the two excitations $B_1$ and $B_2$, the subspace $\mathbb{T}_{p,K}^{(2)}$ is infinite dimensional and there is no straightforward way to diagonalize the Hamiltonian. We discuss how to tackle this scattering problem elsewhere\cite{Vanderstraeten:uq}.

On an intuitive level, it is clear that putting more and more $B$-tensors into the uMPS $\ket{\Psi(A)}$ corresponds to having more particles. However, several complications have to be taken into account. In order to build a consistent framework, we have to check that the vacuum and the elementary excitations are good approximations of true eigenstates of the Hamiltonian. This requires that \textit{e.g.} the Hamiltonian does not (or only very weakly) couples the two-particle space with the single-particle space (or even with the vacuum state, which is expressed by the matrices $K(p)$ in Eq.~\eqref{eq:linearizedtdvpfourier2}. If this is the case, then we can start building a Fock space in which we can construct an effective Hamiltonian describing the interactions between the elementary excitations of the system. The two-particle interaction is computed in already computed in Ref.~\onlinecite{Vanderstraeten:uq} in order to solve the scattering problem. There can also be three-particle couplings, or couplings that do not respect the particle number and transform three excitations into two excitations, and so forth. To know how these excitations couple to physical perturbations of the system, it is also required to identify the excitations with the action of physical operators, such that $\ket{\Phi_{p,K}(B^{(\alpha)}}=\hat{A}^{(\alpha)\dagger}_{p}\ket{\Psi(A)}$. It can be shown that, if $K$ is equal to or larger than the injectivity length $\ell$\cite{2006quant.ph..8197P}, then we can always find an operator $\hat{A}^{(\alpha)\dagger}_{p}$ that creates the excitation by acting on exactly $K$ sites. If $K<\ell$, the corresponding operator might have to act on $\ell$ sites. If we want to identify $\hat{A}^{(\alpha)\dagger}_{p}$ with a creation operator, we should also require that $\hat{A}^{(\alpha)}_{p}$ annihilates the ground state. This requires that $\voperator{E}^{B}_{AA\cdots A}$ (with $K$ matrices $A$) is zero, which is completely different and not compatible with the kind of gauge-fixing conditions that we have been imposing so far. It remains an open question whether we can build such a theory consistently and efficiently. 

\section{Conclusions and outlook}
This manuscript discusses in great detail some recent algorithms to study the dynamics of quantum lattice systems that rely on the concept of the tangent space to the manifold of uMPS. The need for understanding the uMPS tangent space arises naturally when applying the TDVP to the manifold of uMPS, in order to capture time-evolving quantum states within this manifold. However, it then turns out that the tangent space is also very convenient in the formulation of `post-MPS' methods for studying elementary excitations of the quantum lattice Hamiltonian and to compute spectral functions. This situation is reminiscent to mean field theory for fermions (Hartree-Fock) and bosons (Gross-Pitaevskii), where similar constructions have been well developed and used extensively.

However, unlike mean field theory, (u)MPS have a refinement parameter $D$ that can be used to improve the ground state approximation if necessary, and we have in this manuscript refrained from studying whether `post-MPS' also imply corrections to the ground state. Rather, we assume the uMPS description of the ground state to be `quasi'-exact, and we want to obtain a better understanding of the elementary excitations in the system, and the interactions that exist between them. We can easily generalize the tangent space ansatz in order to obtain a more accurate description of elementary excitations, and have given an outlook on how we can start building a new Fock space on top of the MPS vacuum, in which an effective Hamiltonian can be created that describes the quantum dynamics of the elementary excitations in the system. The formalism in this paper should also be applicable to the case of continuous MPS, as well as to other tensor networks for higher dimensional quantum systems. This program should enable us in the long term to construct an effective description of the low-energy behavior of strongly interacting quantum systems, which can be used to describe the complete thermodynamics in the quantum regime.

Near completion of this work, we learned about Ref.~\onlinecite{papersebastian} where related concepts were introduced for MPS on finite lattices and were compared in detail to similar constructions for Hartree-Fock theory.  The contents of this paper and its relation to our manuscript were briefly discussed in Section~\ref{s:analogies}.

\acknowledgements{Inspiring discussions with Ignacio Cirac, Damian Daxler, Andrew Ferris, Ashley Milsted, Bogdan Pirvu, Vid Stojevic, Luca Tagliacozzo, Laurens Vanderstraeten, David Weir, Sebastian Wouters and Valentin Zauner are greatly acknowledged. This work is supported by an Odysseus grant from the FWO Flanders, the FWF grants FoQuS and Vicom, the ERC grants QUERG and QFTCMPS and by the cluster of excellence EXC 201 Quantum Engineering and Space-Time Research.}

\bibliography{paperslibrary,manuallibrary}

\end{document}